\RequirePackage[table]{xcolor}
\documentclass[acmsmall,nonacm]{acmart}
\setcopyright{none}

\usepackage{float}
\usepackage{etex}
\usepackage[utf8]{inputenc}
\usepackage[T1]{fontenc}
\usepackage{multicol}
\usepackage{algpseudocode}
\usepackage{algorithmicx}
\usepackage{moreverb}
\usepackage{mdframed}
\usepackage{graphicx}
\usepackage{placeins}

\newcommand{\tagarray}{\mbox{}\refstepcounter{equation}$(\theequation)$}

\algrenewcommand\algorithmicindent{1.0em}

\DeclareFontEncoding{LS1}{}{}
\DeclareFontSubstitution{LS1}{stix}{m}{n}
\DeclareSymbolFont{arrows1}       {LS1}{stixsf}   {m} {n}
\DeclareMathDelimiter{\Ddownarrow} {\mathrel}{arrows1}{"60}{arrows1}{"60}

\usepackage{syntax}

\makeatletter
\renewcommand{\ulitleft}{\normalfont\bf\ttfamily\syn@ttspace\frenchspacing}
\renewcommand{\ulitright}{}
\renewcommand{\litleft}{`\bgroup\ulitleft}
\renewcommand{\litright}{\ulitright\egroup'}
\makeatother
\usepackage{adjustbox}
\usepackage{multirow,array}
\usepackage{multicol}

\usepackage{subfig}
\usepackage{pgfplots}
\usepackage{pgfplotstable}
\usepackage{tikz}
\usetikzlibrary{shapes.geometric}
\usetikzlibrary{arrows}
\pgfplotsset{compat=1.12}

\usepackage{listings}
\usepackage{xcolor}
\usepackage{paralist}
\usepackage{xspace}
\usepackage{booktabs}
\usepackage{dcolumn}
\newcolumntype{d}{D{.}{.}{0}}

\usepackage{mathtools}
\usepackage{tikz-qtree}
\usepackage{csvsimple}
\usepackage[binary-units]{siunitx}

\usepackage{cleveref}

\newcolumntype{R}[2]{>{\adjustbox{angle=#1,lap=\width-(#2)}\bgroup}l<{\egroup}}

\newcommand{\cell}[1]{{\parbox{1.0cm}{\centering\ #1}}}

\newcommand{\bigCell}[1]{{\parbox{1.2cm}{\centering\ #1}}}

\newcommand{\vCell}[1]{{\parbox{1.9cm}{\centering\ #1}}}

\DeclareSIUnit[number-unit-product={}]{\percent}{\%}
\DeclareSIUnit{\Bit}{Bit}

\newcommand{\result}[3]{
	{	
		\begin{table}[t]
      	\footnotesize
		\caption{#3}
        \label{table:#2}	
		\centering
			\csvreader[head to column names,
			tabular={lSSSS[table-format=4.4,table-align-text-post=false]},
			late after line= \ifcsvprostrequal{\program}{Total}{\\\midrule}{\\},
			table head=\toprule Program &
			 {{\itshape \scmc} (\si{\milli\second})} &
             \vCell{{\itshape \scmc}\\ Solver (\si{\milli\second})}  &
			 {{\itshape \matchc} (\si{\milli\second})} & {Ratio}\\\midrule,			
			table foot=\bottomrule]{#1}{}			
			{\ifthenelse{\equal{\type}{g}}{\cellcolor{lightgray}
                                  group
                                  {\ttfamily \program}}{\ifcsvprostrequal{\program}{Total}{Total}{{\ttfamily\program}}& \totalNI & \solverTimeNI & \matchC & 
				 \ratio\si{\percent}
				}
			} \end{table}
	}		
}

\newcommand{\aincrementalResult}[3]{
	{
		\begin{table}[t]
			\caption{#3\label{table:#2}}
			{\scriptsize\centering
				\csvreader[head to column names,
				tabular=>{\small}lrrrrrrr,
                                late after line= \ifcsvprostrequal{\program}{Average}{\\\midrule}{\\},
				table head=\toprule Program &
				\cell{avg. initial  (\si{\milli\second})} &
				\cell{avg. \mbox{w/o incr.}  (\si{\milli\second})} &
\cell{avg.  incr.  (\si{\milli\second})} &
\cell{\mbox{avg. solv.} initial  (\si{\milli\second})} &
				\cell{\mbox{avg. solv.} \mbox{w/o incr.}  (\si{\milli\second})} &
				\cell{\mbox{avg. solv.}  incr.  (\si{\milli\second})} &
				\cell{Speedup}\\\midrule,
				table foot=\bottomrule]{#1}{}
				{\ifcsvprostrequal{\program}{Average}{Average}{{\ttfamily\program}}& \init & \ni & \i & \inits & \nis & \is & \speedup}}				
		\end{table}
	}		
}

\newcommand{\incrementalResult}[3]{
	{
		\begin{table}[t]
			\caption{#3\label{table:#2}}
			\scriptsize\centering
				\csvreader[head to column names,
				tabular=>{\small}lrrrrrr,
                                late after line= \ifcsvprostrequal{\program}{Average}{\\\midrule}{\\},
				table head=\toprule {Program} &
				\cell{avg. w/o incr.  (\si{\milli\second})} &
\cell{avg.  incr. (\si{\milli\second})} &
\bigCell{\mbox{avg. solv.} w/o incr.  (\si{\milli\second})} &
				\bigCell{\mbox{avg. solv.} incr.  (\si{\milli\second})} &
				\cell{Speedup} \\\midrule,
				table foot=\bottomrule]{#1}{}
                                {\program & \ni & \i & \nis & \is & \speedup}			
		\end{table}
	}		
}

\newcommand{\incrementalResultSec}[3]{
	{
		\begin{table}[t]
			\caption{#3\label{table:#2}}
			\scriptsize\centering
				\csvreader[head to column names,
				tabular=>{\small}lrrrrrr,
                                late after line= \ifcsvprostrequal{\program}{Average}{\\\midrule}{\\},
				table head=\toprule {Program} &
				\cell{avg. w/o incr.  (\si{\second})} &
\cell{avg.  incr. (\si{\second})} &
\bigCell{\mbox{avg. solv.} w/o incr.  (\si{\second})} &
				\bigCell{\mbox{avg. solv.} incr.  (\si{\second})} &
				\cell{Speedup} \\\midrule,
				table foot=\bottomrule]{#1}{}
                                {\program & \ni & \i & \nis & \is & \speedup}			
		\end{table}
	}		
}

\usepackage{threeparttable}
\newcommand{\tcasIncrementalResult}[3]{
	{
		\begin{table}[t]		
			\caption{#3\label{table:#2}}	 	
                  \scriptsize\centering
                  \begin{threeparttable}			
				\csvreader[head to column names,
				tabular=>{\small}lrrrrrrrr,
                                late after line= \ifcsvprostrequal{\program}{Average}{\\\midrule}{\\},
				table head=\toprule {Version} &
				\cell{Error} &
				\cell{Change Type} &
				\cell{StCh \mbox{Position}} &
				\cell{avg. \mbox{w/o incr.}  (\si{\milli\second})} &
\cell{avg. incr.  (\si{\milli\second})} &
\bigCell{\mbox{avg. solv.} w/o incr.  (\si{\milli\second})} &
				\bigCell{\mbox{avg. solv.}  incr.  (\si{\milli\second})} &
				\cell{Speedup} \\\midrule,
				table foot=\bottomrule
				]{#1}{}{\index & \ie & \type & \change & \ni & \i & \nis & \is & \speedup}	                              
                              \begin{tablenotes}
                                \small
                              \item Description of the column
                                \textit{Change Type}:
			"a": relational operator mutation;
			"b": value change;
			"c": logic operator mutation;
			"d": remove logic expression;
			"e": arithmetic expression mutation;
			"f": array index change;
			"g": change in structure of logic expression;
			"h": ternary operand switch.
         \end{tablenotes}			
       \end{threeparttable}
		\end{table}
	}		
}

\newcommand\circleRadius{\fontsize{14.4}{20}\selectfont}
\newcommand{\parnt}[2]{Node #2}
\newcommand{\skipnt}[2]{#2}
\newcommand{\textnt}[2]{node #2}
\newcommand{\skiptn}[2]{#2}
\newcommand{\tablent}[2]{#2}

\newcommand{\snt}[2]{\(\synt{#1}\)}
\newcommand{\refsnt}[2]{{\setlength\fboxsep{0pt}\setlength{\fboxrule}{1pt}\fbox{{\setlength\fboxsep{2pt}\colorbox{lightgray}{\(\synt{#1}\)}}}}}
\newcommand{\nt}[2]{\(\overset{\text{\circleRadius \textcircled{\scriptsize #2}}}{\synt{#1}}\)}
\newcommand{\tn}[2]{\(\overset{\text{\circleRadius \textcircled{\scriptsize #2}}}{\texttt{#1}}\)}

\newcommand{\refnt}[2]{{\setlength\fboxsep{0pt}\setlength{\fboxrule}{1pt}\fbox{{\setlength\fboxsep{4pt}\colorbox{lightgray}{\(\overset{ \text{\circleRadius \textcircled{\scriptsize #2}}}{\synt{#1}}\)}}}}}

\newcommand{\spec}[3]{{\vbox{\hbox{
				\(\overset{\text{\circleRadius \textcircled{\scriptsize #3}}}{\synt{#2}}\)
}\hbox{\setlength\fboxsep{0pt}\setlength{\fboxrule}{1pt}\fbox{{\setlength\fboxsep{2pt}\colorbox{lightgray}{(#1)}}}}}}}

\newcommand{\sidecar}{SiDECAR\xspace}
\newcommand{\scmc}{SCMatchC\xspace}
\newcommand{\matchc}{\textsc{MatchC}\xspace}

\DeclareCaptionLabelFormat{continued}{
		    \ifthenelse{\equal{#1}{Table}}{#1\ }{Fig.\ }#2 (Continued)}
\usepackage[many]{tcolorbox}

\newtcbox{\ml}{enhanced,nobeforeafter,tcbox raise base,boxrule=0.4pt,top=0mm,bottom=0mm,
	right=0mm,left=4mm,arc=1pt,boxsep=2pt,before upper={\vphantom{dlg}},
	colframe=green!50!black,coltext=green!25!black,colback=green!10!white,
	overlay={\begin{tcbclipinterior}\fill[green!75!blue!50!white] (frame.south west)
			rectangle node[text=white,font=\sffamily\bfseries\tiny,rotate=90] {ML} ([xshift=4mm]frame.north west);\end{tcbclipinterior}}}

                  
\begin{document}
\markboth{}{}

\title{Syntax-driven Incremental Program Verification of Matching
  Logic Properties}

\author{Domenico Bianculli}
\affiliation{
	       \institution{University of Luxembourg}
	}
\author{Antonio Filieri}
\affiliation{
	       \institution{Imperial College London}
	}
\author{Carlo Ghezzi}
\author{Dino Mandrioli}
\author[Alessandro M. Rizzi]{Alessandro Maria Rizzi}
\affiliation{
	       \institution{Politecnico di  Milano}
	}

\begin{abstract}
Incrementality is a fundamental design principle to master the
complexity of large, long-lived software systems.  This
principle has been embraced by agile development processes and it lays at the
base of continuous software evolution.  A major challenge in this context is to incrementally re-verify the correctness of software artifacts after every change, focusing the verification efforts only on the parts affected by the change.

We present an approach to the incremental verification of programs
written in KernelC, annotated with properties
expressed in matching logic.  The approach is based on a
syntactic-semantic framework that enables analyzing
code chunks in isolation so that, after a change to a program
fragment, only the part whose semantics is affected by the change is
re-processed. This property is obtained by expressing the language
syntax through an operator precedence grammar and by formalizing its semantics through a
synthesized attribute schema. 

We have implemented our technique in a prototype tool and experimentally evaluated
its effectiveness.
The results show that our approach does
not penalize the efficiency of formal verification and can outperform
program re-verification after changes, depending on the presence and
type of annotations, as well as the position of the change and the
program structure.

 \end{abstract}


\maketitle

\section{Introduction}
\label{secIntroduction}

Incrementality is a fundamental engineering principle that manifests
itself in many situations and in different forms. To
dominate complexity, real-world problems are seldom solved monolithically, but
rather through increments. Additionally, design is often
exploratory in nature: different design decisions are explored (and detracted) before choosing the most desirable ones.  Whenever a change is made in a design
decision it is useful for the designer to just focus on the effect of the change.

We are especially interested in understanding how the effect of
changes may be analyzed incrementally through formal verification. The best way to achieve incrementality is through careful design that tries to anticipate likely changes, encapsulating their possible effects within modules, accessible through well-defined interfaces~\cite{Parnas:1972:CUD:361598.361623}, specified by following the principle of \emph{design by contract}~\cite{Meyer}.

Changes,
however, cannot always be anticipated; moreover, their effect can be
cross-cutting. It is therefore important to complement incrementality ``by design'' with an approach that, for any given change, automatically and dynamically discovers  the minimal extent of new analysis required and  re-uses as much as possible from previous analysis.

To further motivate the need for incremental formal verification, we
mention two possible important applications, which concern agile
development and verification at run time. Agile
development~\cite{AgileManifesto2010} is a widely practiced software
development approach that implements and delivers software through
increments, which may progressively add new functionalities and/or
adjust existing ones through direct control of the
stakeholders~\cite{larman03:iterat}. Development proceeds through
continuous iterations. As for verification, the current
state-of-the-art is based on continuous
testing~\cite{Muslu:2013:MOA:2491411.2491460,Saff:2004:EEC:1007512.1007523}. The
adoption of formal verification into agile development to complement
testing requires a serious effort to make verification
incremental. Incrementality may not only make the verification faster,
but also can help the designer to reason on the effects of changes.
Another area where incrementality is of paramount importance is
\emph{on-the-fly verification} or, more appropriately, \emph{verification during system operation}: in
many cases software maintenance must occur without interrupting --- or
minimizing the suspension of --- system operations.  In this case
verification of the updated version must also satisfy strong real-time
requirements, which may be solved by exploiting incrementality.

Furthermore, in some sense incremental formal verification is the formal counterpart of regression testing in the same way as formal correctness proof is the traditional alternative or complement to testing as the main verification technique. Incremental formal verification, in fact, while concentrating on the system fragment affected by the change, it also implicitly guarantees that the remaining part of the system still behaves as in the previous version.

In this paper, we use the term \emph{program verification} in a broad
sense that needs to be clarified upfront. The term is often used to
denote a proof that a program satisfies a \emph{specification}. In our
work the specification is expressed in terms of logic statements,
embedded in the program as annotations written according to a suitable
syntax. Following Hoare's approach, annotations are used to specify
the initial state of the program (the precondition) and the expected
end state (the postcondition).  Verification is successful if we can
prove that whenever the program execution starts in a state satisfying
the precondition and execution ends, the desired postcondition
holds. The task of verification can be mechanized and modularized by
further annotating the program with intermediate assertions that
summarize the intended behavior of certain program portions. Loop
invariants summarize the intended behavior of loops. Once the
invariant is proved to hold for a loop, it can be used in the
verification of enclosing program segments as a summary of the effect
of the loop. Likewise, if pre and postconditions are provided for a
function, once the function is proved to be correct with respect to
them, they can be used as function summary in the verification of a
program unit that invokes the function.  Program verification à la
Hoare assumes a program to be \emph{fully annotated} with pre and
postconditions and invariants; the verification task is performed by
simple propagation of the program's symbolic state along linear (non
branching) statement sequences. Our approach, however, does not
require programs to be fully annotated. If no annotations at all are
provided, then verification amounts to propagating the symbolic state
and checking that no violating states are entered: it corresponds to
symbolic execution. If the program is partially annotated, assertions
are used to check that the symbolic state satisfies the corresponding
assertion. Missing annotations in general penalize verification
efficiency and may even cause non termination. For example, missing
invariants cause loop unfolding. In summary, the term verification is
used to cover a whole spectrum of activities ranging from
\emph{Hoare-style program verification} to what we may call
\emph{debugging-style verification}.

The specific context we explore in this paper is incremental
verification at the program level --- although the problem can also be
studied at  the model
level)~\cite{Sistla:1996:HIM:242224.242384} --- using a logical
framework for defining the properties to check. More specifically, we
focus on the incremental verification of programs written in
KernelC~\cite{Stefanescu2014183} (a subset of the C programming
language) and annotated with properties expressed in matching
logic~\cite{rosu2010:matching-logic:}. Our choice is motivated by the following reasons:
\begin{itemize}
\item KernelC is a large and meaningful subset of C, so that it can support writing and analyzing non-trivial, realistic, and  large programs;
\item KernelC's semantics is fully formalized in terms of  matching logic which is a prerequisite for formal verification;
\item Matching logic is a nice and effective formalism that integrates verification through Hoare's style annotations and through symbolic execution;
\item KernelC program verification is supported by a publicly
  available tool, namely MatchC~\cite{Stefanescu2014183}, and the accompanying benchmark, useful for various types of experimental evaluations and comparisons.
\end {itemize}

The present work on incremental program verification is based on our previous proposal for a general-purpose, 
\emph{syntactic-semantic}  approach to
incremental verification of various types of system properties such as reliability, efficiency, etc., called \sidecar (Syntax-DrivEn
inCrementAl veRification)~\cite{bfgm-scp2015}.  The \sidecar approach is
rooted into the past work of some of the authors on incremental
syntactic program analysis~\cite{Ghezzi1979}, which is extended
to formal verification using
attribute grammars~\cite{knuth1968} and incremental attribute
evaluation~\cite{Jalili:1985:GIE:3468.3472,sidecar-report}. 

 \sidecar  defines the verification procedure
 through semantic attributes, which are computed and propagated through the
 syntax tree of a program.  \sidecar leverages operator precedence
 grammars~\cite{Floyd1963} which, upon a change in a program, allow
 for re-parsing, and hence semantic re-analysis, to be confined within
 an inner portion of the input that encloses the changed
 part~\cite{barenghi2015parallel}.  This property is the key for an efficient
 incremental verification procedure: since the verification procedure
 is encoded within attributes, their evaluation proceeds
 incrementally, hand-in-hand with parsing of the new version of the
 program.  In our earlier work, we applied \sidecar to reachability
 analysis in simple toy
 programs~\cite{sidecar-report} and to reliability analysis of
 structured workflows~\cite{2014-isola}.
In this paper we show how \sidecar
 can be used to build an incremental verifier for realistic KernelC
 programs  specified in matching logic, by completing and extending our preliminary work~\cite{formalise2015} where the first steps to encode KernelC's semantics in our attribute schema evaluation are reported, and by making it incremental.

  We performed an extensive experimental assessment to show that:
\begin{itemize}
\item
Our syntactic-semantic  approach based on attribute grammars does not penalize the efficiency of formal verification when (re)done from scratch. We do this assessment by comparing the efficiency of verification with \sidecar with the efficiency of the state-of-the-art, non-incremental tool MatchC~\cite{Stefanescu2014183}.
\item The incremental approach supported by \sidecar outperforms program re-verification after changes if the program is fully annotated; instead, benefits range from just significant to striking if annotations are missing. In this latter case, the efficiency gains vary  according to several conditions. Since  verification through symbolic evaluation is intrinsically based on a sequential scan of the program, the gain may depend on the position in the program where the change is made, as well as on the program structure. 
\end{itemize}

To summarize, the contributions of this paper are:
\begin{inparaenum}[1)]
\item the definition of an incremental verification procedure for  KernelC programs annotated with
  matching logic properties, built on top of \sidecar, our
  general syntactic-semantic framework for incremental verification;
\item an extensive experimental evaluation of the benefits and costs
  introduced by our incremental verification procedure, using a large benchmark including non-toy, industrial-size programs. 
  
\end{inparaenum}

\subsubsection*{Reading guide}
The paper is organized in three parts.

The first part
(sections~\ref{sec:background}--\ref{sec:semantics} as well as
appendices~~\ref{sec:construct-details}--\ref{sec:heapFormalization})
makes the reader acquainted with the main concepts used in
the paper: section~\ref{sec:background} provides the reader with some
background concepts on operator-precedence grammars, incremental
parsing, attribute grammars, and matching logic;
section~\ref{sec:semantics} and
appendices~~\ref{sec:construct-details}--\ref{sec:heapFormalization}
introduce KernelC and its formalization in terms of matching logic.

The second part
(sections~\ref{sec:reach-check-thro}--\ref{sec:running-example} and appendix~\ref{sec:attributeEvaluation})
describes and exemplifies how to encode KernelC's semantics in our
attribute schema evaluation: section~\ref{sec:reach-check-thro}
illustrates how to encode a verification procedure for KernelC
programs annotated with matching logic properties using a
syntactic-semantic approach; section~\ref{sec:running-example} and
appendix~\ref{sec:attributeEvaluation} show
the application of this verification procedure to a sample program.

The third part
(sections~\ref{sec:incrementality}--\ref{sec:conclusion} and appendix~\ref{sec:improveIncrementality}) presents our approach for incremental
verification, compares it, also experimentally, with the state of
the art, and discusses open problems:
section~\ref{sec:incrementality}
 and appendix~\ref{sec:improveIncrementality} discuss how to make
the syntactic-semantic verification procedure for KernelC incremental using
\sidecar; section~\ref{sec:evaluation} reports on the experimental
evaluation of our implementation;  section~\ref{sec:related-work}
discusses related work; section~\ref{sec:conclusion} concludes the
paper, illustrating further research directions. 

We remark that, given the large amount of technical details necessary to carry out
such a work, the main body of the paper focuses on the most
relevant and critical issues; additional finer-grain technical details
have been included in appendices.

\section{Background}
\label{sec:background}
\subsection{A syntactic-semantic approach to incremental evaluation}\label{secSidecar}

Upon performing a change on the data represented with a tree 
structure, it is possible to identify the subtree affected by the
change and replace it, without touching the rest of the tree.  Various
techniques have been developed to perform this type of incremental
evaluation; see, for example, the procedure for efficient incremental
parsing of LR and other deterministic context-free languages
in~\cite{Ghezzi1979}.

In this section we outline the theoretical underpinnings for building
tree structures that are particularly well-suited for incremental
evaluation, both upon syntactic and  semantic changes. We
consider Floyd's operator precedence grammars (OPGs)~\cite{Floyd1963},
enriched with semantic annotations defined through a synthesized
attribute schema~\cite{knuth1968}.  We show how the combination of these
two formalisms allows for a simple, bottom-up parsing coupled with a
semantic evaluation procedure, which can be started efficiently
\emph{at any position} of the input stream, without being constrained
to proceed rigorously left-to-right.  This feature naturally lends
itself to incremental---and possibly parallel---evaluation since  changes may occur at any position of the program.

\subsubsection*{Operator Precedence Grammars and their incremental parsing}

Herewith we summarize the essential definitions and properties that are necessary to make the paper self-contained. For more information on formal languages and grammars, we refer the reader to classic textbooks such as~\cite{Salomaa1987}.

A \emph{context-free (CF)} grammar $G$ is a tuple
$G=\langle V_N,V_T,P,S \rangle$, where $V_N$ is a finite set of
\emph{non-terminal} symbols; $V_T$ is a finite set of \emph{terminal} symbols,
disjoint from $V_N$; $P \subseteq V_N \times (V_N \cup V_T)^*$ is
a relation  representing the \emph{productions} of the grammar;
$S \in V_N$ is the \emph{axiom} or start symbol.  Unless otherwise specified, we use the following 
convention: non-terminals  are
shown in \textsf{sans serif font}, such as \synt{A}; terminals  are enclosed
within single quotes, such as \lit{+} or are denoted by lowercase
letters at the beginning of the alphabet $(a,b,c)$; lowercase letters
at the end of the alphabet $(u, v, w, x, y, z )$ denote terminal
strings; 
Greek letters denote strings in $(V_N \cup V_T)^*$;
$\varepsilon$ denotes the empty string.

For a CF grammar $G$, the \emph{immediate derivation} relation, denoted
by $\Rightarrow$, is defined on $(V_N \cup V_T)^*$ as follows:
$\alpha \Rightarrow \beta$ if and only if (\textit{iff})
there exist $\alpha_1, \alpha_2$,
such that
$\alpha = \alpha_1 \synt{A} \alpha_2$ and 
$\beta = \alpha_1 \delta \alpha_2$,
and $(\synt{A}, \delta) \in P$.
Its reflexive and transitive closure
is denoted by $\stackrel{*}{\Rightarrow}$.  The language generated by
a grammar $G$ is the set of strings $L(G)$ =
\{$x \mid S \stackrel{*}{\Rightarrow} x$\}.
Through the rest of the paper we denote a production $(\synt{A}, \delta) \in P$
with the usual notation $\synt{A} \Coloneqq \delta$,
where $\synt{A}$ and $\delta$ are the left hand side (lhs)
and the right hand side (rhs) of $P$, respectively.

Any string generated by a grammar $G$ can be represented by a syntax tree
describing the derivation of the string through repeated application
of the production rules starting from a root node representing the axiom.
A string is recognized to belong to the language if such syntax tree can be constructed.
This recognition is called \emph{parsing}. Different parsing strategies have been studied to build the syntax tree, either bottom-up or top-down.

A production is in \emph{operator form} if its rhs has no adjacent non-terminals; an \emph{operator grammar (OG)} contains only productions in operator form. Any CF grammar admits an equivalent OG~\cite{Salomaa1987}.
A classic example of an operator grammar is the one generating arithmetic expressions, shown in Figure~\ref{fig:cfg}, where \lit{n} stands for any number.

\begin{figure}[tb]
	\centering
	\subfloat[]{\begin{tabular}[b]{p{.5em}>{$}p{.5em}<{$}l}
			\synt{S} & \Coloneqq & \synt{E} $\mid$ \synt{T}\\
			\synt{E} & \Coloneqq & \synt{E} \lit{+} \synt{T} $\mid$ \synt{T} \lit{+} \synt{T}\\
			\synt{T} & \Coloneqq & \synt{T} \lit{*} \lit{n} $\mid$ \lit{n} \\
		\end{tabular}
		\label{fig:cfg}
	}\qquad
	\subfloat[]{
		\begin{tabular}[b]{p{0.5em}@{\hspace{10pt}}|>{$}p{0.5em}<{$}>{$}p{0.5em}<{$}>{$}p{0.5em}<{$}}
			& \lit{n}        & \lit{*}        & \lit{+} \\ \hline
			\lit{n} &          & \gtrdot  & \gtrdot \\
			\lit{*} & \doteq   &          & \\
			\lit{+} & \lessdot & \lessdot & \gtrdot \\
		\end{tabular}
		\label{fig:opm}
	}\caption{Example of an operator grammar and its operator precedence  matrix}
	\label{fig:grammars}
\end{figure}

R.\W.\ Floyd took inspiration from the traditional notion of precedence
between arithmetic operators and defined a broad class of
languages, called \emph{operator-precedence languages (OPLs)} or \emph{Floyd
  languages}, which are characterized by the fact that the shape of their
syntactic tree is solely determined by binary relations between terminals
that are consecutive, or become consecutive, 
up to a possible non-terminal in between, after a bottom-up
reduction step. OPLs are generated by OPGs, which are formalized as follows.

Let $G$ be an OG, $\synt{A} \in V_N$, and $B \in V_N \cup \{\varepsilon\}$;
$\mathcal{L}_G(\synt{A})$ is defined as $\mathcal{L}_G(\synt{A})= \{ a \in V_T \mid \synt{A} \stackrel{*}{\Rightarrow}
Ba\alpha\}$ and is named the \emph{left terminal set} of $\synt{A}$;
symmetrically, $\mathcal{R}_G(\synt{A})= \{ a \in V_T \mid \synt{A} \stackrel{*}{\Rightarrow}\alpha aB\}$ is the \emph{right terminal set}.
Three binary relations on $V_T$, called \emph{operator precedence relations}, are defined:

\begin{eqnarray}
\begin{array}{rl}
 \nonumber \text{equal-in-precedence: } & a\doteq b  \text{ iff }   \exists \synt{A}\Coloneqq\alpha aBb\beta, B\in V_N\cup\{\varepsilon\} \\
   \text{takes-precedence: } & a\gtrdot b  \text{ iff }  \exists \synt{A}\Coloneqq\alpha \synt{D}b\beta, \synt{D}\in V_N \text{ and } a\in \mathcal{R}_G(\synt{D}) \\
 \nonumber    \text{yields-precedence: }& a\lessdot b  \text{ iff }  \exists \synt{A}\Coloneqq\alpha a\synt{D}\beta, \synt{D}\in V_N \text{ and } b\in \mathcal{L}_G(\synt{D})
 \end{array}
\end{eqnarray}
For an OG $G$, the \textit{operator precedence matrix} (OPM)
$M=\mathit{OPM}(G)$ is a $|V_T| \times |V_T|$ matrix, in which an
entry $m_{a,b}$ represents the set of operator precedence relations
holding between  $a$ and $b$.
\begin{definition}\label{defFloydGr}
An operator grammar $G$ is an \emph{operator precedence} grammar\footnote{Operator precedence grammars are also often called \emph{Floyd grammars} in honor of the inventor.}
iff $M=\mathit{OPM}(G)$ is a \textit{conflict-free}
matrix, i.e., for each  $a,b \in V_T$, $|m_{a,b}|\leq 1$.
\end{definition}

Notice that, despite the adopted symbols, operator
precedence relations are neither transitive nor reflexive nor
symmetric. Intuitively, they drive the parsing of OPLs since the
rhs of any production is enclosed within a pair
$\langle \lessdot, \gtrdot \rangle$, and $\doteq$ holds between
consecutive terminals occurring within such a pair.  
This pair
determines a rhs to be replaced, with a shift-reduce algorithm, by
the corresponding lhs.
For instance, consider the grammar given in figure~\ref{fig:cfg}, whose
OPM is displayed in figure~\ref{fig:opm}, and the input string \lit{n +
  n * n}.  By adding at the beginning and at the end of this string
the conventional symbol \lit{\#}, which yields precedence to all
symbols in $V_T$ and over which all symbols in $V_T$ take precedence, we
obtain the following labeling of the string:
$\lit{\#}\lessdot \lit{n} \gtrdot \lit{+} \lessdot \lit{n} \gtrdot
\lit{*} \doteq \lit{n} \gtrdot \lit{\#}$.
This labeled string displays two occurrences of the rhs \lit{n} (each
of them enclosed in a $\langle \lessdot, \gtrdot \rangle$ pair); each
of them can be reduced, in any order, to the corresponding lhs
\synt{T}. Since for OPGs non-terminals are unessential to compute
precedence relations, we now obtain the new precedence relations
$\lit{\#} \lessdot \synt{T} \lit{+} \lessdot \synt{T} \lit{*} \doteq
\lit{n} \gtrdot \lit{\#}$,
which display the rhs $\synt{T} \lit{*} \lit{n}$, which in turn is
reduced to $\synt{T}$.  We obtain
$\lit{\#} \lessdot \synt{T} \lit{+} \synt{T} \gtrdot \lit{\#}$, which
is reduced to $\synt{E}$ and finally to $\synt{S}$. A detailed
description of parsing algorithms for OPGs can be found
in~\cite{parsingTechniques}. Notice that the precedence relations of
the grammar of figure~\ref{fig:grammars} impose that the \lit{*}
operator is reduced before  \lit{+}, so that the resulting
syntax tree reflects the usual semantics of arithmetic expressions,
where multiplication has precedence over addition.

The main reason for the choice of OPGs is that, unlike more commonly used grammars that support deterministic parsing, they
enjoy the \textit{local parsability property}, i.e., the possibility
of starting the parsing from any arbitrary point of the sentence to be
analyzed, independent of the context within which the sentence is
located. Since the parsing of an OPL sentence is driven by precedence
relations, a partial syntax tree corresponding to a derivation
$\synt{N} \stackrel{*}{\Rightarrow} x$ can be deterministically
built\footnote{Every OPG can be tranformed into en equivalent one in
  Fischer normal form~\cite{fischer1969some}, where no two productions
  have the same rhs (as in the case of figure~\ref{fig:cfg}), so that for
  every reduction the lhs to replace the rhs is univocally defined.}
(in linear time) in a bottom-up way by using only a pair of single
characters, say, $\langle a,b\rangle$ as the context of $x$ (notice
that necessarily $a$ yields precedence to the first character of $x$
and the last character of $x$ takes precedence over $b$).

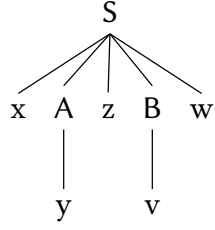
\begin{figure}[tb]
	\centering
	\begin{tikzpicture}[scale=1.2]
		\Tree [.{\synt{S}} {x} [.{\synt{A}} {y} ] {z}  [.{\synt{B}} {v} ] {w} ]
	\end{tikzpicture}
	\caption{The syntax tree of sentence $xyzvw$}
	\label{fig:syntax-tree}
\end{figure}

Consider now the case where, after building a syntax tree for a
given  program, one or more parts of the input
are changed.  Thanks to the local parsability property, only the
changed parts should be re-parsed. Afterwards, the newly built subtrees
must be merged with the global syntax tree using the \emph{matching
  condition}. This condition is satisfied when, after parsing a substring, it is possible to identify the correct nesting
point of its syntax subtree within the global one.

To intuitively illustrate the notion of matching condition,
consider a sentence $xyzvw$ and let its
corresponding syntax tree be the one displayed in
figure~\ref{fig:syntax-tree}. Imagine now that within the string
$y = y_1y_2y_3$, the substring $y_2$ is replaced by $u$, and that
within string $v = v_1v_2v_3$ the substring $v_2$ is replaced by $s$.
The new parsing could be restarted independently, possibly even in
parallel, at the beginning of $u$ and $s$ respectively (by looking for
suitable $\lessdot$, $\doteq$, and $\gtrdot$ within them and at their
left and right).  Suppose now that during the reparsing of the input,
two independent derivations are obtained, respectively,
$\synt{A} \stackrel{*}{\Rightarrow} y_1uy_3$ and
$\synt{B} \stackrel{*}{\Rightarrow} v_1sv_3$. Such a situation
satisfies the \emph{matching condition}, since at this point
we can stop re-parsing, and simply replace (i.e., \emph{match}) in the
original syntax tree the subtree corresponding to the derivation
$\synt{A} \stackrel{*}{\Rightarrow} y$ with the one corresponding to
the derivation $\synt{A} \stackrel{*}{\Rightarrow}y_1uy_3$
and similarly for the subtree rooted in \synt{B}.
As a result, the overall cost of reparsing is proportional to the length of
$y_1uy_3v_1sv_3$, which in many practical cases may be considerably
shorter than the whole $xy_1uy_3zv_1sv_3w$.
The above example also shows that the local parsability property naturally supports both incremental and parallel parsing, thus enabling managing multiple changes and even increasing the efficiency of parsing from scratch.

A complete and detailed description of a parallel but not incremental parsing algorithm exploiting the local parsability property of OPGs is given in~\cite{barenghi2015parallel}.
In this paper our attention is mainly focused on incremental evaluation of single changes, without exploiting parallelism; such a feature however, is already  present in our tool and, where appropriate, we will add a few hints on the further improvements that can be obtained by enabling it.

The local parsability property has a price in terms of
generative power. More specifically, while LR grammars---traditionally
used to describe and parse programming languages but not featuring
this property---can generate all  deterministic languages, OPGs
cannot.  This limitation, however, is more of theoretical interest
than of real practical impact. In fact, large parts of the grammars of
many computer languages are operator
precedence~\cite[p. 271]{parsingTechniques}. Complete OPGs
have been given for Prolog~\cite{bosschere1996} and
Algol~68~\cite{meertens81:operat-prior-gramm-for-algol}; recently,
OPG-based (parallel) parsers for JSON, XML, Lua have been developed~\cite{barenghi2015parallel}.
Moreover, in many practical cases an
OPG can be obtained with minor adjustments to a non
operator-precedence grammar~\cite{Floyd1963}.

In summary, OPGs appear to be a natural choice to support our
syntactic-semantic approach to incremental and possibly parallel
evaluation.

\subsubsection*{Attribute Grammars and  incremental semantic evaluation}

Attribute Grammars (AGs) have been proposed by Knuth as a formalism to
represent semantics of programming languages~\cite{knuth1968}. AGs
extend CF grammars by associating \emph{attributes} and semantic
functions to the production rules of a CF grammar; attributes define
the ``meaning'' of the corresponding nodes in the syntax tree of a
sentence generated by the grammar.

Attributes can be of two types: \emph{synthesized} attributes
describe  a bottom-up information flow,
from descendent nodes (of a syntax tree) to their ancestors;
\emph{inherited} attributes instead describe  a top-down information flow, from ancestors to
descendants.  In this paper we are only interested in synthesized
attributes, whose evaluation can go hand-in-hand with bottom-up
parsing. It has been proved that synthesized attributes are semantically
complete, i.e., any inherited attribute can be translated into
a (set of) synthesized attribute(s) reproducing the same
information~\cite{knuth1968}.

An attribute grammar that has only synthesized attributes is called an
\emph{S-attributed grammar}. Such a grammar can be obtained from a
context-free grammar $G=\langle V_N, V_T, P, S \rangle$ by adding a
finite set of attributes $\mathit{SYN}$ and a set $\mathit{SF}$ of
semantic functions. Each symbol $\synt{X} \in V_N$ has a set of (synthesized) attributes $\mathit{SYN}(\synt{X})$;
$\mathit{SYN} = \bigcup_{\synt{X} \in V_N} \mathit{SYN}(\synt{X})$.  We use the
symbol $\mathcal{A}$ to denote a generic element of $\mathit{SYN}$; we
assume that each $\mathcal{A}$ takes values from a corresponding domain
$T_\mathcal{A}$.  The set $\mathit{SF}$ consists of functions, each of which is
associated with a production $p$ in $P$.  For each attribute $\mathcal{A}$
of the left hand side of $p$, a function $f_{p\mathcal{A}} \in \mathit{SF}$
synthesizes the value of $\mathcal{A}$ based on (a subset of) the
attributes of the nonterminals occurring in the right hand side of
$p$.
For example, the grammar in figure~\ref{fig:cfg} can be extended to an attribute grammar that computes the value of an expression, in the following way: the set of attributes is defined as $\mathit{SYN}= \mathit{SYN}(\synt{E}) \cup \mathit{SYN}(\synt{T}) = \{\mathit{value}\} \cup  \{\mathit{value}\} =  \{\mathit{value}\}$, with $T_{\mathit{value}}=\mathbb{N}$; in other words, we associate an attribute \textit{value} over the naturals, to both non-terminals \synt{E} and \synt{T}. The set of semantic functions $\mathit{SF}$ is defined as below, where semantic functions are enclosed in braces next to each production:\\[5pt]
\makebox[\linewidth][c]{\begin{small}
\begin{tabular}[]{p{.5em}>{$\;}p{.5em}<{$}l@{\hspace{5pt}}>{$\{}l<{$}@{\hspace{2pt}}>{$}c<{$}@{\hspace{2pt}}>{$}l<{$}}
	\synt{S}   & \Coloneqq & \synt{E}     & \mathit{value}(\synt{S})& =  &  \mathit{value}(\synt{E})\}\\
	\synt{S}   & \Coloneqq & \synt{T}     & \mathit{value}(\synt{S})& =  &  \mathit{value}(\synt{T})\}\\
      \synt{E\textsubscript{0}} & \Coloneqq & \synt{E\textsubscript{1}} \lit{+} \synt{T} & \mathit{value}(\synt{E\textsubscript{0}})& =  &  \mathit{value}(\synt{E\textsubscript{1}}) + \mathit{value}(\synt{T})\}\\      
      \synt{E} & \Coloneqq & \synt{T\textsubscript{1}} \lit{+} \synt{T\textsubscript{2}} & \mathit{value}(\synt{E})& =  &  \mathit{value}(\synt{T\textsubscript{1}}) + \mathit{value}(\synt{T\textsubscript{2}})\}\\   
      \synt{T\textsubscript{0}} & \Coloneqq & \synt{T\textsubscript{1}} \lit{*} \lit{n} & \mathit{value}(\synt{T\textsubscript{0}})& =  &  \mathit{value}(\synt{T\textsubscript{1}}) * \mathit{eval}(\lit{n})\}\\
      \synt{T}   & \Coloneqq & \lit{n}     & \mathit{value}(\synt{T})& =  &  \mathit{eval}(\lit{n})\} \\
\end{tabular}
\end{small}
}
\\[5pt]
The $+$ and $*$ operators appearing within braces correspond to the standard arithmetic operators (addition and multiplication); $eval(\cdot)$ evaluates its input as a number. Notice also that, within a production, different occurrences of the same grammar symbol are labeled by distinct subscripts.

During a bottom-up syntax analysis, semantic actions may be
performed in conjunction with a reduction and the computed attribute values for each node of the tree are kept for future use in case of change. 
When a change is made, it is only necessary to compute the new values of attributes for the new nodes of the new subtree produced by the incremental parser and then propagate these values to the tree root by recomputing the attributes of the nodes encountered along the path towards the root.

\begin{figure}[tb]
	\centering
	\begin{tikzpicture}
\begin{scope}
 \tikzset{every node/.style={isosceles triangle, draw,anchor=apex, shape border rotate=90, isosceles triangle stretches}}
 \node[minimum height=35mm,minimum width=8cm] (out) at (4,4) {};
 \node[minimum height=15mm,minimum width=2.5cm] (in) at (4.2,2) {};
 \node[minimum height=7.5mm,minimum width=10mm] (in2) at (4.5,1.25) {};
\end{scope}
 \node (S) at (4,4.3) {$\mathcal{A}_S$};
 \node (M) at (3.75,2.2) {$\mathcal{A}_M$};
 \node (N) at (4.2,1.3) {$\mathcal{A}_N$};
 \node (K) at (4.5,2.75) {$\mathcal{A}_K$};
 \node (P) at (2.5,1.5) {$\mathcal{A}_P$};
 \node (Q) at (5.5,1.5) {$\mathcal{A}_Q$};
 \node (t) at (4,0) {$xw^\prime z$};
 \draw[-latex] (t.east) -- (in.right corner);
 \draw[-latex] (t.west) -- (in.left corner);
 \draw[densely dashed] (in2.apex) -- (in.apex);
 \draw[densely dashed] (in.apex) -- (K);
 \draw[densely dashed] (K) -- (out.apex);
 \draw[o-] (P) -- (in.apex);
 \draw[o-] (Q) -- (in.apex);
\end{tikzpicture}

 	\caption{Incremental evaluation of semantic attributes}
	\label{fig-attributes01}
\end{figure}
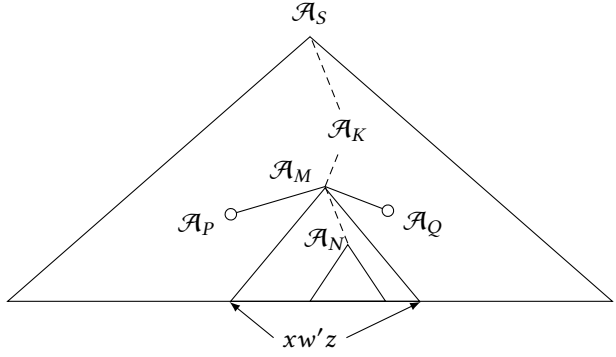

For example, suppose that after replacing the substring $w$ with
$w^\prime$ in the input string $xwz$,
incremental re-parsing builds a derivation $\synt{N}
\stackrel{*}{\Rightarrow} x w^\prime z$, with the same non-terminal
\synt{N} as in $\synt{N} \stackrel{*}{\Rightarrow} xwz$, so that the
matching condition is satisfied.
Assume also that \synt{N} has an attribute $\mathcal{A}_N$. 
Two situations may occur related to the computation of $\mathcal{A}_N$:
\begin{asparaenum}
\item The $\mathcal{A}_N$ attribute associated with the new subtree rooted
  in \synt{N} has the same value as before the change. In this case,
  all the remaining attributes in the rest of the tree will not be
  affected, and no further analysis is needed.
\item The new value of $\mathcal{A}_N$ is different from the one it had
  before the change.  In this case (see figure~\ref{fig-attributes01})
  only the attributes on the path from \synt{N} to the root \synt{S}
  (e.g., $\mathcal{A}_M, \mathcal{A}_K, \mathcal{A}_S$) may change and
  they need to
  be recomputed. The values of the other attributes not on the path
  from \synt{N} to the root (e.g., $\mathcal{A}_P$ and $\mathcal{A}_Q$) do not
  change and thus there is no need to recompute them. 

\end{asparaenum}

We emphasize that the effort spent in evaluating the semantic
attributes is proportional only to the size of (the data structure
that has been affected by) the change.  Notice also that often the
semantic evaluation may cost orders of magnitude more than the cost of
pure parsing (as it is the case of the application developed in this
paper); hence, the global gain obtained from incrementality may become
quite tangible. Finally, since synthesized attributes depend only on the attributes of their descendants in the syntax-tree, their computation can be parallelized "hand in hand" with parallel bottom-up parsing.

\subsection{Matching Logic}\label{secMatchingLogic}

Matching logic~\cite{rosu2010:matching-logic:} is a system to
formalize the semantics, and reason about the properties, of programs
written in a von-Neumann language, such as C. It is inspired by the traditional Floyd-Hoare method to prove
program correctness. In essence, matching logic integrates a formal
operational semantics based on the notion of state and its
transformation during program execution with an axiom system that
relates the preconditions holding before executing a piece of code
and the postconditions holding after its execution.  This section
 introduces the basic concepts and notation of matching logic, which
are then used throughout the paper. For a detailed and comprehensive
description of matching logic and its proof system the reader can refer 
to~\cite{rosu2010:matching-logic:,rosu-stefanescu-2012-oopsla}.
Please notice, however, that slight  notational changes appear in the various published versions of matching logic; hereafter we take the liberty of providing a unified notation.

Matching logic defines a basic element called
\emph{configuration}~\cite{rosu-serbanuta-2010-jlap}, which
represents the code to be executed at a given point of the program
execution and the current state of the abstract machine executing the
program.  More precisely, a configuration is a set of \emph{cells},
where each cell represents a specific semantic item whose content
depends both on the purpose of the analysis and on the features of the
programming language. For example, cells may represent the current
memory allocation table, the residual code to be executed, the I/O
output buffer, the next instruction pointer, the state of the
activation record.  Cells are denoted with labeled angle-brackets, as
in $\langle \dots \rangle_\mathit{label}$; a configuration is denoted
with a pair of outer angle brackets enclosing a number of cells. For
example, the following configuration for a C-like program fragment:
\begin{gather}\label{eqExampleConfigurationSimplified}
\left\langle \left\langle \texttt{x=2; y=x} \right\rangle _{k}\left\langle \texttt{x}\mapsto 5, \texttt{y}\mapsto 4 \right\rangle _{\mathit{env}}\right\rangle
\end{gather}
contains two cells, labeled as $k$ and $\mathit{env}$. Cell $k$
represents the residual program code to be executed, in the form of a
list of statements; in this case there are two statements: the
assignment of value $\texttt{2}$ to variable $\texttt{x}$ and the
assignment of the value of $\texttt{x}$ to $\texttt{y}$. Cell
$\mathit{env}$ contains a representation of the current memory state
as a map from program variables to values; in this case variable $\texttt{x}$
is bound to value 5 and $\texttt{y}$ to 4.

Matching logic formulae, also called \emph{configuration patterns}
(or \emph{patterns}), are defined as follows.
Let $\mathit{Var}$ be an infinite set of logical variables.
A matching logic formula has the form $\exists X.(\pi \wedge \varphi)$\footnote{From a syntactic point of
	view, reference~\cite{rosu-stefanescu-2012-oopsla} introduces a dummy symbol
	$\square$ of type ``configuration'' so that by
	writing $\exists X.((\square = \pi) \wedge \varphi)$ in place of $\exists X.(\pi \wedge \varphi)$ we obtain a syntactically-complete FOL\textsubscript{=} formula. For simplicity, we leave the inclusion of the $\square$ symbol and subsequent ‘=’ as implicit in our formulae.}, where:

\begin{itemize}
	\item $X$ is a subset of $\mathit{Var}$;
\item $\pi$, called the \emph{pattern structure} (or \emph{basic pattern}),
		generalizes a configuration by allowing
logical variables in $\mathit{Var}$;\item $\varphi$, called the \emph{pattern constraint}, is a formula in an arbitrary first-order logic with equality (FOL\textsubscript{=})  with variables in $\mathit{Var}$.
\end{itemize}
Notice that in matching logic, differently from Hoare logic, there is
a distinction between program variables (appearing in the program
code) and logical variables (defined in
$\mathit{Var}$). To better highlight this distinction, we use a
\texttt{monospaced font} to denote program variables (e.g.,
\texttt{x}) and  \textit{italics} to denote logical variables
(e.g., $a$). On the other hand, similarly to Hoare logic, free
variables in specifications act as proof parameters, and they are
expected to be bound during the proof.

A configuration pattern defines (intensionally) a set of
configurations through the concept of
\emph{matching}:
a configuration $\gamma$ \emph{matches} a configuration pattern
$\psi\equiv\exists X.(\pi \wedge \varphi)$ iff there is a substitution $\tau$ of
concrete values to the variables in $\mathit{Var}$ such that $\gamma$
is equal to $\pi$ after applying $\tau$ and the predicate obtained by
applying $\tau$ to $\varphi$ holds.
For example, the following configuration pattern
\begin{equation}\label{eqExamplePatternSimplified}
\exists a. \left\langle \left\langle \texttt{x=2; y=x} \right\rangle _{k}\left\langle \texttt{x}\mapsto a, \texttt{y}\mapsto b, 
\dots \right\rangle _{\mathit{env}}\right\rangle \wedge a\geq 4
\end{equation}
matches any configuration in which the residual program statements
to execute are ``\texttt{x=2; y=x}'' and the program variable
\texttt{x} has a value greater than or equal to 4.  In this
configuration, the logical variable $b$ is free, hence it has to be
bound by the proof context; the symbol ``$\dots$'' captures the other
elements of the environment.  We use the notation
``$\left\langle\cdot\right\rangle_{\mathit{label}}$'', taken
from~\cite{rosu-stefanescu-2012-oopsla}, to represent an empty cell.
Configuration~\eqref{eqExampleConfigurationSimplified}
matches configuration pattern~\eqref{eqExamplePatternSimplified} with
$\tau=\{a \mapsto 5, b \mapsto 4\}$.
Other examples of configuration patterns matched by configuration~\eqref{eqExampleConfigurationSimplified}
are:
\begin{align*}
& \left\langle \left\langle \texttt{x=2; y=x} \right\rangle _{k}\left\langle \texttt{x}\mapsto 
\mathit{5} \dots \right\rangle _{\mathit{env}}\right\rangle  \\
\exists a. & \left\langle \left\langle \texttt{x=2; y=x} \right\rangle _{k}\left\langle \texttt{x}\mapsto 
\mathit{a} \dots \right\rangle _{\mathit{env}}\right\rangle  \wedge \mathit{a}\geq 4\\
\exists a, b. & \left\langle \left\langle \texttt{x=2;} \ldots \right\rangle _{k}\left\langle \texttt{x}\mapsto 
\mathit{a}, \texttt{y}\mapsto \mathit{b}  \right\rangle _{\mathit{env}}\right\rangle  \wedge \mathit{a} > \mathit{b}
\end{align*} 

Given a programming language,
it is possible to define its semantics by specifying 
how the program's abstract state changes by executing each instruction.
Matching logic captures this aspect through the concept of
\emph{reachability rule}, which  defines a  transition
$\psi\Ddownarrow\psi^\prime$ between two patterns $\psi$ and
$\psi^\prime$.
Intuitively, a rule $\psi\Ddownarrow\psi^\prime$ states that every
configuration $\gamma$ that matches pattern $\psi$, after the execution of the next instruction (as defined in
the $k$ cell of $\gamma$), becomes a
configuration $\gamma^\prime$ that 
matches pattern $\psi^\prime$
\footnote{We recall that, as usual, only the free variables of the patterns are shared in a rule.
For example, the following rule correctly describes the patterns before and after the statement, but incorrectly specifies the intended semantics, since the bound variables in the two existential quantifiers are unrelated: 
$
\exists a.\left\langle \left\langle \texttt{x+=1;} \right\rangle _{k}\left\langle \texttt{x}\mapsto 
\mathit{a} \right\rangle _{\mathit{env}}\right\rangle \Ddownarrow
\exists a. \left\langle \left\langle \cdot \right\rangle _{k}\left\langle \texttt{x}\mapsto 
\mathit{a} + 1 \right\rangle _{\mathit{env}}\right\rangle
$. 
The following reachability rule instead correctly specifies the intended semantics, since the $\mathit{a}$ in the left pattern is the same as  $\mathit{a}$ in the right pattern:
$
	 \left\langle \left\langle \texttt{x+=1;} \right\rangle _{k}\left\langle \texttt{x}\mapsto 
	 \mathit{a} \right\rangle _{\mathit{env}}\right\rangle \Ddownarrow
	 \left\langle \left\langle \cdot \right\rangle _{k}\left\langle \texttt{x}\mapsto 
	 \mathit{a} + 1 \right\rangle _{\mathit{env}}\right\rangle
$.
}.

We express reachability rules using the same notation as the
$\mathbb{K}$ framework~\cite{rosu-serbanuta-2010-jlap}, where the
lhs of a rule is above a horizontal line and the rhs is
below the line; parts of the rule without a line are the same on both
sides of the rule.
In this notation, we also unify the pattern constraints $\varphi$ and $\varphi^\prime$
into a single constraint for the reachability rule,
which expresses the conjunction of $\varphi$ and $\varphi^\prime$.
For sake of clarity, we omit in this notation the existential quantifiers before the rule patterns,
avoiding situations in which a logical variable contained in both patterns of a reachability rule
is existentially quantified.
For example, the reachability rule:
\begin{equation}\label{eq:ex-ml}
\left\langle 
\left\langle \frac{\texttt{x=2; y=x}}{2; \texttt{y=x}}\right\rangle_{k}
\left\langle \texttt{x}\mapsto \frac{\mathit{a}}{2},
\texttt{y}\mapsto 4\dots \right\rangle _{\mathit{env}}
\right\rangle 
\land \mathit{a} \geq 0
\end{equation}
allows the transition from a configuration matching
pattern~\eqref{eqExamplePatternSimplified} to another configuration
where the residual program \texttt{x=2; y=x} has been partially
interpreted (leaving the value of the expression, the integer value 2,
in cell $k$\footnote{Notice that the integer value $2$ remains in cell
  $k$ even after evaluating the assignment statement; this is
  consistent with the standard C semantics, and is required to
  support, for example, multiple assignment statements like
  \texttt{y=x=2}.})  and the previous value of $\texttt{x}$ (i.e.,
$a$) is replaced by the integer number 2 (cell $\mathit{env}$).  In
general, reachability rules can be used to express the semantics of
each instruction of a programming language.
These mechanisms will be described in detail in
Section~\ref{sec:semantics} to specify the semantics of KernelC.

Besides defining the semantics of a programming language, reachability
rules can be used for specifying \emph{function contracts} (e.g., pre and
postconditions) on the behavior of a program fragment, e.g., a
function: in fact, a rule would map the configuration pattern on the
lhs (e.g., defining a function precondition) to the pattern on the rhs
(defining a postcondition). For example, a reachability rule that
defines a contract for a function \texttt{sqrt} computing the square
root of a positive integer $x$ can be defined as follows:
\begin{equation}\label{eq:sqrt}
\left\langle \left\langle \frac{\texttt{sqrt(}x\texttt{)}}{y} \right\rangle _{k}
\right\rangle 
\land x \geq 0
\land y*y \leq x
\land (y+1)*(y+1) > x
\end{equation}
In this example, configurations contain only cell $k$. In the
pattern on the rhs, variable $y$ represents the value computed
by function  \texttt{sqrt} after its termination\footnote{Note that it is possible to have logical
	variables to represent integer values in cell $k$, like $x$ and $y$.}. 
The pattern constraint  contains both the precondition
($x \geq 0$) and the postcondition
($y*y \leq x \land (y+1)*(y+1) > x$).

Assuming that the semantics of a programming language is defined
through reachability rules, it is possible to verify the correctness
of program execution by means of \emph{reachability checking over
  configuration patterns}.  Let $\vartheta, \vartheta^\prime$ be some
configuration patterns, and $\mathcal{R}$ be the set of reachability
rules that define the semantics of a program fragment; we define the
\emph{reachability relation} $\leadsto$ between patterns over
$\mathcal{R}$ as follows.  We say that
$\mathcal{R} \vdash \vartheta \leadsto \vartheta^\prime$ iff for each
configuration $\gamma$ matching $\vartheta$, there exists a
configuration $\gamma^\prime$ matching $\vartheta^\prime$ and there is
a reachability rule $\psi\Ddownarrow\psi^\prime \in
\mathcal{R}$ such that 
$\gamma$ and $\gamma^\prime$ match $\psi$ and $\psi^\prime$
respectively and $\gamma^\prime$ is the configuration obtained
applying $\psi\Ddownarrow\psi^\prime$ to $\gamma$.
Similarly, we can define $\stackrel{*}{\leadsto}$ and $\stackrel{+}{\leadsto}$ as the
reflexive-transitive and transitive closure of $\leadsto$,
respectively.

In other words, given a program characterized by a set of reachability
rules $\mathcal{R}$, $\mathcal{R} \vdash \vartheta \stackrel{*}{\leadsto} \vartheta^\prime$
corresponds to a program execution, where $\vartheta$ is the configuration
pattern representing the state before the program execution, and
$\vartheta^\prime$ is the one representing the state after the program
execution.  The reachability relation can also express a Hoare
style of program (fragment) verification, in which $\vartheta$ expresses the
program preconditions and $\vartheta^\prime$ the program postconditions.
For example, verifying the correctness of the example above is
equivalent to checking whether this reachability relation,
which represents a proof obligation, holds:
\begin{equation}\label{eq:gcdReachability}
\mathcal{R}_{\texttt{sqrt}} \vdash
\left\langle \left\langle \texttt{sqrt(}x\texttt{)} \right\rangle _{k}  \right\rangle \land x \geq 0
\stackrel{*}{\leadsto}
\exists y. \left\langle \left\langle y \right\rangle _{k}
\right\rangle 
\land y*y \leq x
\land (y+1)*(y+1) > x
\end{equation}
where $\mathcal{R}_{\texttt{sqrt}}$ is the set containing the
reachability rules that describes the semantics of the implementation
of function \texttt{sqrt}.
On the basis of such proof obligation it is possible to derive the following reachability relation:
\begin{multline}\label{eq:gcdReachability1}
\mathcal{R}_{\texttt{sqrt}} \vdash
\left\langle \left\langle \texttt{y = sqrt(x)} \right\rangle _{k} 
\left\langle \texttt{x} \mapsto a, \texttt{y} \mapsto b \right\rangle _{\mathit{env}} 
\right\rangle \land a \geq 0
\stackrel{*}{\leadsto}\\
\exists c. \left\langle \left\langle \cdot \right\rangle _{k}
\left\langle \texttt{x} \mapsto a, \texttt{y} \mapsto c \right\rangle _{\mathit{env}} 
\right\rangle 
\land c*c \leq a
\land (c+1)*(c+1) > a
\end{multline}

Reachability properties can be expressed by adding suitable
\textit{annotations} to the program code as it is now customary in
various programming languages. These annotations may also specify loop
invariants, which are a key point in Hoare style verification. Proofs are then built by applying matching logic's
axiomatization, which consists of eight axioms and inference
rules~\cite{rosu-stefanescu-2012-oopsla}.

The verification of reachability properties is supported by
MatchC~\cite{Stefanescu2014183}, a tool implemented on top of the
$\mathbb{K}$ framework~\cite{rosu-serbanuta-2010-jlap}, which is a rewriting-based formalism for developing and analyzing programming languages.  
The $\mathbb{K}$ framework in turn uses Maude~\cite{Clavel:2007:MHL:1808998} as rewriting system.
To check that every possible execution reaches a desired configuration
pattern $\vartheta^\prime$, a set of reachable configurations is first
initialized with the configuration pattern $\vartheta$; then, all the
reachability rules in $\mathcal{R}$ are applied on all the 
reachable configurations matching the pattern, discovering new reachable
configurations. The process terminates when a fixed point is reached
or as soon as a configuration pattern violating $\vartheta^\prime$ survives
the next iteration (i.e., the execution cannot leave it). Notice that
reachability rules have to encode also the execution of invalid
operations, e.g., a division by zero, by moving the execution into a
sink error configuration that cannot be left.  Given that program
correctness is an undecidable problem, the tool cannot always provide
a final answer. In particular, in the case of loops, if they are
annotated by suitable invariants, the tool tries to exploit them; otherwise it unfolds them, on the condition
that an upperbound for the number of unfolding steps is provided (to
avoid the risk of nontermination).

\subsection*{Reading Guide}
The next two sections describe 
how KernelC and its matching logic formal semantics can be used in
\sidecar to   build an incremental verifier for realistic KernelC
programs specified in matching logic. More precisely:
\begin{itemize}
\item In  Section~\ref{sec:semantics} we show how the syntax of
  KernelC has been
 reshaped into an operator precedence grammar. We remark that
 ``reshaping the language syntax'' is a fairly normal and easy
 practice whenever one has to adapt the original BNF of a programming
 language to the format of a given parser generator (e.g., by removing
 shift-reduce conflicts when using an LALR parser like
 Bison~\cite{Levine:2009:FB:1696439} or to make it suitable for LL parsing). Section~\ref{sec:semantics} also includes a
 summary of the semantics of KernelC expressed using matching logic.
\item In the key Section~\ref{sec:reach-check-thro} we show how the
  language semantics has been implemented  as a suitable attribute
  schema, to be used within \sidecar. We remark that the equivalence
  between the original KernelC definition and our own implementation
  can be verified by a --- tedious but natural --- inspection of the
  various attribute definitions;  it is further confirmed by our
  experiments, as will be shown in
  Section~\ref{sec:evaluation}.
\end{itemize}
After these two sections, we describe how KernelC
programs specified in matching logic can be verified using \sidecar,
either from scratch (through an example 
Section~\ref{sec:running-example}) or incrementally (Section~\ref{sec:incrementality}).

\section{KernelC and its Semantics}
\label{sec:semantics}
KernelC 
supports several types of expression (assignment,
pointer referencing and de-referencing, structure member, arithmetic
and logic operators, ternary conditional, function call), the basic
control structures (sequences, \texttt{while}, \texttt{if/else},
\texttt{return}), 
the data types integer, void, struct and
corresponding pointers, arrays, and some functions from the standard library
(\texttt{malloc/free} and basic I/O). An excerpt of the KernelC
grammar (in operator precedence form) is shown in
figure~\ref{fig:grammar}\footnote{The full operator precedence KernelC grammar is available at~\url{https://hub.docker.com/r/arizzi/scmatchc/}.}, where question marks indicate optional
elements.  The non-terminal \synt{Annotation} (not expanded in the
grammar) is used to define function contracts and loop invariants,
further discussed in Section~\ref{sec:loops}~and~\ref{sec:functions}.

\begin{figure}[tb]
\begin{grammar}
	<program> ::= <global\_decl>
	
	<global\_decl> ::= `struct' IDENTIFIER `{' <struct\_field\_list> `}' `;' <global\_decl>? |\\
	<parameter> `;' <global\_decl>? | <function\_def>
	
	<struct\_field\_list> ::= <parameter> `;' <struct\_field\_list>?  	
	
	<compound> ::= <compound\_decl> | <compound\_stm>
	
	<compound\_decl> ::=  <parameter> `;' <compound>
	
	<parameter> ::= <type> IDENTIFIER | <ptr\_type> `*' <id>
	
	<parameter\_list> ::= <parameter> | <parameter> `,' <parameter\_list>
	
	<type> ::= `int' | `void' | `struct' IDENTIFIER
	
	<ptr\_type> ::= <ptr\_type> `*' | <type>
	
	<type2> ::= `(' ( <type> | <ptr\_type> ) `)'  
	
	<id> ::= IDENTIFIER  
	
	<compound\_stm> ::= <Annotation>? `while' `('<exp>`)' `{' <compound> `}' <compound\_stm> |\\
	`if' `('<exp>`)' `{' <compound> `}' `else' `{' <compound> `}' <compound\_stm> |\\
	`if' `('<exp>`)' `{' <compound> `}'  <compound\_stm> |\\
	`return'? <exp> `;' <compound\_stm>	| <stm>
	
	<stm> ::= `return'? <exp> `;' | `{' <compound\_decl> `}' |\\
	`if' `('<exp>`)' `{' <compound> `}' `else' `{' <compound> `}' |\\
	`if' `('<exp>`)' `{' <compound> `}'
	
	<argument\_exp\_list> ::= <exp> `,' <argument\_exp\_list> | <exp> `,' <exp>

	<exp> ::= <unary\_exp> (`='|`+='|`-='|`*='|`/='|`\%=') <exp> | <cond\_exp>
	
	<cond\_exp> ::= <log\_or\_exp> | <log\_or\_exp> `?' <exp> `:' <cond\_exp>
	
	<log\_or\_exp> ::= <log\_or\_exp> `||' <log\_and\_exp> | <log\_and\_exp>
	
	<log\_and\_exp> ::= <log\_and\_exp> `\&\&' <eq\_exp> | <eq\_exp>
	
	<eq\_exp> ::= <eq\_exp> (`=='|`!=') <relat\_exp> | <relat\_exp>
	
	<relat\_exp> ::= <relat\_exp> (`>'|`<'|`>='|`<=') <add\_exp> | <add\_exp>
	
	<add\_exp> ::= <add\_exp> (`+'|`-') <mult\_exp> | <mult\_exp>
	
	<mult\_exp> ::= <mult\_exp> (`*'|`/'|`\%') <cast\_exp> | <cast\_exp>
	
	<cast\_exp> ::= `(' <ptr\_type> `)' <cast\_exp> | <unary\_exp>
	
	<function\_def> ::= <ptr\_type> IDENTIFIER <function\_def2> 
	
	<function\_def2> ::= `(' <parameter\_list>? `)' <Annotation>? `{' <compound>? `}' <global\_decl>?
	
	<postfix\_exp> ::= <id> | Constant <postfix\_exp1>? | `('<exp>`)' |\\
	IDENTIFIER ( <postfix\_exp2> | <postfix\_exp1> )   
	
	<postfix\_exp1> ::= (`.'|`->') IDENTIFIER ( <postfix\_exp1> )?
	
	<postfix\_exp2> ::= `('(<exp>|<argument\_exp\_list>)?`)' <postfix\_exp1>?
	
	<unary\_exp> ::= <postfix\_exp> | `sizeof' <type2> | <cast\_exp> |\\
	(`+'|`-'|`*'|`!'|`sizeof') <unary\_exp>
	
\end{grammar}
\caption{Excerpt of the KernelC grammar}
\label{fig:grammar}
\vspace{-2mm}
\end{figure}
 
We present the operational semantics of KernelC
constructs in the form of Matching Logic reachability rules. These rules correspond to the original semantics defined in~\cite{Stefanescu2014183}, with minor syntactic changes.
We adopt the same assumptions and
restrictions as~\cite{Stefanescu2014183} to resolve the ambiguity
inherited from the corresponding C semantics and to simplify
reachability verification: expressions are evaluated left to right to
avoid non-determinism; integers have infinite precision; structure
fields can only be of type integer; I/O primitives can only read and
write integers.

KernelC semantics is defined using the following Matching Logic configuration cells:
\begin{itemize}
	\item $k$: the residual program code to be executed;
	\item $\mathit{env}$: a mapping of variable identifiers to their values;
	\item $\mathit{stack}$: the call stack of the active functions;
	\item $\mathit{estack}$: the stack of the local scopes for each active function in $\mathit{stack}$;
	\item $\mathit{mem}$: the heap state, represented as a map from addresses to dynamically-allocated objects;
	\item $\mathit{ma}$: a map from dynamically-allocated objects
          to the size of the memory allocated for them;
	\item $\mathit{in}$: standard input, represented as a list of
          integers in which values are accessed left to right;
	\item $\mathit{out}$: standard output, represented as a list
          of integers to which values can only be appended.
\end{itemize}

In the rest of the section we describe  semantics  by
presenting, for each type of syntactic construct, the structure of the
corresponding cells and the associated reachability rules.  Due to
space reasons, we omit the description of input/output and provide
only a short overview to memory management; the description of the
complete semantics for these constructs is available in
Appendices~\ref{sec:io}~and~\ref{sec:heapFormalization}, respectively.

\subsection{Basic constructs}
\label{sec:basic-constructs}

\begin{table}[tb]
 \caption{Reachability rules for the basic constructs of KernelC\label{eq:basicConstructs}}
  \centering
	    \begin{tabular}{cl}
        \toprule
        Construct & Reachability rule\\ \midrule			
		\textit{Typed variable declaration} & 
$\displaystyle \left\langle
		\left\langle \frac{\texttt{type x;}}{\cdot} \dots \right\rangle _{k}
		\left\langle \frac{\cdot}{\texttt{x} \mapsto (\mathit{type}, \mathit{\bot})}\dots \right\rangle _{\mathit{env}}
		\right\rangle$\\[15pt]
		\textit{Typed variable evaluation} & 
$\displaystyle \left\langle
		\left\langle \frac{\texttt{x}}{(t,a)} \dots \right\rangle _{k}
		\left\langle \texttt{x} \mapsto \mathit{(t,a)}\dots \right\rangle _{\mathit{env}}
		\right\rangle$\\[15pt]	
		\textit{Assignment} & 
$\displaystyle \left\langle
		\left\langle \frac{\texttt{x = }a}{a} \dots \right\rangle _{k}
		\left\langle \texttt{x} \mapsto \frac{b}{a}\dots \right\rangle _{\mathit{env}}
		\right\rangle$\\[15pt]
		\textit{Selection (\texttt{then} branch)} & 
$\displaystyle \left\langle
		\left\langle \frac{\texttt{if(}c\texttt{) \{T\} else \{E\}}}{\texttt{T}} \dots \right\rangle _{k}
		\right\rangle \land c\neq 0$\\[15pt]
		\textit{Selection (\texttt{else} branch)} & 
$\displaystyle \left\langle
		\left\langle \frac{\texttt{if(}c\texttt{) \{T\} else \{E\}}}{\texttt{E}} \dots \right\rangle _{k}
		\right\rangle \land c = 0$\\[15pt]
		\textit{Selection (branch taken)} & 
$\displaystyle \left\langle
		\left\langle \frac{\texttt{if(}c\texttt{) \{T\}}}{\texttt{T}} \dots \right\rangle _{k}
		\right\rangle \land c\neq 0$\\[15pt]
		\textit{Selection (branch not taken)} & 
$\displaystyle \left\langle
		\left\langle \frac{\texttt{if(}c\texttt{) \{T\}}}{\cdot} \dots \right\rangle _{k}
		\right\rangle \land c = 0$\\[15pt]
        \bottomrule
        \end{tabular}
\end{table}

Table~\ref{eq:basicConstructs} shows the reachability rules related to
KernelC constructs for declaration, expression evaluation, assignment, and choice.

A variable declaration adds to $\mathit{env}$ a mapping from
the identifier used in the declaration to an undefined value
(conventionally denoted with $\mathit{\bot}$)\footnote{The case in which the variable identifier is already
	declared is not discussed here.}.
Expressions composed by a single variable are evaluated replacing the
variable with its value from $\mathit{env}$.
We maintain in cells $k$ and $\mathit{env}$ pairs $(\mathit{type}$,$\mathit{variable})$
keeping track of the specific variable type, where \textit{type} is any of the types supported by Kernel C, e.g., \texttt{int} or \texttt{struct listNode*}.

We use the logical variable $t$ to represent a generic
type, which remains unknown up to the application of the reachability rule to a configuration pattern which binds $t$ to a certain \textit{type}.
To simplify the notation,
throughout the rest of the paper we omit the type 
for a plain C \texttt{int} integer logical variable.

More complex expressions
are evaluated recursively, preserving the precedence among the
operators defined in the grammar of KernelC.  Evaluating an assignment
statement results in updating the entry in $\mathit{env}$
corresponding to the lhs with the new value obtained evaluating the
rhs of the assignment. As already mentioned in
Section~\ref{secMatchingLogic}, the assignment statement in cell $k$
is replaced by the new value of the assigned variable.

Evaluating a selection statement introduces a branch in the execution,
accounting for the satisfaction or violation of the selection
condition. This corresponds to the introduction of two guarded
reachability rules, which determine the next statements to execute
based on the boolean evaluation of the selection condition.  The
evaluation of a sequence of statements corresponds to evaluating each
statement in the sequence, in the order prescribed by the sequence
itself.

Note that for any reachability rule, unless otherwise specified, every
logical variable contained in it has a local scope within the rule
itself and is different from the variables with the same name in other
reachability rules.

\subsection{Loops}
\label{sec:loops}

KernelC only provides  \texttt{while} loops. The semantics of a loop
construct corresponds to its one-step unwinding, as described in
rule~\eqref{eq:loop}. Notice that rule~\eqref{eq:loop} is applied before
evaluating the loop condition \texttt{C}; the latter is evaluated
before executing the selection statement. The unwinding is recursively
applied until the condition cannot be satisfied (possibly leading to
nontermination\footnote{For practical reasons, the total number of execution steps can be bound by the user to guarantee the (possibly inconclusive) termination of the verification process.}).
\begin{equation}
\label{eq:loop}
\left\langle
\left\langle \frac{\texttt{while(C)\{T\}}}{\texttt{if(C)\{T while(C)\{T\}\}}} \dots \right\rangle _{k}
\right\rangle
\end{equation}
For example, for the  code snippet \verb~while(t>0){c=c+t; t=t-1;}~, 
the corresponding loop unrolling reachability rule is:
\begin{equation}
\label{eq:loopEx}
\left\langle
\left\langle \frac{\texttt{while(t>0)\{c=c+t; t=t-1;\}}}{\texttt{if(t > 0)\{c=c+t; t=t-1; while(t>0)\{c=c+t; t=t-1;\}\}}} \dots \right\rangle _{k}
\right\rangle
\end{equation}
Let us consider the following configuration, characterizing the state
of the abstract machine at the beginning of the loop:
\begin{equation}
\label{eq:loopEx1}
\left\langle
\left\langle \texttt{while(t>0)\{c=c+t; t=t-1;\}} \right\rangle _{k}
\left\langle \texttt{t} \mapsto 1, \texttt{c} \mapsto 0\right\rangle _{\mathit{env}}
\right\rangle
\end{equation}
By applying rule~\eqref{eq:loopEx} we obtain the configuration:
\begin{equation}
\label{eq:loopEx2}
\left\langle
\left\langle \texttt{if(t>0)\{c=c+t;t=t-1; while(t>0)\{c=c+t;t=t-1;\}\}} \right\rangle _{k}
\left\langle \texttt{t} \mapsto 1, \texttt{c} \mapsto 0 \right\rangle _{\mathit{env}}
\right\rangle
\end{equation}
The execution of the \texttt{if} statement leads to the following configuration:
\begin{equation}
\label{eq:loopEx5}
\left\langle
\left\langle \texttt{while(t>0)\{c=c+t;t=t-1;\}} \right\rangle _{k}
\left\langle \texttt{t} \mapsto 0, \texttt{c} \mapsto 1 \right\rangle _{\mathit{env}}
\right\rangle
\end{equation}
Although similar to configuration~\eqref{eq:loopEx1}, in this
configuration the program variable \texttt{t} has value 0, which will
cause the \texttt{if} branch to be skipped in the next loop iteration.

Rule~\eqref{eq:loop} executes loops through a series of successive
unwinding steps. However, KernelC supports also the concept of a
\textit{loop invariant}, as already proposed in \textit{Hoare
  logic}. The loop invariant is represented by an annotation $A$
placed before the loop body; the invariant is checked and used to
execute the loop (instead of the unwinding steps).
Annotation $A$ has the form of the following  configuration pattern
\begin{equation}
\label{eq:loopConfig}
\left\langle
\left\langle \texttt{while(C)\{T\}} \dots \right\rangle _{k}
\varPsi
\right\rangle
\land \varphi
\end{equation}
with $\varPsi$ representing the rest of the basic pattern
and $\varphi$ the pattern constraint.

An example of an invariant for the above example loop is that the value
of \texttt{t} is non-negative and that \texttt{c} contains the sum of
the numbers between $i$ and the initial value of \texttt{t} (denoted
with $h$); this invariant is formalized by the following
configuration pattern:

\begin{multline}
\label{eq:loopConfigEx}
\exists i,j.
\left\langle
\left\langle \texttt{while(t>0)\{c=c+t; t=t-1;\}} \dots \right\rangle _{k}
\left\langle \texttt{t}\mapsto i, \texttt{c} \mapsto j \right\rangle_{\mathit{env}}
\right\rangle\\
\land i \geq 0
\land j = \frac{h(h+1) - i(i+1)}{2}
\end{multline}
The loop invariant is checked to see if it holds after every iteration
of the loop.
\begin{equation}
\label{eq:loopConfig1}
\exists X.
\left\langle
\left\langle \texttt{while(C)\{T\}} \dots \right\rangle _{k}
\varPsi
\right\rangle
\land \varphi
\stackrel{+}{\leadsto}
\exists X^\prime.
\left\langle
\left\langle \texttt{while(C)\{T\}} \dots \right\rangle _{k}
\varPsi^\prime
\right\rangle
\land \varphi^\prime
\end{equation}

This check is performed by evaluating
formula~\eqref{eq:loopConfig1}, in which the lhs is the invariant itself
and the rhs is another instance of the invariant, obtained by
replacing any bound variable in $\varPsi$ and $\varphi$ with a new
variable, thus $\varPsi^\prime$ and $\varphi^\prime$ are obtained respectively from $\varPsi$ and $\varphi$ by renaming every logical variable occurrence.
This corresponds to the application of the \textit{while
  rule} of Hoare logic.

To check the invariant, any execution 
from the configuration pattern 
$
\exists X.
\left\langle
\left\langle \texttt{if(C)\{T\}} \right\rangle _{k}
\varPsi
\right\rangle
\land \varphi$
must reach the target configuration pattern 
$
\exists X^\prime.
\left\langle
\left\langle \cdot \right\rangle _{k}
\varPsi^\prime
\right\rangle
\land \varphi^\prime$.
Notice that as part of this check, one iteration of the loop is
performed and thus  the \texttt{while} loop is reduced to an
\texttt{if} statement.
For the  example above, checking  the invariant corresponds to evaluating the following formula,
given $\mathit{LOOP} \equiv \texttt{while(t>0)\{c=c+t;t=t-1;\}}$:
\begin{multline}
\label{eq:loopConfig1Ex}
\exists i,j.
\left\langle
\left\langle \mathit{LOOP} \right\rangle _{k}
\left\langle \texttt{t} \mapsto i, \texttt{c} \mapsto j \right\rangle _{\mathit{env}}
\right\rangle
\land i \geq 0
\land j = \frac{h(h+1) - i(i+1)}{2}
\stackrel{+}{\leadsto}\\
\exists l,m.
\left\langle
\left\langle \mathit{LOOP} \right\rangle _{k}
\left\langle \texttt{t} \mapsto l, \texttt{c} \mapsto m \right\rangle _{\mathit{env}}
\right\rangle
\land l \geq 0
\land m = \frac{h(h+1) - l(l+1)}{2}
\end{multline}
In this case, the initial configuration pattern is:
\begin{equation}
\label{eq:loopConfig2Ex}
\exists i, j.
\left\langle
\left\langle \texttt{if(t>0)\{c=c+t;t=t-1;\}} \right\rangle _{k}
\left\langle \texttt{t} \mapsto i, \texttt{c} \mapsto j  \right\rangle _{\mathit{env}}
\right\rangle
\land i \geq 0
\land j = \frac{h(h+1) - i(i+1)}{2}
\end{equation}
and the target configuration pattern is:
\begin{equation}
\label{eq:loopConfig3Ex}
\exists l, m.
\left\langle
\left\langle \cdot \right\rangle _{k}
\left\langle \texttt{t} \mapsto l, \texttt{c} \mapsto m \right\rangle _{\mathit{env}}
\right\rangle
\land l \geq 0
\land m = \frac{h(h+1) - l(l+1)}{2}
\end{equation}

After checking that a loop invariant is satisfied, it
can be used in the following reachability rule\footnote{The notation $\frac{\psi}{\psi^\prime}$ is a generalization of the one used in rule~\eqref{eq:ex-ml} for expressing the pattern structure.}:
\begin{equation}
\label{eq:loopInvariant}
\left\langle
\left\langle \frac{\texttt{while(C)\{T\}}}{\texttt{assume(!C);}} \dots \right\rangle _{k}
\frac{\varPsi}{\varPsi^\prime}
\right\rangle
\land \varphi \land \varphi^\prime
\end{equation}
which changes the current configuration replacing bound variables in the pattern with fresh ones,
while still satisfying the invariant.
In this rule, function \texttt{assume} (whose semantics is provided in rule~\eqref{eq:loopInvariant1}),
is used to discard from the obtained configuration pattern all the configurations
in which the loop condition \texttt{C} evaluates to \texttt{false}.
\begin{equation}
\label{eq:loopInvariant1}
\left\langle
\left\langle \frac{\texttt{assume(}c\texttt{);}}{\cdot} \dots \right\rangle _{k}
\right\rangle
\land c\neq 0
\end{equation}

In the case of our example, rule~\eqref{eq:loopInvariant} is
instantiated as:
\begin{multline}
\label{eq:loopInvariantEx}
\left\langle
\left\langle \frac{\texttt{while(t>0)\{c=c+t;t=t-1;\}}}{\texttt{assume(!(t>0));}} \dots \right\rangle _{k}
\left\langle \texttt{t} \mapsto \frac{i}{l}, \texttt{c} \mapsto \frac{j}{m} \right\rangle _{\mathit{env}}
\right\rangle\\
\land i \geq 0 \land l \geq 0 \land j = \frac{h(h+1) - i(i+1)}{2}
\land m = \frac{h(h+1) - l(l+1)}{2}
\end{multline}
By combining it with the semantics of \texttt{assume}, we obtain the following reachability rule:
\begin{multline}
\label{eq:loopInvariantEx1}
\left\langle
\left\langle \frac{\texttt{while(t>0)\{c=c+t;t=t-1;\}}}{\texttt{assume(!(t>0));}} \dots \right\rangle _{k}
\left\langle \texttt{t} \mapsto \frac{i}{l}, \texttt{c} \mapsto \frac{j}{m} \right\rangle _{\mathit{env}}
\right\rangle\\
\land i \geq 0 \land l = 0 \land j = \frac{h(h+1) - i(i+1)}{2}
\land m = \frac{h(h+1)}{2}
\end{multline}

Notice that the constraint $l = 0$ is derived from the conjunction of
the condition in rule~\eqref{eq:loopInvariantEx} ($l \geq 0$) with the
semantics of \texttt{assume} over the condition $l \leq 0$.
Thus $l(l+1)$ has value $0$.

\subsection{Functions}
\label{sec:functions}

Supporting function invocation requires formalizing the call stack
and the local scope of each active function. This is accomplished by
introducing cells $\mathit{stack}$ and $\mathit{estack}$.

When a function \texttt{f} is invoked, its identifier is pushed to
$\mathit{stack}$, the current environment described in $\mathit{env}$
is pushed to $\mathit{estack}$,
and the new current environment becomes a mapping from the formal parameters of \texttt{f} to
their values at the invocation\footnote{In this way we can handle only
  local variables.  Although global variables are not supported by
  KernelC, it would be possible to manage them by adding another cell
  $\mathit{global}$, with the same structure of $\mathit{env}$,
  holding global variables. Differently from $\mathit{env}$, such cell
  would not be modified by function calls.  This addition would only
  require slight changes to the rules evaluating and storing
  variables.}. This is formalized in rule~\eqref{eq:functionCall}, where
\texttt{F} represents the body of the function, which is inlined at
the invocation site in cell $k$ and followed by the placeholder
\textit{END} (representing the end of the function body).

\begin{equation}
\label{eq:functionCall}
\left\langle
\left\langle \frac{\textit{f}(\mathit{arg}_1,~\dots)}{\texttt{F}~\textit{END}} \dots \right\rangle _{k}
\left\langle \frac{E}{\texttt{param1} \mapsto \mathit{arg}_1, \dots} \right\rangle _{env}
\left\langle \frac{\cdot}{E}\dots \right\rangle _{estack}
\left\langle \frac{\cdot}{\mathit{f}}\ldots \right\rangle _{stack}
\right\rangle
\end{equation}

\noindent For example, let us consider the following function:

\verb|int sum(int a, int b) { return a + b; }|

\noindent Its associated reachability rule is the following:

\begin{equation}
\label{eq:functionCallEx}
\left\langle
\left\langle \frac{\textit{sum}(i,j)}{\texttt{return a + b;}~\textit{END}} \dots \right\rangle _{k}
\left\langle \frac{E}{\texttt{a} \mapsto i, \texttt{b} \mapsto j} \right\rangle _{env}
\left\langle \frac{\cdot}{E}\dots \right\rangle _{estack}
\left\langle \frac{\cdot}{\mathit{sum}}\ldots \right\rangle _{stack}
\right\rangle
\end{equation}

Let us assume the following start configuration:
\begin{equation}
\label{eq:functionCallEx1}
\left\langle
\left\langle \textit{sum}(\texttt{x},\texttt{x}) \right\rangle _{k}
\left\langle \texttt{x} \mapsto 2 \right\rangle _{env}
\left\langle \cdot \right\rangle _{estack}
\left\langle \cdot \right\rangle _{stack}
\right\rangle
\end{equation}
Before executing the function call, we have to evaluate the actual parameters
to be passed to function \texttt{sum}, evaluating the expression \texttt{x} thus getting:
\begin{equation}
\label{eq:functionCallEx2}
\left\langle
\left\langle \textit{sum}(2,2) \right\rangle _{k}
\left\langle \texttt{x} \mapsto 2 \right\rangle _{env}
\left\langle \cdot \right\rangle _{estack}
\left\langle \cdot \right\rangle _{stack}
\right\rangle
\end{equation}
By applying rule~\eqref{eq:functionCallEx}, we get
\begin{equation}
\label{eq:functionCallEx3}
\left\langle
\left\langle \texttt{return a + b;}~\textit{END} \right\rangle _{k}
\left\langle \texttt{a} \mapsto 2, \texttt{b} \mapsto 2 \right\rangle _{env}
\left\langle (\texttt{x} \mapsto 2) \right\rangle _{estack}
\left\langle \mathit{sum} \right\rangle _{stack}
\right\rangle;
\end{equation}
continuing the execution, we end up with: 
\begin{equation}
\label{eq:functionCallEx4}
\left\langle
\left\langle \texttt{return }4\texttt{;}~\textit{END} \right\rangle _{k}
\left\langle \texttt{a} \mapsto 2, \texttt{b} \mapsto 2 \right\rangle _{env}
\left\langle (\texttt{x} \mapsto 2) \right\rangle _{estack}
\left\langle \mathit{sum} \right\rangle _{stack}
\right\rangle
\end{equation}

The return of a function, corresponding to the evaluation of the
\texttt{return} construct, is formalized as shown in
rule~\eqref{eq:functionReturn}: all the code from the return point to
the placeholder \textit{END} is consumed, the function identifier is
popped out of $\mathit{stack}$ and the environment $\mathit{env}$ is
restored to its state before the function invocation (by popping out from
$\mathit{estack}$), the returned value \texttt{i} is added to the
residual code in place of the function invocation. Notably, the
evaluation of recursive functions may lead to nontermination.

\begin{equation}
\label{eq:functionReturn}
\left\langle
\left\langle \frac{\texttt{return}~\mathit{i}~\dots~\textit{END}}{\mathit{i}} \dots \right\rangle _{k}
\left\langle \frac{\dots}{E} \right\rangle _{env}
\left\langle \frac{E}{\cdot}\dots \right\rangle _{estack}
\left\langle \frac{\mathit{f}}{\cdot}\ldots \right\rangle _{stack}
\right\rangle
\end{equation}

In the example above, the application of reachability rule~\eqref{eq:functionReturn} would result
in the following configuration:

\begin{equation}
\label{eq:functionCallEx5}
\left\langle
\left\langle \cdot \right\rangle _{k}
\left\langle \texttt{x} \mapsto 2 \right\rangle _{env}
\left\langle \cdot \right\rangle _{estack}
\left\langle \cdot \right\rangle _{stack}
\right\rangle
\end{equation}

A function can be annotated with a function specification (i.e.,
function contract) that represents a reachability rule defining how the program
state changes after the function execution. This rule, which can be used to summarize a function call, has the form:
\begin{equation}
\label{eq:functionRule}
\left\langle
\left\langle \frac{\textit{f}(\mathit{arg}_1,~\dots)}{i} \dots \right\rangle _{k}
\frac{\varPsi}{\varPsi^\prime}
\right\rangle
\land \varphi
\end{equation}

For the \texttt{sum} function above, let us assume that the function is annotated
with a specification that states that the result of the function is 
 the sum of the two arguments. The behavior of the function can then
 be summarized by the 
 following reachability rule:
 \begin{equation}
\label{eq:functionRuleEx}
\left\langle
\left\langle \frac{\textit{sum}(o, p)}{q} \dots \right\rangle _{k}
\right\rangle
\land q = o + p
\end{equation}

Starting from rule~\eqref{eq:functionRule}, one can generate the
configuration patterns representing the state at the beginning of the
function invocation:
\begin{equation}
\label{eq:precondition}
\exists X.
\left\langle
\left\langle \textit{f}(\mathit{arg}_1,~\dots) \right\rangle _{k}
\varPsi^\prime
\right\rangle
\land \varphi
\end{equation}
and at the end:
\begin{equation}
\label{eq:postcondition}
\exists X^\prime.
\left\langle
\left\langle i \right\rangle _{k}
\varPsi
\right\rangle
\land \varphi
\end{equation}

The function contract can then be checked by checking that 
pattern~\eqref{eq:postcondition} is always reached starting from
pattern~\eqref{eq:precondition}.

 \subsection{Memory Management}
\label{sec:memory}

KernelC supports dynamic allocation of memory objects.
These objects can be either integers or sequences of integers,
possibly constituting a \texttt{struct}.

The cell $\textit{mem}$ of a KernelC configuration is used to represent
dynamically allocated objects, with a map between locations and
objects.  For example
$\left\langle l \mapsto O \right\rangle_\textit{mem}$ represents a
heap allocation in which the memory object $O$ is at location $l$.
Object $O$ can be an integer, a structured type (i.e., a sequence of
integers or a \texttt{struct}), or an abstract data type like a list.
Each addressable memory location contains one infinite-precision
integer\footnote{Notice that the actual size of the memory location
  may vary depending on the size of the integer stored in the
  location.}; structured objects span over multiple locations.

Structured types can have two different representations: a compact one,
in which the whole object is associated to its first location; a flat
one, in which the object is split into contiguous locations, each holding
an integer value.  For example, let us assume that the sequence
$[2,3]$ is stored starting from memory location $p$.  The content of
the cell $\mathit{mem}$ using the compact notation would be
$\left\langle p \mapsto [2,3] \right\rangle_\textit{mem}$;  using the
flat notation for the same memory location, we would get
$\left\langle p \mapsto 2, p + 1 \mapsto 3
\right\rangle_\textit{mem}$.

The deallocation of an object through a pointer to its first location
(similarly to the \texttt{free} operation in C) is supported by
keeping track of the size of each object.  More precisely, the number
of blocks allocated for a structured object is saved in the KernelC
configuration cell $\textit{ma}$, which contains a map $l \mapsto n$
associating to each location allocated in cell $\textit{mem}$ the
number of blocks occupied by the associated object\footnote{We further extend the operations supported in~\cite{Stefanescu2014183} to catch additional errors related to memory allocation. These extensions are described in details in Appendix~\ref{sec:heapFormalization}.}.

 \section{Semantic evaluation}
\label{sec:reach-check-thro}

We now describe how reachability checking over matching
logic configuration patterns can be performed by evaluating semantic
attributes.  

\subsection{Overview}
\label{sec:overview}

As we mentioned earlier, our approach can handle program verification at any level of precision, ranging from Hoare-style program verification, when the program is fully annotated, to debugging-style verification if few or no annotations are provided. In fact, a distinguishing feature of Matching Logic is the integration of symbolic interpretation to evaluate program semantics with program property proofs in classical Hoare's style through the key concept of reachability rule. For instance, a reachability rule can formalize the effect of a single loop iteration, which can be iterated through loop unfolding, but can also be used to prove a loop invariant if such an invariant is supplied. 

To  encompass the entire spectrum of kinds of program verification, we introduce the notion of  \emph{verification task}, which is attached to any single program unit in a typical compositional, syntax-directed style.  A verification task is a triple
$\mathit{vt}=\langle C_r, R_v, c_t \rangle$, where:

\begin{itemize}
\item  $C_r$ is
a set of configuration patterns, initialized with the singleton
$\{c_s\}$, where $c_s$ is a
configuration pattern representing the initial state of the program as
defined by the precondition;
\item $R_v$ is a set of reachability rules specifying the program
  behavior;
\item $c_t$ is a configuration pattern representing the desired state
  after the program execution, as defined by the postcondition.
\end{itemize}

Before presenting the algorithm for performing a
verification task, we introduce the concept of \emph{final}
configuration pattern. A configuration pattern is \emph{final} if one
of these conditions is satisfied:
\begin{itemize}
\item its cell $k$ is empty, i.e., there is no residual program code
  to be  executed;
\item its cell $k$ is not empty but contains only an integer value
  resulting from the evaluation of a statement;
\item it represents a \emph{sink} error configuration, i.e., a configuration
	characterized by a special token in cell $k$ from which
	no further steps can be performed.
\end{itemize}
Notice that $c_t$ in a verification task is always final.

\begin{figure}
	\begin{scriptsize}
\begin{multicols}{2}
			\begin{algorithmic}[1]				
				\Function{Process}{$v_t=\langle C_r, R_v, c_t \rangle$}
\Repeat\label{line:beginBigLoop}
				\State $\mathit{Changed}\leftarrow\mathit{false}$
				\For{$c_i \in C_r$}
				\State $\mathit{temp}\leftarrow \emptyset$
				\For{$r_i \in R_v$}\label{line:forallRuleBegin}
				\If{$\mathit{isApplicable}(r_i,c_i)$}	\label{line:isApplicable}	
				\State$c^\prime \leftarrow\mathit{ApplyRule}(r_i, c_i)$\label{line:applyRule}
				\If{$\mathit{isSat}(c^\prime)$}\label{line:isSat}
				\State $\mathit{temp}\leftarrow\mathit{temp}\cup\mathit{c^\prime}$
				\EndIf
				\EndIf			
				\EndFor\label{line:forallRuleEnd}
				\If{$\mathit{temp}\neq \emptyset$}\label{line:tempNotNullBegin}
				\State$\mathit{Changed}\leftarrow\mathit{true}$
				\State$C_r\leftarrow C_r\cup \mathit{temp}\smallsetminus \{c_i\}$		
				\EndIf\label{line:tempNotNullEnd}
				\EndFor
				\Until{$\neg\mathit{Changed}$}\label{line:endBigLoop}
				\State\textbf{return }$\langle C_r, R_v, c_t \rangle$
				\EndFunction
				\Function{Check}{$v_t=\langle C_r, R_v, c_t \rangle$}				
                             \If{$\textsc{CheckNonFinal}(C_r)$}\label{line:cnf}				
				\State\textbf{return }$\mathit{false}$
				\EndIf
				\For{$c_i \in C_r$}\label{line:beginSC}
\If{$\neg\mathit{isContained}(c_i, c_t)$}\label{line:isContained}
\State\textbf{return }$\mathit{false}$\label{line:reportError}
				\EndIf
\EndFor \label{line:endSC}
				\State\textbf{return }$\mathit{true}$ \label{line:retTrue}
				\EndFunction
                                \Statex
				\Function{CheckNonFinal}{$C_r$}
				\For{$c_i \in C_r$}\label{line:beginFinalLoop}
				\If{$\mathit{isNonFinal}(c_i)$}\label{line:fixedPoint}
				\State\textbf{return }$\mathit{true}$\label{line:delay}
				\EndIf	
				\EndFor \label{line:endFinalLoop}
				\State\textbf{return }$\mathit{false}$
				\EndFunction
			\end{algorithmic}
		\end{multicols}
\end{scriptsize}
	\caption{The \textsc{Process} and \textsc{Check} algorithms for verification tasks}
	\label{fig:algo}
	\vspace{-2mm}
\end{figure}

A verification task is performed using the \textsc{Process} and
\textsc{Check} algorithms shown in figure~\ref{fig:algo}. The latter
takes as input a verification task $\mathit{vt}$ (resulting from the
processing by the former), and returns a boolean value stating
whether the unit in $\mathit{vt}$ satisfies its specification.  
These algorithms use a number of auxiliary functions:
\begin{itemize}
\item
  function
$\mathit{isApplicable}(r,c)$ takes as input a reachability rule $r$ of
the form $\psi\Ddownarrow\psi^\prime$ and a pattern $c$, and returns
true iff there exists a configuration $\gamma$ such that $\gamma$
matches both $c$ and $\psi$ without considering their constraints,;
its purpose is to avoid the application of expensive operations using
the structure of the configuration pattern;
\item
  function $\mathit{ApplyRule}(r,c)$
applies rule $r$ to pattern $c$ (corresponding to abstractly executing
the next statement in cell $k$ of $c$) and returns the new
configuration reached by the abstract execution; \item 
function $\mathit{isSat(c)}$ takes a configuration pattern $c$,
transforms it into a logic formula with integer arithmetics and
unspecified functions, and checks the satisfiability of this formula with an SMT solver; if the formula is satisfiable it means that there
exists at least one configuration satisfying the constraints in the
configuration pattern $c$;
\item
function
$\mathit{isNonFinal}(c)$ checks whether the configuration pattern $c$ is
non-final;
\item
function $\mathit{isContained}(c_1,c_2)$ takes two configuration
patterns and uses an SMT solver to check whether all the
configurations matching the first configuration pattern also match the
second.
\end{itemize}

Algorithm~\textsc{Process} performs a verification task by iterating
(lines~\ref{line:beginBigLoop}--\ref{line:endBigLoop}) through the set
$C_r$ (initially containing only $c_s$) until reaching a fixed point,
i.e., when $C_r$ does not change anymore. For each configuration
pattern $c_i \in C_r$, the algorithm applies all the rules in $R_v$
(lines~\ref{line:forallRuleBegin}--\ref{line:forallRuleEnd}) that are
applicable to $c_i$. At each iteration, the application of a rule
$r_i \in R_v$ to a pattern $c_i$ results in a new configuration
pattern $c^\prime$; if the constraints in $c^\prime$ are satisfiable,
$c^\prime$ is added to an auxiliary list $\mathit{temp}$.  When all
the applicable rules have been applied to $c_i$, if at least one new
configuration pattern has been discovered (meaning that list
$\mathit{temp}$ is not empty), $c_i$ is removed from $C_r$, while the
newly-generated configuration patterns stored in $\mathit{temp}$ are
added to $C_r$
(lines~\ref{line:tempNotNullBegin}--\ref{line:tempNotNullEnd}).  At
each iteration, $C_r$ represents the maximal front of reachable
configuration patterns from the initial configuration.

Algorithm~\textsc{Check} analyzes whether a verification task
obtained through the \textsc{Process} algorithm satisfies its specification.
In the verification task provided by \textsc{Process}, the configuration
patterns $C_r$ must be final; a non-final configuration pattern in
$C_r$ would indicate an incorrect input program (e.g., a program in
which some function definitions are missing).
Function \textsc{CheckNonFinal} is invoked (line~\ref{line:cnf}) to detect
non-final configuration patterns (through the loop at
lines~\ref{line:beginFinalLoop}--\ref{line:endFinalLoop}); in case a
non-final configuration pattern is found, the algorithm is aborted
returning $\mathit{false}$.  The final loop
(lines~\ref{line:beginSC}--\ref{line:endSC}) checks whether for each
configuration pattern $c_r \in C_r$, for all configuration $\gamma$
matching $c_r$, $\gamma$ also matches $c_t$; the algorithm returns
true iff this check is successful.

To better understand our reachability checking procedure, 
let us consider the following simple example:
\begin{lstlisting}[escapechar=~]	
	int sum(int t) 
	~\ml{$\left\langle \left\langle \frac{\texttt{sum(}x\texttt{)}}{y} \right\rangle _{k}	\right\rangle \land x \geq 0 \land y = x(x+1)/2$}\label{line:decAnnot}~
	{
	  int c;
	  c=0;
	  ~\ml{$ \exists i,j. \left\langle \left\langle \texttt{while(t>0)\{c=c+t;t=t-1;\}} \right\rangle _{k} \left\langle \texttt{t} \mapsto i, \texttt{c} \mapsto j \right\rangle _{\mathit{env}} \right\rangle \land i \geq 0 \land j = (h(h+1)-i(i+1))/2$} \label{line:decLoopAnnot}~
	  while(t>0){
	    c=c+t;
	    t=t-1;
	  }
	  return c;
	}
	\end{lstlisting}

This code snippet consists of a function which sums
the first \texttt{t} non-zero natural numbers.
The function contract (line~\ref{line:decAnnot}) specifies 
that the input parameter must be non-negative and the returned value must be the sum of the first \texttt{t} numbers.
The loop is annotated with an invariant (line~\ref{line:decLoopAnnot})
specifying that program variable \texttt{t} must be non-negative
and $c$ must contain the sum of the natural numbers between the current value of \texttt{t} and the initial value of \texttt{t}. 
The specifications provided as matching logic configuration patterns and
reachability rules are enclosed in a box like  \ml{specification}.

For this example, our reachability checking procedure would consider
two verification tasks: one for the function contract and one for the
loop invariant. Here, for brevity, we only illustrate how to perform
the verification task corresponding to the loop invariant.

Notice that the loop code (and its loop invariant) are the same
 as the ones  examined in Section~\ref{sec:loops}; following the same reasoning, we have to check formula~\eqref{eq:loopConfig1Ex}.
The three components of the verification task for the loop invariant
are:
\begin{itemize}
\item $C_r=\{
\exists i,j.
\left\langle
\left\langle \texttt{if(t>0)\{c=c+t;t=t-1;\}} \right\rangle _{k}
\left\langle \texttt{t} \mapsto i, \texttt{c} \mapsto j  \right\rangle _{\mathit{env}}
\right\rangle
\land i \geq 0
\land j = (h(h+1) - i(i+1))/2
\}$ (the same as formula \eqref{eq:loopConfig2Ex});
\item $R_v$ is the set containing the reachability rules  describing
  the semantics of the program; in this case, it includes the
  reachability rules for the \texttt{if} statement,
expression (variable) evaluation, assignment and comparison operations
(see Section~\ref{sec:semantics});
\item $c_t=\exists l,m.
\left\langle
\left\langle \cdot \right\rangle _{k}
\left\langle \texttt{t} \mapsto l, \texttt{c} \mapsto m \right\rangle _{\mathit{env}}
\right\rangle
\land l \geq 0
\land m = (h(h+1) - l(l+1))/2
$ (the same as formula \eqref{eq:loopConfig3Ex}).
\end{itemize}

According to algorithm \textsc{Process}, the first iteration of the application of  all rules in $R_v$ to
the configuration patterns in $C_r$ will derive the following configuration patterns:
\begin{equation}
\label{eq:vtExA}
\exists i,j.
\left\langle
\left\langle \texttt{c=c+t;t=t-1;} \right\rangle _{k}
\left\langle \texttt{t} \mapsto i, \texttt{c} \mapsto j \right\rangle _{\mathit{env}}
\right\rangle
\land i > 0
\land j = (h(h+1) - i(i+1))/2
\end{equation}
and
\begin{equation}
\label{eq:vtExB}
\exists i,j.
\left\langle
\left\langle \cdot \right\rangle _{k}
\left\langle \texttt{t} \mapsto i, \texttt{c} \mapsto j \right\rangle _{\mathit{env}}
\right\rangle
\land i = 0
\land j = h(h+1)/2.
\end{equation}
Both patterns are satisfiable and are added to the set $C_r$. A second
iteration of rules application will derive from  configuration
pattern~\eqref{eq:vtExA} another pattern:
\begin{equation}
\label{eq:vtExA1}
\exists i.
\exists j.
\left\langle
\left\langle \cdot \right\rangle _{k}
\left\langle \texttt{t} \mapsto i-1, \texttt{c} \mapsto j+i \right\rangle _{\mathit{env}}
\right\rangle
\land i > 0
\land j = (h(h+1) - i(i+1))/2.
\end{equation}
This iteration will reach the fixed point, with $C_r$ containing 
configuration patterns~\eqref{eq:vtExB} and~\eqref{eq:vtExA1}.  Both
configuration patterns are final, so algorithm~\textsc{Process} terminates.

Then, algorithm~\textsc{Check} ensures that both
patterns~\eqref{eq:vtExB} and~\eqref{eq:vtExA1} match pattern $c_t$.
Configuration pattern \eqref{eq:vtExB} trivially matches $c_t$ for
$l=i \land m=j$. Configuration pattern~\eqref{eq:vtExA1} matches
pattern $c_t$ for $l=i-1$ and $m=j+i$, since $i>0$ entails that
$i+1>0$, which corresponds to $i \geq 0$ of the constraint in $c_t$.
Then, by substituting $l$ and $m$ into the second constraint, we
obtain $j+i = (h(h+1) - i(i-1))/2$. Since from the constraint of
pattern~\eqref{eq:vtExA1} $j = (h(h+1) - i(i+1))/2$, we have
that $(h(h+1) - i(i+1))/2 + i = (h(h+1) - i(i-1))/2$. By
simplifying the last expression, we obtain $i - i^2 = i - i^2$.
Finally, algorithm \textsc{Check} returns true since both checks are
successful.

\subsection{Semantic evaluation for reachability checking}
\label{sec:attriubte-description}

We now present our semantic procedure, which encodes
the approach presented in Section~\ref{sec:overview} using the
evaluation of semantic attributes.

We associate the following attributes  to each node in the parse tree:
\begin{itemize}
 \item $K$, the program	fragment underlying the subtree rooted in the node.
\item $R$, the set of instances of matching logic reachability
        rules defining the semantics of the code fragment(s) contained
        in $K$, and instantiated for this code. Note that we use the
        reachability rules defined in Section~\ref{sec:semantics} as
        templates which are instantiated on the actual code of the
        program to be verified.
 \item $V$, a set of pending \emph{verification tasks}.
 \item $C$, a boolean flag indicating whether the specification can be satisfied,
	 according to the information available at the node.
 \item Two attributes $S$ and $F$ to handle \texttt{struct} definitions.
\end{itemize}

\begin{figure}
	\begin{scriptsize}
\begin{multicols}{2}
			\begin{algorithmic}[1]
				\Function{CompAttribute}{$a_1, ... a_n$}\label{line:cattr}				
\State $K \leftarrow \mathit{compose}(a_1.K, \ldots, a_n.K)$\label{line:k}
				\State $R_{\mathit{temp}} \leftarrow \mathit{gen}(K)$\label{line:rt}				
				\State $R \leftarrow \bigcup_{i=1}^{n}a_i.R \cup R_{\mathit{temp}}$\label{line:r}
				\State $V_{\mathit{temp}} \leftarrow \bigcup_{i=1}^{n}a_i.V$\label{line:vt}	
				\State $C \leftarrow \bigwedge_{i=1}^{n}a_i.C$\label{line:ct}	
				\If{$\mathit{hasContract}()$}\label{line:beginContract}	
				\State $V_{\mathit{temp}} \leftarrow V_{\mathit{temp}} \cup \mathit{genVT}(R)$\label{line:genVT}		
				
				\EndIf\label{line:endContract}	
				\State $V \leftarrow \emptyset$				
				\For{$v_i \in V_{\mathit{temp}}$}\label{line:beginFor}
				\State $v_i^\prime \leftarrow v_i$ \State $v_i^\prime.R_v \leftarrow v_i^\prime.R_v \cup R_{\mathit{temp}}$\label{line:vip}
				\State  $v_i^\prime \leftarrow\textsc{Process}(v_i^\prime)$\label{line:Process}
				\State $\mathit{res} \leftarrow\textsc{Check}(v_i^\prime)$\label{line:Check}		
				\If{$\mathit{res} = \mathit{delay}$}\label{line:beginV}				
				\State $V \leftarrow V \cup v_i^\prime$\label{line:v}	
				\EndIf\label{line:endV}		
				\If{$\mathit{res} = \mathit{false}$}\label{line:beginC}					
				\State $C \leftarrow \mathit{false}$\label{line:setFalse}	
				\EndIf\label{line:endC}					
				\EndFor\label{line:endFor}
				\If{$\mathit{isRootNode}() \land V \neq \emptyset$}\label{line:beginCheck}
				\State $C \leftarrow \mathit{false}$
\EndIf\label{line:endCheck}
				\State $a_0 \leftarrow (K,R,V,C)$			
				\State\textbf{return }$a_0$	
				\EndFunction
			\end{algorithmic}
		\end{multicols}
\end{scriptsize}
	\caption{The \textsc{CompAttribute} algorithm for semantic attribute evaluation}
	\label{fig:algo2}
	\vspace{-2mm}
\end{figure}

Our syntax-driven reachability checking procedure computes the
attributes of each node in the parse tree according to
algorithm~\textsc{CompAttribute}, shown in figure~\ref{fig:algo2}.
Since  attributes are evaluated during parsing,
~\textsc{CompAttribute} is called by the parser as it
proceeds bottom-up.  The algorithm takes as input the attributes
$a_1, \ldots, a_n$ computed for all the $n$ children of the node.
Each $a_i$ is a tuple containing the attributes of a child
node.  Node attributes are denoted by a dot notation; e.g., attribute $K$  of the first child is $a_1.K$; when clear from the context, 
we refer to attribute $a_i.K$ of $\mathit{node}_i$ with the shortcut $K_i$.

Attribute $K$ is built by suitably composing the 
children's $K$ attributes
(line~\ref{line:k})\footnote{For the leaves of the parse tree, $K$ is directly defined by the
corresponding terminal symbol.}.

Attribute $R$ (lines~\ref{line:rt}--\ref{line:r}) is the union of the children's $R$ attributes and the reachability
rules created for the node using its code (computed by invoking
function $\mathit{gen}(K)$).
For the leaf nodes, $R$ is obtained by instantiating the reachability
rules according to specific templates associated with each possible
production of the grammar.

Attribute $V$ contains the set of verification tasks available in each
node. A verification task $\langle C_r, R_v, c_t \rangle$
is generated as soon as an annotated loop, an annotated function, or
the definition of the \texttt{main} function is encountered: $C_r$ and
$c_t$ are generated according to the node attribute $K$, while $R_v$
is set to the value of the attribute $R$.   Since a verification task is instantiated as soon as the definition of
a verification unit is found, some of the 
required reachability rules may be
missing; this situation will be described below (Section~\ref{sec:synt-based-eval}).

Attribute $C$ is initially computed as the
conjunction of the corresponding attributes $C$
of the children (line~\ref{line:ct}),
since if one child does not satisfy a verification
task, the entire program is incorrect.
If a verification task is not satisfied,
attribute $C$ is set to false (lines~\ref{line:beginC}--\ref{line:endC}).
If some verification tasks are still not completed 
in the root node (lines~\ref{line:beginCheck}--\ref{line:endCheck}),
attribute $C$ is also set to $\mathit{false}$.

We now outline how program verification (in the generalized meaning
defined in the context of matching logic) is performed by exploiting
these attributes.  Essentially, given a program $P$ to verify, its
correctness is determined by the value of attribute $C$ of the root
node $r$ of $P$ parse tree.  The verification procedure starts by
building the parse tree of $P$ and then computes the attributes of the
internal nodes in a bottom-up way, until reaching node $r$.  However,
the application of algorithm \textsc{CompAttribute} on node $r$
requires the value of the attributes $a_1, \ldots, a_n$ of the
children of $r$.  Thus, algorithm \textsc{CompAttribute} has to be
recursively applied to the children of the root node, until the leaves
of the parse tree are reached and no further call to
\textsc{CompAttribute} is required.

In other words, program (interpretation and) verification is performed
through a bottom-up attribute evaluation; however, completing these
evaluation requires, once the root node is reached, to execute a
second traversal\footnote{In the next sections we will show how some
  optimizations may anticipate the execution of some verification
  (sub)tasks already during the construction of the parse tree and the
  evaluation of its attributes.} of the tree, in a top-down fashion.

The following two subsections provide some technical details and examples about our syntax-driven verification approach.

\subsubsection{Syntax-based evaluation of verification
  tasks}\label{sec:synt-based-eval}
As mentioned above, some of the required reachability rules may be
missing when  a verification task is instantiated, since the
instantiation happens as soon as the definition of a verification unit
is found. This is illustrated by verification of the following snippet, which includes  function \texttt{sum} from the example in
Section~\ref{sec:overview} and its generalization \texttt{anySum},
which accepts as parameter any integer (not
necessarily positive):
\begin{lstlisting}[escapechar=~,mathescape]	
int sum(int t) 
~\ml{$\left\langle \left\langle \frac{\texttt{sum(}x\texttt{)}}{y} \right\rangle _{k}	\right\rangle \land x \geq 0 \land y = x(x+1)/2$}~
{
  int c;
  c=0;
  ~\ml{$ \exists i,j. \left\langle \left\langle \texttt{while(t>0)\{c=c+t;t=t-1;\}} \right\rangle _{k} \left\langle \texttt{t} \mapsto i, \texttt{c} \mapsto j \right\rangle _{\mathit{env}} \right\rangle \land i \geq 0 \land j = (h(h+1)-i(i+1))/2$} ~
  while(t>0){
    c=c+t;
    t=t-1;
  }
  return c;
}

int anySum(int t)
~\ml{$\left\langle \left\langle \frac{\texttt{anySum(}x\texttt{)}}{y} \right\rangle _{k}	\right\rangle \land x \geq 0  \land y = x(x+1)/2 \lor x < 0  \land y = x(x-1)/2$}~
{
  int i;
  i = t;
  if(i < 0){
    i = -i;
  }
  return sum(i);~\label{line:result}~
}

\end{lstlisting}

\begin{figure}
	\centering	
	\begin{tikzpicture}[scale=0.8,level distance=40pt]
\Tree[.\refnt{function_def}{1}
	\nt{type}{2}
	\tn{IDENTIFIER}{3} [.\nt{function_def2}{4}
	\nt{parameter}{5}   \nt{compound_stm}{6}  \refnt{function_def}{7}
	]
	]				
	\end{tikzpicture}	
	\caption{Sample parse tree}	
	\label{fig:sampleParseTree}
\end{figure}
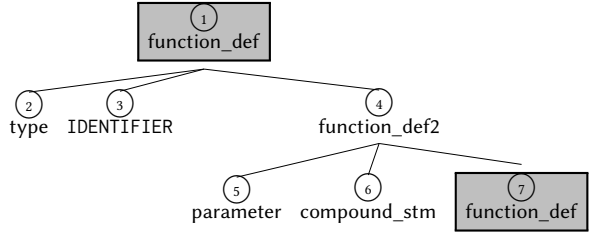

A sketch of the parse tree is shown in
Figure~\ref{fig:sampleParseTree}; the definitions of functions
\texttt{sum} and  \texttt{anySum} are  contained in nodes \skipnt{function_def}{1} and
\skipnt{function_def2}{4}, respectively,  When following a bottom-up parsing strategy,
\textnt{function_def2}{4} is evaluated before \textnt{function_def}{1}.  This means that, when the verification task
associated with the \texttt{anySum} function is performed, the
definition of function \texttt{sum} is not available.  However,
to verify \texttt{anySum}, the semantics
of \texttt{sum} (invoked at line~\ref{line:result}) is required.
In such a situation, we define the  verification task as
\emph{incomplete}.
The idea is that access to missing semantic information interrupts the verification task. The interrupted task is propagated upwards the tree, until the missing information needed to resume  verification becomes available.

To deal with delayed verification tasks, we have to modify the
\textsc{Check} algorithm in figure~\ref{fig:algo}, described in
Section~\ref{sec:overview}.  While the original formulation assumed
that $R_v$ contains the complete  semantics, and thus the
\textsc{Process} algorithm could complete the execution of the
verification task, the modified version must also handle the case in
which $R_v$ is not sufficient to complete the execution of the
verification task in \textsc{Process}.  In this case, a
partially-processed verification task can be supported by having the
function \textsc{Check} return a
$\mathit{true}$/$\mathit{false}$/$\mathit{delay}$ answer. A
\textit{delay} answer indicates that the semantic information currently available
in $R_v$ is incomplete and does not  allow algorithm~\textsc{Process} to complete
the verification task. In this case,
algorithm~\textsc{Check} cannot determine whether the specification is
satisfied or violated. 
The change to algorithm \textsc{Check} can be accomplished by modifying
the definition of the \textsc{CheckNonFinal} function as shown in
figure~\ref{fig:algo1}: the function will now return $\mathit{delay}$
(line~\ref{line:delay-1}, which corresponds to line~\ref{line:delay}
in figure~\ref{fig:algo})
if a non-final configuration is found.

\begin{figure}
	\begin{center}
		\begin{scriptsize}
			\begin{algorithmic}[1]				
				\Function{CheckNonFinal}{$C_r$}
				\For{$c_i \in C_r$}
				\If{$\mathit{isNonFinal}(c_i)$}
				\State\textbf{return }$\mathit{delay}$\label{line:delay-1}
				\EndIf	
				\EndFor 
				\State\textbf{return }$\mathit{false}$
				\EndFunction
			\end{algorithmic}
		\end{scriptsize}
	\end{center} 
	
	\caption{The \textsc{CheckNonFinal} function changed to handle
		delayed verification tasks}
	\label{fig:algo1}
	\vspace{-2mm}
\end{figure}

To compute attribute $V$, we first
collect the incomplete verification tasks from children nodes and save
them in the temporary variable $V_{\mathit{temp}}$
(line~\ref{line:vt} in figure~\ref{fig:algo2}).  If the current node is annotated with a
specification, we generate a new verification task with function
$\mathit{genVT}$
(lines~\ref{line:beginContract}--\ref{line:endContract}).  We then
update the reachability rules of every verification task, adding the
new rules generated in the node (line~\ref{line:vip}), and then we
execute them through the \textsc{Process} function (line~\ref{line:Process}). 
Only the verification tasks that cannot be fully performed
are retained (lines~\ref{line:beginV}--\ref{line:endV}).

\subsubsection{Example of attributes computation}

To exemplify the \textsc{CompAttribute} algorithm, consider the following simple annotated \texttt{while} loop:
\begin{lstlisting}[escapechar=~]	
~\ml{$ \exists i,j. \left\langle \left\langle \texttt{while(t>0)\{c=c+t;t=t-1;\}} \right\rangle _{k} \left\langle \texttt{t} \mapsto i, \texttt{c} \mapsto j \right\rangle _{\mathit{env}} \right\rangle \land i \geq 0 \land j = (h(h+1)-i(i+1))/2$} \label{line:tloopannot}~
while(t>0){
	c=c+t;
	t=t-1;
}
\end{lstlisting}
whose parse tree fragment is:\\
\begin{center}
\begin{tikzpicture}[scale=0.8,level distance=40pt]
\Tree[.\refnt{stm}{9}
\tn{Annotation}{0} \tn{`while'}{1}  \refnt{relat_exp}{2} [.\refnt{compound_stm}{8} \refnt{exp}{3} \tn{`;'}{4} [.\refnt{stm}{7} \refnt{exp}{5} \tn{`;'}{6} ] ]
]			
\end{tikzpicture}
\end{center}

The corresponding attributes are:
\begin{itemize}
\item $K_2=\{\texttt{t > 0}\}$, $K_3=\{\texttt{c = c + t}\}$,
  $K_5=\{\texttt{t = t - 1}\}$, $K_7=\{\texttt{t = t - 1;}\}$,\\
$K_8=\{\texttt{c = c + t; t = t - 1;}\}$,
$K_9=\{\texttt{while(t > 0)\{c = c + t; t = t - 1;\}}\}$;
\item $R_2 = \{\eqref{eq:attributeEx}\}$, $R_7 = \{\eqref{eq:attributeEx1}\}$,
$R_8 = \{\eqref{eq:attributeEx1}, \eqref{eq:attributeEx2}\}$, $R_9=R_2 \cup R_8 \cup \{\eqref{eq:loopInvariantEx1},
\eqref{eq:iteTrue}, \eqref{eq:iteFalse}\}$;
\item $V_2=V_3=V_5=V_7=V_8=V_9=\emptyset$\footnote{The verification
task set attribute of \textnt{stm}{9} would be
 $V_T=\{\langle \{\eqref{eq:loopConfig2Ex}\}, R_9,
\eqref{eq:loopConfig3Ex} \rangle\}$.  However, notice that the newly
generated verification task is completely processed by algorithm
\textsc{Process} (line \ref{line:Process}) inside the node, since all
the necessary reachability rule instances are available, thus
$V_9=\emptyset$.};
\item $C_2 = C_3 = C_5 = C_7 = C_8 = C_9 = \mathit{true}$;
\end{itemize}
where the additional reachability rules \eqref{eq:attributeEx}--\eqref{eq:iteFalse} are defined below:

\begin{flalign}
\label{eq:attributeEx}
&\left\langle
\left\langle \frac{\texttt{t > 0}}{j} \right\rangle _{k}
\left\langle \texttt{t} \mapsto i \right\rangle _{\mathit{env}}
\right\rangle
\land j = i > 0\\
\label{eq:attributeEx1}
&\left\langle
\left\langle \frac{\texttt{t = t - 1;}}{\cdot} \right\rangle _{k}
\left\langle \texttt{t} \mapsto \frac{i}{j} \right\rangle _{\mathit{env}}
\right\rangle
\land j = i - 1\\
\label{eq:attributeEx2}
&
\left\langle
\left\langle \frac{\texttt{c = c + t;}}{\cdot} \right\rangle _{k}
\left\langle \texttt{c} \mapsto \frac{i}{j} \texttt{t} \mapsto l \right\rangle _{\mathit{env}}
\right\rangle
\land j = i + l\\
\label{eq:iteTrue}
&\left\langle
\left\langle \frac{\texttt{if(}b\texttt{) \{c=c+t; t=t-1;\}}}{\texttt{c=c+t; t=t-1;}} \dots \right\rangle _{k}
\right\rangle \land b\neq 0\\
\label{eq:iteFalse}
&\left\langle
\left\langle \frac{\texttt{if(}b\texttt{) \{c=c+t; t=t-1;\}}}{\cdot} \dots \right\rangle _{k}
\right\rangle \land b = 0
\end{flalign}

\section{A Complete Example}
\label{sec:running-example}

\begin{figure}[p]
\begin{lstlisting}[escapechar=~]	
#include <stdlib.h>
#include <stdio.h>	
struct listNode {~\label{line:beginStruct}~
 int val;~\label{line:val}~
 struct listNode *next;~\label{line:next}~
};~\label{line:endStruct}~	
struct listNode* swap(struct listNode* x){
 struct listNode* p;~\label{line:pDecl}~
 if(x != 0){~\label{line:validListComp}~
  if(x->next->val - x->val <0){~\label{line:ltComp}~
   p = x->next;~\label{line:pAssign}~
   x->next = p->next;~\label{line:xNextAssign}~
   p->next = x;~\label{line:pNextAssign}~
   return p;~\label{line:retP}~
  }~\label{line:InnerIfEnd}~
 }~\label{line:OuterIfEnd}~
 return x;~\label{line:retX}~
}	
struct listNode* min(struct listNode* x)~\label{line:beginMin}~	
~\ml{${\left\langle \left\langle \frac{\texttt{min(}i\texttt{)}}{\mathit{j}} \dots \right\rangle _{k} \left\langle \frac{i \mapsto \mathit{list}(A)}{j \mapsto \mathit{list}(A^{\prime})} \dots \right\rangle _{\mathit{mem}}\right\rangle}$}\label{line:beginMinAnnot}~
~\ml{${\land i\neq 0 \land j \neq 0 \land \mathit{listPermutation}(A,A^\prime) \land \mathit{leq}([\mathit{head}(A^\prime)],A^\prime)}$}\label{line:endMinAnnot}~	
{
 if(x->next != 0) {
  x->next = min(x->next);
 }
 return swap(x);~\label{ln:min:swap}~	
}~\label{line:endMin}~	
struct listNode* sort(struct listNode* x)	
~\ml{${\left\langle \left\langle \frac{\texttt{sort(}i\texttt{)}}{\mathit{j}} \dots \right\rangle _{k} \left\langle \frac{i \mapsto \mathit{list}(A)}{j \mapsto \mathit{list}(A^{\prime})} \dots \right\rangle _{\mathit{mem}}\right\rangle\land \mathit{listPermutation}(A,A^\prime) \land \mathit{isSorted}(A^\prime)}$}\label{line:sortAnnot}~	
{  
 struct listNode* p;
 struct listNode* y;
 if(x == 0){
  return x;
 }
 y = min(x);
 x = y->next;
 y->next = 0;
 p = y;
~\ml{$ \exists i,l,v,B,C.	    \langle	    \left\langle\texttt{while(x != 0)\{\ldots\}} \dots \right\rangle _{k}	    \left\langle\texttt{x}\mapsto i, \texttt{y}\mapsto j, \texttt{p}\mapsto l\right\rangle_{\mathit{env}}$}\label{ln:sort-inv-b}~
~\ml{$\left\langle j \mapsto \mathit{lseg}(l, B), l \mapsto[v,0], i \mapsto \mathit{list}(C)  \right\rangle_{\mathit{mem}}	    \rangle$}\label{ln:sort-mem}~
~\ml{$\land \mathit{listPermutation}(A, B@[v]@C) \land \mathit{leq}([v],C) \land	   \mathit{leq}(B,[v]) \land \mathit{isSorted}(B)$}\label{ln:sort-inv-e}~
 while(x != 0){
  p->next = min(x);
  p = p->next;
  x = p->next;
  p->next = 0;
 }
 return y;
}
\end{lstlisting}
	\caption{List sort example (see
          Fig.~\ref{exampleListSwapSortLegend} for the functions used in the
          \ml{specifications})}
        \label{exampleListSwapSort}
\end{figure}

\begin{figure}
	\begin{footnotesize}
          \begin{itemize}
	\item	$\mathit{list}(A)$ 
		creates a heap object for a \texttt{struct listNode} list,
		containing the elements of the list $A$.
\item $\mathit{lseg}(i, A)$
		creates a heap object for a fragment of a
                \texttt{struct listNode} list, which contains the
                elements of $A$ up to (and including) the first one whose
                \texttt{next} field address is equal to variable $i$.
\item $\mathit{listPermutation}(A, B)$ indicates whether
		 $A$ is a permutation of $B$.
		\item $\mathit{leq}(A, B)$ indicates whether 
		every element of $A$ is less than or equal to every
		element of $B$.
		\item ${[}i{]}$ is a list constructor.
		\item $\mathit{head}(A)$ returns the first element
		of list $A$.
		\item $\mathit{isSorted}(A)$
                  indicates whether $A$ is sorted.
		\item $A @ B$ represents the list concatenation
                  function.
	\end{itemize}
	\end{footnotesize}
	\caption{Signatures and description of the functions and
          predicates used in
          the \ml{specifications} of the example in Fig.~\ref{exampleListSwapSort}; $A$ and $B$ are
      lists, $i$ is an integer}
	\label{exampleListSwapSortLegend}
\end{figure}

\lstset{basicstyle=\normalsize}

In this section we present a complete example of application of our
reachability checking procedure on a sorting program, shown in
figure~\ref{exampleListSwapSort}, which contains  function
\texttt{sort} that takes as input a pointer to a list and returns a
pointer to its sorted rearrangement.  Function \texttt{sort} invokes function \texttt{min},
which takes as input a pointer to a list and returns a pointer to a
list which is a permutation of the input list and in which the lowest
value is at the beginning of the list.  Function \texttt{min}, in turn, invokes
function \texttt{swap}.  

Functions \texttt{sort} and \texttt{min} are annotated with a matching
logic specification, while function \texttt{swap} is not; the
functions and predicates used in the \ml{specifications} of
Fig.~\ref{exampleListSwapSort} are described in
Fig.~\ref{exampleListSwapSortLegend}.
The specification of
function \texttt{min}
(lines~\ref{line:beginMinAnnot}--\ref{line:endMinAnnot}) states that
upon execution of the function, the heap content changes from a list
$A$ whose first element is pointed to by parameter $i$ (as indicated
by the mapping $i \mapsto \mathit{list}(A)$ in cell $\mathit{mem}$) to
a new list $A^\prime$, whose first element is pointed by pointer $j$
(specified by the mapping $j \mapsto \mathit{list}(A^\prime)$ in cell
$\mathit{mem}$).  List $A^\prime$ has the same elements as list $A$
(specified by the predicate $\mathit{listPermutation}(A,A^\prime)$)
and the first element of $A^\prime$ is less than or equal to any other
element of $A^\prime$ (specified by the predicate
$\mathit{leq}([\mathit{head}(A^\prime)],A^\prime)$). 

The specification of function \texttt{sort}
(line~\ref{line:sortAnnot}) states that upon the invocation of the
function, list $A$, whose first element is pointed to by parameter
$i$, is transformed into list $A^\prime$, whose first element is
pointed to by the returned value $j$. List $A^\prime$ has the same
elements as list $A$ and it is sorted (as specified by the predicate
$\mathit{isSorted}(A^\prime)$).  The function works by repeatedly
removing the minimum element from the list pointed to by \texttt{x},
and appending it to the list pointed to by \texttt{y}; this behavior
is specified by the loop invariant at lines~\ref{ln:sort-inv-b}--\ref{ln:sort-inv-e}.
More precisely, the invariant specifies that:
\begin{itemize}
\item 
the elements of the  list $A$ are contained either in the list
pointed to by  \texttt{x}  or in the one pointed to by
\texttt{y} (as indicated by the mappings in the cell \emph{mem}
(line~\ref{ln:sort-mem}) and by the predicate
$\mathit{listPermutation}(A, B@[v]@C)$),
where $v$ is the element pointed by \texttt{p},
which is the last element of the sorted list segment $B@[v]$;
\item the list pointed by \texttt{y} is sorted
(as indicated by the predicate
$\mathit{leq}(B,[v]) \land \mathit{isSorted}(B)$);
\item   the last element
of the list pointed by \texttt{y} is less than or equal to all the
elements of the list pointed by \texttt{x} (specified by predicate
$\mathit{leq}([v],C)$). 
\end{itemize}

Notice that there is a bug in function \texttt{swap}: field
\texttt{x->next} is dereferenced at line~\ref{line:ltComp} without
first performing a \texttt{NULL} pointer check.

\begin{figure}	
	\centering
	\begin{tikzpicture}[scale=0.6,level distance=60pt]
	\Tree[.\refnt{compound_decl}{28}
	[.\nt{parameter}{36} \edge[roof]; \#\ref{line:pDecl} ]
	[.\refnt{compound_stm}{43}
	\tn{`if'}{44}  
	[.\refnt{eq_exp}{46} \edge[roof]; \#\ref{line:validListComp} ] [.\refnt{stm}{49}
	\tn{`if'}{50}  [.\refnt{relat_exp}{51}  \edge[roof]; \#\ref{line:ltComp} ]   [.\refnt{compound_stm}{52} [.\refnt{exp}{87} \edge[roof]; \#\ref{line:pAssign} ] [.\refnt{compound_stm}{94}
	[.\refnt{exp}{95} \edge[roof]; \#\ref{line:xNextAssign} ]
	[.\refnt{compound_stm}{99}
	[.\refnt{exp}{100} \edge[roof]; \#\ref{line:pNextAssign} ] [.\refnt{stm}{101} \edge[roof]; \#\ref{line:retP} ]
	]
	]	 ]
	] [.\nt{stm}{53} \edge[roof]; \#\ref{line:retX} ]
	]
	]
	\end{tikzpicture}
	\caption{Simplified abstract syntax tree of the \texttt{swap}
          function. Triangle-shaped leaves correspond to subtrees not
          expanded; the numbers on them 
 refer to the corresponding line numbers in Fig.~\ref{exampleListSwapSort}.}
	\label{fig:swapParseTree}
\end{figure}
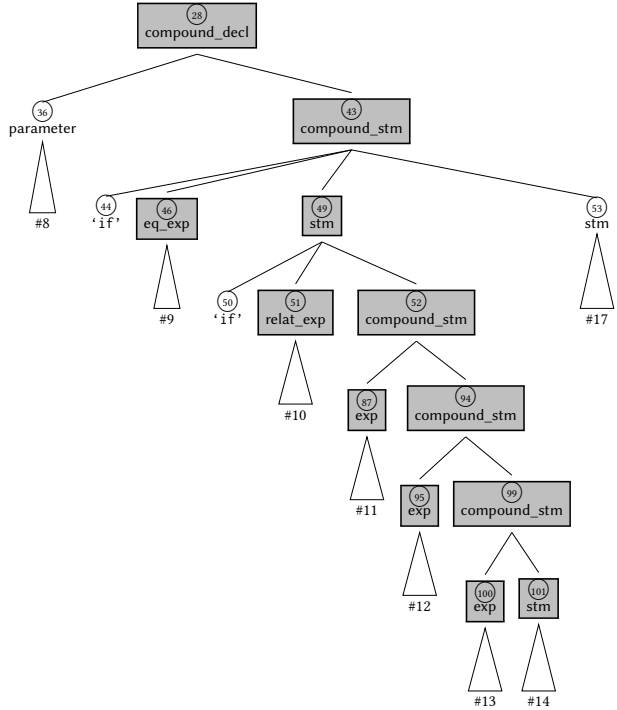

A sketch of the abstract syntax tree of the example is shown in
figure~\ref{fig:global-tree}. For space reasons, we only describe the
application of our approach to function \texttt{swap}; its
(simplified) abstract syntax tree is shown in
figure~\ref{fig:swapParseTree}, in which the triangle-shaped leaves
correspond to subtrees not expanded.
Throughout the example we use the abbreviations for code/environment
snippets shown in figure~\ref{eq:abbrv}.

\begin{figure}[hbt]
\begin{footnotesize}
\noindent
\begin{minipage}[t]{.45\textwidth}
\begin{equation*}
\begin{array}{lll}
m				& \equiv & a \mapsto \mathit{list}(A) \\
m^\prime		& \equiv & a \mapsto [v,d], d \mapsto \mathit{list}(B)  \\
m^{\prime\prime}			& \equiv & a \mapsto [v,d]  \\
e_0		& \equiv & \texttt{x} \mapsto (\mathit{listNode*}, a)\\
e					& \equiv & \texttt{x} \mapsto (\mathit{listNode*}, a), \texttt{p} \mapsto \mathit{undef}\\
p					& \equiv &  A\neq \mathit{nil} \land a\neq 0 \land A=[v]@B \\
p^\prime	& \equiv & p \land d = 0\\
\mathit{COND}            & \equiv &                                  \text{read(read(\lstinline[basicstyle=\ttfamily]!x->next!)\lstinline[basicstyle=\ttfamily]!->val!)~\lstinline[basicstyle=\ttfamily]!-!}\\
& & \text{read(\lstinline[basicstyle=\ttfamily]!x->val!)~\lstinline[basicstyle=\ttfamily]!< 0!}\\
\mathit{COND1}            & \equiv & \text{\lstinline*x != 0*}\\
\mathit{ASSIGN}			& \equiv &
                                           \text{\lstinline[basicstyle=\ttfamily]!x->next = !~read(\lstinline[basicstyle=\ttfamily]!p->next!)}\\
& & \text{\lstinline[basicstyle=\ttfamily]!; p->next = x;!}\\
\end{array}
\end{equation*}
\end{minipage}
\hfill
\begin{minipage}[t]{.45\textwidth}
\begin{equation*}
\begin{array}{lll}
\mathit{THEN1}            & \equiv & \text{\lstinline[basicstyle=\ttfamily]!if(!COND\lstinline[basicstyle=\ttfamily]!)\{!~THEN~\lstinline[basicstyle=\ttfamily]!\}!}\\
\mathit{THEN2}			& \equiv & \text{\lstinline[basicstyle=\ttfamily]!x->next = min(x->next); !}\\
\mathit{OTH}			& \equiv & \text{\lstinline[basicstyle=\ttfamily]!\{x->next = min(x->next);\} !}~\mathit{OTH1}\\
\mathit{OTH1}			& \equiv & \text{\lstinline[basicstyle=\ttfamily]!return swap(x);!}\\
\mathit{OTH2}			& \equiv & \text{\lstinline[basicstyle=\ttfamily]!\{!} \mathit{THEN1} \text{\lstinline[basicstyle=\ttfamily]!\} return x;!}\\
\mathit{OTH3}			& \equiv & \text{\lstinline[basicstyle=\ttfamily]!-!}\mathit{read}(\text{\lstinline[basicstyle=\ttfamily]!x->val!}) \text{\lstinline[basicstyle=\ttfamily]!)\{!}\mathit{THEN}\text{\lstinline[basicstyle=\ttfamily]!\} return x;!} \\
\mathit{SWAP}			& \equiv & \text{\lstinline[basicstyle=\ttfamily]!struct listNode* p; !~}\text{\lstinline[basicstyle=\ttfamily]!if(!}\mathit{COND1}\text{\lstinline[basicstyle=\ttfamily]!)!~}\mathit{OTH2}\\
\mathit{MIN}			& \equiv & \text{\lstinline|if(x->next!=0) |~}\mathit{OTH}\\
\mathit{EM}				& \equiv & \left\langle e \right\rangle _{\mathit{env}}
\left\langle m^{\prime\prime} \dots \right\rangle _{mem}\\
V_{t\mathit{MIN}}		& \equiv & \left\langle\{\eqref{eq:step1}\}, R_{29}, \eqref{eq:stopMin}\right\rangle\\
\end{array}
\end{equation*}
\end{minipage}
\end{footnotesize}
\caption{List of abbreviations}
\label{eq:abbrv}
\end{figure}
 
\begin{figure}
	\centering		
	\tikzset{every tree node/.style={align=center,anchor=north}}
	\begin{tikzpicture}[scale=0.6,level distance=1.5cm]
	\Tree[.\nt{program}{1}
	[.\nt{global_decl}{2}
	$\ldots$	
	[.\nt{struct_field_list}{5} \edge[roof]; \refsnt{\texttt{struct listNode} definition}{} ]
	[.\nt{global_decl}{8} 
	$\ldots$
	[.\spec{\texttt{swap} definition}{function_def}{9}
	[.\nt{function_def2}{26}
	$\ldots$
	[.\nt{compound_decl}{28} \edge[roof]; \refsnt{\texttt{swap} body}{}  ]
	  [.\spec{\texttt{min} definition}{function_def}{29}
	[.\nt{function_def2}{130} 
	$\ldots$	 
	[.\nt{compound_stm}{132} \edge[roof]; \refsnt{\texttt{min} body}{} ]
	[.\nt{function_def}{133} \edge[roof]; \refsnt{\texttt{sort} definition}{} ]	  
	]
	]	 ]		
	]	
	]
	]
	]
	\end{tikzpicture}
	\caption{Sketch of the abstract syntax tree of the example in Fig.~\ref{exampleListSwapSort}}
	\label{fig:global-tree}
\end{figure}
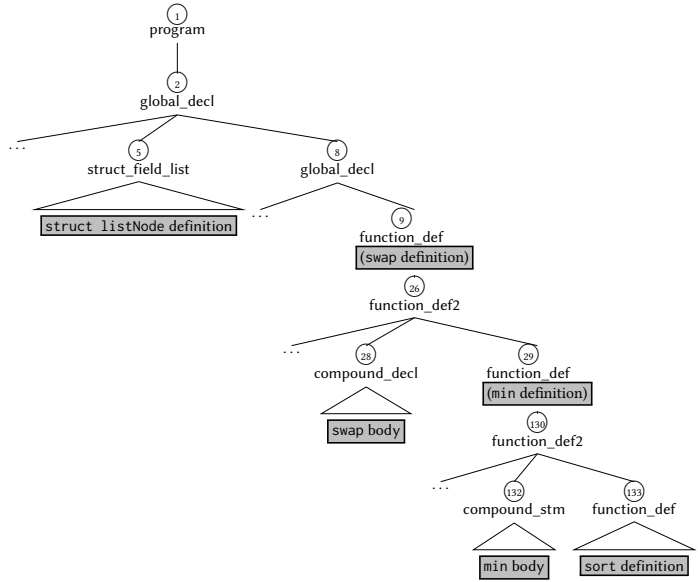

\subsection{Attribute evaluation}
The final values of all the attributes at the end of their evaluation
and the reachability rules referenced by them are reported,
respectively, in table~\ref{tb:KattributeValues} and in table~\ref{tb:reachabilityRules} whereas the details of the
step are explained in Appendix~\ref{sec:attributeEvaluation},  node by node. Here we  only illustrate the generation of the verification task for function \texttt{min} which is the core of the attribute of the corresponding subtree.

\paragraph[Verification Task Generation]{\parnt{function_def}{29} (figure~\ref{fig:global-tree}): verification task generation}
The specification of function \texttt{min}
(lines~\ref{line:beginMinAnnot}--\ref{line:endMinAnnot}) is translated
into the initial configuration pattern~\eqref{eq:startMin}:
\begin{equation}
\label{eq:startMin}
\exists a .
\left\langle
\left\langle \mathit{MIN}  \right\rangle_{k}
\left\langle \texttt{x} \mapsto (\mathit{listNode*},a) \right\rangle _{\mathit{env}}
\left\langle a \mapsto \mathit{list}(A) \right\rangle _{mem}
\right\rangle
\land p
\end{equation}
and target configuration pattern~\eqref{eq:stopMin}:
\begin{equation}
\label{eq:stopMin}
\exists a^{\prime} .
\left\langle
\left\langle \mathit{a^{\prime}}  \right\rangle_{k}
\left\langle a^{\prime} \mapsto \mathit{list}(A^{\prime}) \right\rangle _{mem}
\right\rangle
\land \mathit{listPermutation}(A,A^\prime) \land \mathit{leq}(\mathit{head}(A^\prime),A^\prime).
\end{equation}

The corresponding verification task, $\mathit{vt}_{29}=\left\langle\{\eqref{eq:startMin}\},
R_{29}, \eqref{eq:stopMin}\right\rangle$, is produced in \textnt{function_def}{29} by  function $\mathit{genVT}$  of the
\textsc{CompAttribute} procedure (figure~\ref{fig:algo2}, line~\ref{line:genVT}).

\begin{table}
	\caption{Values of the attributes for the example in
		Fig.~\ref{exampleListSwapSort} \label{tb:KattributeValues}}
	\centering
	\footnotesize	
\begin{tabular}{c}
\subfloat[Attributes: $K$ (list of code tokens), $R$
          (set of reachability rules), $V$ (set of verification
          tasks), $C$ (verification failed flag)]{\begin{tabular}{lllll}
		\toprule
		Node & $K$ & $R$ & $V$ & $C$\\
		\midrule			 	
		\tablent{exp}{100} & $\{\texttt{p->next = x;}\}$ & 
		$\{\eqref{eq:assignPnext}, \eqref{eq:evalX}, \eqref{eq:evalP}, \eqref{eq:memWrite}\}$ & $\emptyset$ & $\mathit{true}$ \\
		\tablent{stm}{101} & $\{\texttt{return p;}\}$ & $ \{\eqref{eq:evalP}, \eqref{eq:return}\}$ & $\emptyset$ & $\mathit{true}$ \\
		\tablent{exp}{95} & $\{\texttt{x->next = }\mathit{read}(\texttt{p->next})\texttt{;}\}$ & $\{\eqref{eq:assignXnext}, \eqref{eq:evalX}, \eqref{eq:evalP}, \eqref{eq:memWrite},  \eqref{eq:memRead}\}$ & $\emptyset$ & $\mathit{true}$\\
		\tablent{exp}{87} & $\{\texttt{p = }\mathit{read}(\texttt{x->next})\texttt{;}\}$ & $\{\eqref{eq:pAssign}, \eqref{eq:evalX}, \eqref{eq:memRead}\}$ & $\emptyset$ & $\mathit{true}$\\
		\tablent{compound\_stm}{99} & $\{\texttt{p->next = x; return p;}\}$ & $\{\eqref{eq:assignPnext}, \eqref{eq:evalX}, \eqref{eq:evalP}, \eqref{eq:memWrite}, \eqref{eq:return}\}$ & $\emptyset$ & $\mathit{true}$ \\
		\tablent{compound\_stm}{94} & $K_{95}K_{99}$ & $R_{95} \cup R_{99}$ & $\emptyset$ & $\mathit{true}$ \\
		\tablent{compound\_stm}{52} & $K_{87}K_{94}$ & $R_{87} \cup R_{94}$ & $\emptyset$ & $\mathit{true}$ \\
		\tablent{relat_exp}{51} & $\{\texttt{COND}\}$ & $\{\eqref{eq:doSub}, \eqref{eq:cmpTrue}, \eqref{eq:cmpFalse}, \eqref{eq:zero}, \eqref{eq:evalX}\}$ & $\emptyset$ & $\mathit{true}$ \\
		\tablent{stm}{49} & $\{\texttt{if(} \mathit{COND} \texttt{)\{} \mathit{THEN} \texttt{\}}\}$ & $\{\eqref{eq:ifTrue}, \eqref{eq:ifFalse}\} \cup R_{51} \cup R_{52}$ & $\emptyset$ & $\mathit{true}$ \\
		\tablent{eq_exp}{46} & $\{\texttt{x != 0}\}$ & $\{\eqref{eq:neqTrue}, \eqref{eq:neqFalse}, \eqref{eq:evalX}\}$ & $\emptyset$ & $\mathit{true}$ \\
		\tablent{compound\_stm}{43} & $\{\texttt{if(} \mathit{COND1} \texttt{)\{} \mathit{THEN1} \texttt{\} return x;}\}$ & $\begin{array}{l}\{\eqref{eq:if1True}, \eqref{eq:if1False}, \eqref{eq:return}, \eqref{eq:evalX}\}\\\cup R_{46} \cup R_{49}\end{array}$ & $\emptyset$ & $\mathit{true}$ \\
		\tablent{compound\_decl}{28} & $\{\texttt{struct listNode* p; }K_{43}\}$ & $\{\eqref{eq:defineP}\} \cup R_{43}$ & $\emptyset$ & $\mathit{true}$ \\
		\tablent{function_def}{9} & $\{(\mathit{SWAP})\}$ & $\{\eqref{eq:SwapCall}\} \cup R_{28} \cup R_{29}$ & $V_{t\mathit{MIN}}$ & $\mathit{true}$ \\
		\midrule
		\tablent{compound\_stm}{132} & $\{\mathit{MIN}\}$ & $\begin{array}{l}\{\eqref{eq:if2True}, \eqref{eq:if2False}, \eqref{eq:return}, \eqref{eq:memRead},\\ \eqref{eq:memWrite}, \eqref{eq:assignXnext} \cup R_{46}
		\}\end{array}$ & $\emptyset$ & $\mathit{true}$ \\
		\tablent{function_def}{29} & $\{\}$ & $\{\eqref{eq:MinCall}, \eqref{eq:MinCallError}\} \cup R_{132} \cup R_{133}$ & $V_{t\mathit{MIN}}$ & $\mathit{true}$ 
\\			
		\bottomrule
	\end{tabular}
}
\\
\subfloat[Attributes: $N$ (description of \texttt{struct} field), $S$
  (offset of \texttt{struct} field), $F$ ( \texttt{struct} fields description)]{\begin{tabular}{llll}
		\toprule
		Node & $N$ & $S$ & $F$ \\
		\midrule	
		\tablent{parameter}{13} & $(\texttt{next},
                                     \mathit{listNode*})$ & N/A & N/A \\				
		\tablent{struct_field_list}{7} & & $1$ & $\{(\texttt{next}, \mathit{listNode*}, 0)\}$\\
		\tablent{parameter}{6} & $(\texttt{val}, \mathit{int})$ & N/A & N/A \\			
		\tablent{struct_field_list}{5} & & $2$ & $\{(\texttt{val}, \mathit{int}, 0),(\texttt{next}, \mathit{listNode*}, 1)\}$\\					
		\bottomrule
	\end{tabular}		
}\end{tabular}

\end{table}

 \begin{table}
	\caption{Reachability rule instances\label{tb:reachabilityRules}}
	\centering
			\begin{tabular}{cll}
			\toprule
			Node & Reachability rule & \\ \midrule			
			\tablent{exp}{100} & 
			$
			\left\langle
			\left\langle \frac{\texttt{p->next=}i\texttt{;}}{\mathit{write}(\texttt{p->next}, i)} \dots \right\rangle _{k}
			\right\rangle
			$ & \tagarray\label{eq:assignPnext}\\[15pt]	
			\tablent{exp}{100} & 
			$
			\left\langle
			\left\langle \frac{\texttt{x}}{(\mathit{typeX},i)} \dots \right\rangle _{k}
			\left\langle \texttt{x} \mapsto (\mathit{typeX},i)\dots \right\rangle _{\mathit{env}}
\right\rangle
			$ & \tagarray\label{eq:evalX}\\[15pt]	
			\tablent{exp}{100} & 
			$
			\left\langle
			\left\langle \frac{\texttt{p}}{(\mathit{typeP},i)} \dots \right\rangle _{k}
			\left\langle \texttt{p} \mapsto (\mathit{typeP},i)\dots \right\rangle _{\mathit{env}}
\right\rangle
			$ & \tagarray\label{eq:evalP}\\[15pt]	
			\tablent{exp}{100} & 
			$
			\left\langle
			\left\langle \frac{\mathit{write}(i,j)}{\cdot} \dots \right\rangle _{k}
			\left\langle i \mapsto \frac{h}{j} \dots \right\rangle _{mem}
			\right\rangle
			$ & \tagarray\label{eq:memWrite}\\[15pt]	
			\tablent{stm}{101}  &
			$
			\left\langle
			\left\langle \frac{\texttt{return }(\mathit{t},i)\texttt{;}}{(\mathit{t},i)} \dots \right\rangle _{k}
			\right\rangle
			$ & \tagarray\label{eq:return}\\[15pt]	
			\tablent{exp}{95} &
			$
			\left\langle
			\left\langle \frac{\mathit{read}((t,i))}{(t,j)} \dots \right\rangle _{k}
			\left\langle i \mapsto j \dots \right\rangle _{mem}
			\right\rangle
			$ & \tagarray\label{eq:memRead}\\[15pt]
			\tablent{exp}{95} &
			$
			\left\langle
			\left\langle \frac{\texttt{x->next=}i\texttt{;}}{\mathit{write}(\texttt{x->next}, i)} \dots \right\rangle _{k}
			\right\rangle
			$ & \tagarray\label{eq:assignXnext}\\[15pt]
			\tablent{exp}{87} &
			$
			\left\langle
			\left\langle \frac{\texttt{p=}(t,i)\texttt{;}}{\cdot} \dots \right\rangle _{k}
			\left\langle \texttt{p} \mapsto \frac{(s,j)}{(s,i)} \dots\right\rangle _{\mathit{env}}
			\right\rangle
			$ & \tagarray\label{eq:pAssign}\\[15pt]
			\tablent{relat_exp}{51} &
			$
			\left\langle
			\left\langle \frac{i\texttt{-}j}{l} \dots \right\rangle _{k}
			\right\rangle
			\land l = i - j
			$ & \tagarray\label{eq:doSub}\\[15pt]
			\tablent{relat_exp}{51} &
			$
			\left\langle
			\left\langle \frac{i\texttt{<}j}{l} \dots \right\rangle _{k}
			\right\rangle
			\land i<j \land l\neq 0
			$ & \tagarray\label{eq:cmpTrue}\\[15pt]
			\tablent{relat_exp}{51} &
			$
			\left\langle
			\left\langle \frac{i\texttt{<}j}{l} \dots \right\rangle _{k}
			\right\rangle
			\land i\geq j \land l=0
			$ & \tagarray\label{eq:cmpFalse}\\[15pt]
			\tablent{relat_exp}{51} &
			$
			\left\langle
			\left\langle \frac{\texttt{0}}{i} \dots \right\rangle _{k}
			\right\rangle
			\land i=0
			$ & \tagarray\label{eq:zero}\\[15pt]
			\tablent{stm}{49} &
			$
			\left\langle
			\left\langle \frac{\texttt{if(}i\texttt{)\{}\mathit{THEN}\texttt{\}}}{\mathit{THEN}} \dots \right\rangle _{k}
			\right\rangle
			\land i\neq 0
			$ & \tagarray\label{eq:ifTrue}\\[15pt]
			\tablent{stm}{49} &
			$
			\left\langle
			\left\langle \frac{\texttt{if(}i\texttt{)\{}\mathit{THEN}\texttt{\}}}{\cdot} \dots \right\rangle _{k}
			\right\rangle
			\land i= 0
			$ & \tagarray\label{eq:ifFalse}\\[15pt]
\tablent{eq_exp}{46} &
			$
			\left\langle
			\left\langle \frac{i\texttt{!=}j}{l} \dots \right\rangle _{k}
			\right\rangle
			\land i\neq j \land l\neq 0
			$ & \tagarray\label{eq:neqTrue}\\[15pt]
			\tablent{eq_exp}{46} &
			$
			\left\langle
			\left\langle \frac{i\texttt{!=}j}{l} \dots \right\rangle _{k}
			\right\rangle
			\land i=j \land l=0
			$ & \tagarray\label{eq:neqFalse}\\[15pt]
			\tablent{compound\_stm}{43} &
			$
			\left\langle
			\left\langle \frac{\texttt{if(}i\texttt{)\{THEN1\}}}{THEN1} \dots \right\rangle _{k}
			\right\rangle
			\land i\neq 0
			$ & \tagarray\label{eq:if1True}\\[15pt]			
			\bottomrule
		\end{tabular}	
\end{table}
\begin{table}
	\ContinuedFloat
	\caption{Reachability rule instances}
	\centering	
	\begin{tabular}{cll}
		\toprule
		Node & Reachability rule & \\ \midrule	
        	\tablent{compound\_stm}{43} &
			$
			\left\langle
			\left\langle \frac{\texttt{if(}i\texttt{)\{THEN1\}}}{\cdot} \dots \right\rangle _{k}
			\right\rangle
			\land i= 0
			$ & \tagarray\label{eq:if1False}\\[15pt]
			\tablent{compound\_decl}{28} &
			$
			\left\langle
			\left\langle \frac{\texttt{struct listNode* p;}}{\cdot} \dots \right\rangle _{k}
			\left\langle \frac{\cdot}{\texttt{p} \mapsto (\textit{listNode}*, \textit{undef})} \dots \right\rangle _{\mathit{env}}
\right\rangle
			$ & \tagarray\label{eq:defineP}\\[15pt]
			\tablent{function_def}{9} &
			$
			\left\langle \left\langle \frac{\texttt{swap(}(\mathit{listNode*}, i)\texttt{)}}{\mathit{SWAP}} \dots \right\rangle _{k}
\left\langle \frac{\cdot}{\texttt{x} \mapsto i} \dots \right\rangle _{\mathit{env}}
			\right\rangle
			$ & \tagarray\label{eq:SwapCall}\\[15pt]
			\tablent{compound\_stm}{132} &
			$
			\left\langle
			\left\langle \frac{\texttt{if(}i\texttt{)\{}\mathit{THEN2}\texttt{\}}}{\mathit{THEN2}} \dots \right\rangle _{k}
			\right\rangle
			\land i\neq 0
			$ & \tagarray\label{eq:if2True}\\[15pt]
			\tablent{compound\_stm}{132} &
			$
			\left\langle
			\left\langle \frac{\texttt{if(}i\texttt{)\{}\mathit{THEN2}\texttt{\}}}{\cdot} \dots \right\rangle _{k}
			\right\rangle
			\land i= 0
			$ & \tagarray\label{eq:if2False}\\[15pt]
			\tablent{function_def}{29} &
			$
			\begin{array}{l}
			\left\langle \left\langle \frac{\texttt{min(}(\mathit{listNode*}, i)\texttt{)}}{\mathit{j}} \dots \right\rangle _{k} 
\left\langle \frac{i \mapsto \mathit{list}(A)}{j \mapsto \mathit{list}(A^{\prime})} \right\rangle _{mem}
			\left\langle \texttt{x} \mapsto \frac{i}{j} \dots \right\rangle _{\mathit{env}}\right\rangle
			\\
			\land i\neq 0 \land j \neq 0 \land \mathit{listPermutation}(A,A^\prime) \land \mathit{leq}(\mathit{head}(A^\prime),A^\prime)
			\end{array}
			$ & \tagarray\label{eq:MinCall}\\[15pt]
			\tablent{function_def}{29} &
			$
			\left\langle 
			\left\langle \frac{\texttt{min(}(\mathit{listNode*}, i)\texttt{)}}{\mathit{ERROR}} \dots \right\rangle _{k} 
\left\langle \frac{\texttt{x} \mapsto i}{\cdot} \dots \right\rangle _{\mathit{env}}
			\right\rangle
			\land i=0
			$ & \tagarray\label{eq:MinCallError}\\[15pt]			
			\tablent{global_decl}{2} &
			$
			\left\langle
			\left\langle \frac{(\mathit{listNode*},i)\texttt{->val}}{i} \dots \right\rangle _{k}
			\right\rangle
			$ & \tagarray\label{eq:structVal}\\[15pt]
			\tablent{global_decl}{2} &
			$
			\left\langle
			\left\langle \frac{(\mathit{listNode*},i)\texttt{->next}}{(\mathit{listNode*},j)} \dots \right\rangle _{k}
			\right\rangle
			\land j=i+1
			$ & \tagarray\label{eq:structNext}\\[15pt]
			\tablent{global_decl}{2} &
			$
			\left\langle
			\left\langle \frac{\texttt{sizeof(struct listNode)}}{2} \dots \right\rangle _{k}
			\right\rangle
			$ & \tagarray\label{eq:sizeof}\\[15pt]
			\bottomrule
	\end{tabular}
\end{table}

\subsection{Symbolic evaluation}
\label{sec:exampleVerificationTask}

We illustrate the evaluation of the verification task
for function \texttt{min}, as executed through the \textsc{Process}
and \textsc{Check} algorithms shown in figure~\ref{fig:algo}.  The
evaluation of the verification task starts at \textnt{function_def}{29}
with configuration pattern~\eqref{eq:startMin}, which captures the
state in which \texttt{min} is invoked: \texttt{x} points to a
non-empty list $A$.  The configuration patterns reached through the
symbolic execution are shown in table~\ref{tb:configurationPatterns}.

\begin{table}
	\caption{Configuration patterns reached through the symbolic
		execution of the example in Fig.~\ref{exampleListSwapSort}\label{tb:configurationPatterns}}
	\centering	
		\begin{tabular}{ll}
			\toprule
			$
			\exists a .
			\left\langle
			\left\langle \texttt{if(}\mathit{read}((\mathit{listNode*},a)\texttt{->next})\texttt{!=NULL)}\mathit{OTH}\right\rangle_{k}
			\left\langle e_0 \right\rangle _{\mathit{env}}
			\left\langle m \dots \right\rangle _{mem}
			\right\rangle
			\land p
			$ & \tagarray\label{eq:step1}\\[15pt]
$
			\exists a,v,d,B .
			\left\langle
			\left\langle \texttt{if(}(\mathit{listNode*},d)\texttt{!=}0{)}\mathit{OTH}\right\rangle_{k}
			\left\langle e_0 \right\rangle _{\mathit{env}}
			\left\langle m^\prime \dots \right\rangle _{mem}
			\right\rangle
			\land p
			$ & \tagarray\label{eq:step2}\\[15pt]
			$
			\exists a,v,d,B .
			\left\langle
			\left\langle \texttt{return swap(x);} \right\rangle_{k}
			\left\langle e_0 \right\rangle _{\mathit{env}}
			\left\langle m^{\prime\prime} \dots \right\rangle _{mem}
			\right\rangle
			\land p^\prime
			$ & \tagarray\label{eq:step12b}\\[15pt]
$
			\exists a,v,d,B .
			\left\langle
			\left\langle \texttt{return }(\text{\texttt{if(}}\mathit{COND1}\text{\lstinline[basicstyle=\ttfamily]!)!~}\mathit{OTH2}) \right\rangle_{k}
			\left\langle e \right\rangle _{\mathit{env}}
			\left\langle m^{\prime\prime} \dots \right\rangle _{mem}
			\right\rangle
			\land p^\prime
			$ & \tagarray\label{eq:step15b}\\[15pt]
$
			\exists a,v,d,f,B.
			\left\langle
			\left\langle \texttt{return }(\texttt{if(}f\texttt{)}\mathit{OTH2}) \right\rangle_{k}
			\left\langle e \right\rangle _{\mathit{env}}
			\left\langle m^{\prime\prime} \dots \right\rangle _{mem}
			\right\rangle
			\land p^\prime \land f=1
			$ & \tagarray\label{eq:step18b}\\[15pt]
			$
			\exists a,v,d,g,B .
			\left\langle
			\left\langle \texttt{return }(\texttt{if(}\mathit{read}((\mathit{listNode*}, d) \texttt{->val})\mathit{OTH3} \right\rangle_{k}
\mathit{EM}
			\right\rangle
			\land p^\prime \land g=a+1
			$ & \tagarray\label{eq:step22b}\\[15pt]
			$
			\exists a,v,d,r,B .
			\left\langle
			\left\langle \texttt{return }(\texttt{if(}\mathit{read}(r)\mathit{OTH3} \right\rangle_{k}
			\mathit{EM}	
			\right\rangle
			\land p^\prime \land r=0
			$ & \tagarray\label{eq:step23b}\\[15pt]
			\bottomrule
		\end{tabular}	
\end{table}

The first operation is the replacement of variable \texttt{x} with its value $a$ by reachability rule instance~\eqref{eq:evalX},
which yields configuration pattern~\eqref{eq:step1}.
The evaluation of the task in \textnt{function_def}{29} must stop here, since no further rule instances are
available to process this configuration pattern:
the \textsc{Process} algorithm exits from the loop  and returns.
Since configuration pattern~\eqref{eq:step1} is non-final,
the execution of the \textsc{Check} algorithm will return a
$\mathit{delay}$ answer,
terminating the \textsc{Check} algorithm;
the verification task will be resumed in \textnt{global\_decl}{2}.

\begin{figure}	
	\centering
	\begin{tikzpicture}[scale=0.2]
	\tikzstyle{every node}+=[inner sep=0pt]
	\draw [black] (7.1,39.5) circle (3);
	\draw (7.1,39.5) node {\eqref{eq:step1}};
	\draw [black] (15.9,39.5) circle (3);
	\draw (15.9,39.5) node {\eqref{eq:step2}};
	\draw [black] (24.7,34.2) circle (3);
	\draw (24.7,34.2) node {\eqref{eq:step12b}};
	\draw [black] (24.7,44.8) circle (3);
	\draw (24.7,44.8) node {\ldots};
	\draw [black] (33.5,34.2) circle (3);
	\draw (33.5,34.2) node {\eqref{eq:step15b}};
	\draw [black] (42.3,34.2) circle (3);
	\draw (42.3,34.2) node {\eqref{eq:step18b}};
	\draw [black] (51.1,34.2) circle (3);
	\draw (51.1,34.2) node {\eqref{eq:step22b}};
	\draw [black] (59.9,34.2) circle (3);
	\draw (59.9,34.2) node {\eqref{eq:step23b}};
	\draw [black] (10.1,39.5) -- (12.9,39.5);
	\fill [black] (12.9,39.5) -- (12.1,40) -- (12.1,39);
	\draw (16,44) node [left] {\eqref{eq:structNext},\eqref{eq:memRead}};
\draw [black] (18.47,37.95) -- (22.13,35.75);
	\fill [black] (22.13,35.75) -- (21.7,36.59) -- (21.19,35.73);
	\draw (21,35) node [left] {\eqref{eq:neqFalse}};
	\draw [black] (18.47,41.05) -- (22.13,43.25);
	\fill [black] (22.13,43.25) -- (21.19,43.27) -- (21.7,42.41);
	\draw (21,44) node [left] {\eqref{eq:neqTrue}};
	\draw [black] (27.7,34.2) -- (30.5,34.2);
	\fill [black] (30.5,34.2) -- (29.7,34.7) -- (29.7,33.7);
	\draw (24,39) node [right] {\eqref{eq:evalX},\eqref{eq:SwapCall},\eqref{eq:defineP}};
	\draw [black] (36.5,34.2) -- (39.3,34.2);
	\fill [black] (39.3,34.2) -- (38.5,34.7) -- (38.5,33.7);
	\draw (43,29) node [left] {\eqref{eq:evalX},\eqref{eq:neqFalse}};
	\draw [black] (45.3,34.2) -- (48.1,34.2);
	\fill [black] (48.1,34.2) -- (47.3,34.7) -- (47.3,33.7);
	\draw (54,39) node [left] {\eqref{eq:if1False},\eqref{eq:evalX},\eqref{eq:structNext},\eqref{eq:memRead}};
	\draw [black] (54.1,34.2) -- (56.9,34.2);
	\fill [black] (56.9,34.2) -- (56.1,34.7) -- (56.1,33.7);
	\draw (57,30) node [left] {\eqref{eq:structVal}};
	\end{tikzpicture}
	\caption{Steps of the symbolic execution for the example in
          Fig.~\ref{exampleListSwapSort} (nodes are labeled with the
          configuration patterns from
          Table~\ref{tb:configurationPatterns}, edges are labeled with
        the reachability rule instances from Table~\ref{tb:reachabilityRules})}
	\label{fig:symbolicExecutionGraph}
\end{figure}
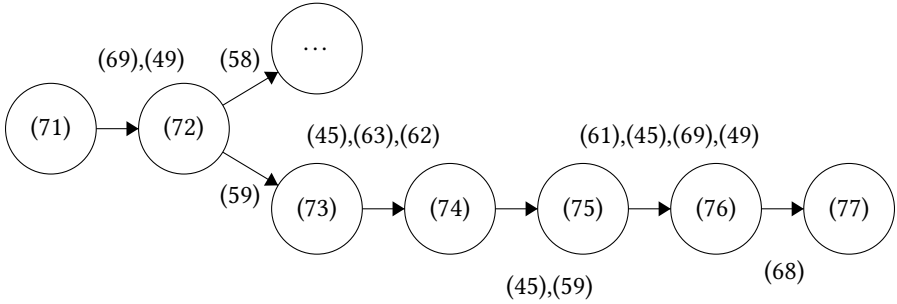

Figure~\ref{fig:symbolicExecutionGraph} depicts the remaining steps in
the symbolic execution: nodes represent the traversed configuration
patterns, while edges are labeled with the applied  reachability rule
instances.

 First reachability rule instance~\eqref{eq:structNext} is applied and
the address of \texttt{next} field is computed.  Then, reachability
rule instance~\eqref{eq:memRead} reads the value of \texttt{next}
field, reaching configuration pattern~\eqref{eq:step2}\footnote{Note
  that the list object $A$ in $\mathit{mem}$ is split into a first
  object $[v]$ containing the first element and a second list object
  $B$ containing the tail of list $A$.}.  At this stage, both
reachability rule instances~\eqref{eq:neqTrue} and \eqref{eq:neqFalse}
could be applied; we only consider the case where rule instance \eqref{eq:neqFalse} is applied, corresponding to the execution of line~\ref{ln:min:swap} in
Fig.~\ref{exampleListSwapSort}.

To execute the call to function \texttt{swap}, first the value of \texttt{x}
is retrieved  through reachability rule instance~\eqref{eq:evalX};
the actual function call is performed by applying 
reachability rule instance~\eqref{eq:SwapCall}.
The first statement of function \texttt{swap} defines variable
\texttt{p}.  Reachability rule instance~\eqref{eq:defineP} executes
this statement and yields configuration pattern~\eqref{eq:step15b}.
To execute the outer \texttt{if} statement of function \texttt{swap},
variable \texttt{x} is evaluated by reachability rule
instance~\eqref{eq:evalX}; reachability rule
instance~\eqref{eq:neqFalse} is applied, resulting in configuration
pattern~\eqref{eq:step18b} \footnote{In this case reachability rule
  instance~\eqref{eq:neqTrue} cannot be applied, meaning that the
  \texttt{then} branch cannot be followed.}.  The \texttt{then} branch
of the \texttt{if} statement is not entered through reachability rule
instance~\eqref{eq:if1False}; hence, the inner \texttt{if} of function
\texttt{swap} is reached.  Variable \texttt{x} is evaluated through
reachability rule instance~\eqref{eq:evalX}; then, the actual address
of the \texttt{next} field is obtained through reachability rule
instance~\eqref{eq:structNext}; the memory read is then performed
through reachability rule instance~\eqref{eq:memRead}.  Starting from
configuration pattern~\eqref{eq:step22b}, reachability rule
instance~\eqref{eq:structVal} computes the \texttt{val} offset,
yielding configuration pattern~\eqref{eq:step23b}.  In this state, the
memory \emph{read} cannot be performed, since the address to be read
is zero.  The sink error configuration is reached and the verification
must stop.  Since no further configuration patterns are reachable from
the sink error configuration, algorithm \textsc{Process} exits,
returning the performed verification task.

A subsequent invocation of the \textsc{Check} algorithm on the
returned verification task would yield $\mathit{false}$: indeed,
algorithm \textsc{Check} would report the verification failure, since
all sink error configurations are final and do not match the target
configuration.

 \section{Incremental Reachability Checking}
\label{sec:incrementality}

Our approach for reachability checking is \emph{intrinsically
  incremental} because:
\begin{itemize}
\item it is syntactically incremental, since it assumes that
  the input program is described by means of an operator precedence
  grammar, which allows for limiting  the parsing process only to a
  fraction of the input, thanks to its  peculiar properties  described
  in Section~\ref{secSidecar};
\item our semantic attribute evaluation method relies only on
  synthesized attributes, meaning that the semantics of a certain node depends
only on its descendants. Upon changes of the semantic attributes  of the
nodes in a subtree, it suffices to recompute the attributes of the
ancestors along the path to the root.
\end{itemize}

This can lead to an incremental verification procedure once verification is encoded as syntax-directed attribute evaluation.

In the remaining of the section we show an application of incremental
verification to the example presented in
Section~\ref{sec:running-example} and also discuss how to speed up the
incremental evaluation process.

\subsection{Incremental Verification: Application to the Example}

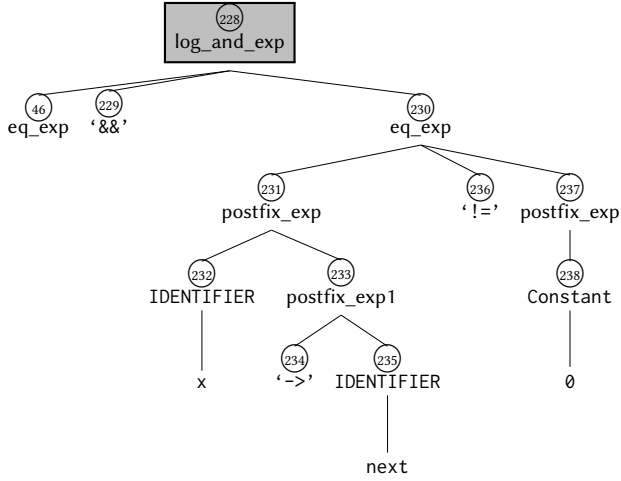
\begin{figure}[tb]
	\centering
	\begin{tikzpicture}[scale=0.8,level distance=40pt]	
	\Tree[.\refnt{log_and_exp}{228}
	\nt{eq_exp}{46} \tn{`\&\&'}{229} [.\nt{eq_exp}{230}
	[.\nt{postfix_exp}{231}
	[.\tn{IDENTIFIER}{232} \texttt{x} ] [.\nt{postfix_exp1}{233}
	\tn{`->'}{234} [.\tn{IDENTIFIER}{235} \texttt{next} ]
	]
	] \tn{`!='}{236} [.\nt{postfix_exp}{237}
	[.\tn{Constant}{238} \texttt{0} ]
	]
	]
	]
	\end{tikzpicture}	
	\caption{Subtree affected by the change at line~\ref{line:validListComp} of the example in Fig.~\ref{exampleListSwapSort}}
	\label{fig:affectedSubtree}
\end{figure}

We consider a change occurring at line~\ref{line:validListComp} of the
\texttt{sort} program shown in (figure~\ref{exampleListSwapSort}), which
replaces 
the original (and faulty) condition \verb~x!=0~ with
\verb|x!=0 && x->next!=0|; figure~\ref{fig:affectedSubtree}  shows the
portion of the syntax tree enclosing the change.

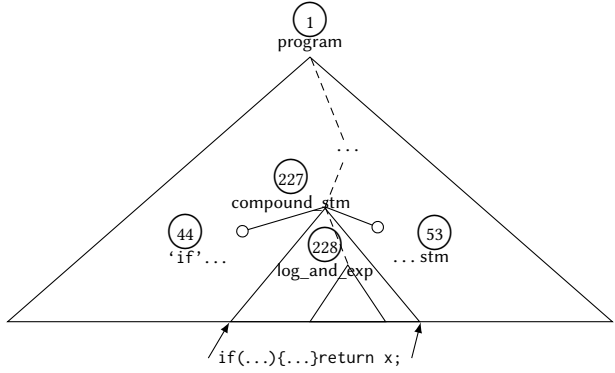
\begin{figure}[tb]
	\centering
	\begin{tikzpicture}
\begin{scope}
	\tikzset{every node/.style={isosceles triangle, draw,anchor=apex, shape border rotate=90, isosceles triangle stretches}}
	\node[minimum height=35mm,minimum width=8cm] (out) at (4,4) {};
	\node[minimum height=15mm,minimum width=2.5cm] (in) at (4.2,2) {};
	\node[minimum height=7.5mm,minimum width=10mm] (in2) at (4.5,1.25) {};
	\end{scope}
	\node (S) at (4,4.3) {\scriptsize \nt{program}{1}};
	\node (M) at (3.75,2.2) {\scriptsize \nt{compound_stm}{227}};
	\node (N) at (4.2,1.3) {\scriptsize \nt{log_and_exp}{228}};
	\node (K) at (4.5,2.75) {\scriptsize $\ldots$};
	\node (P) at (2.5,1.5) {\scriptsize \tn{`if'}{44}$\ldots$};
	\node (Q) at (5.5,1.5) {\scriptsize $\ldots$\nt{stm}{53}};
	\node (t) at (4,0) {\scriptsize \texttt{if(}$\ldots$\texttt{)\{}$\ldots$\texttt{\}return x;}};
	\draw[-latex] (t.east) -- (in.right corner);
	\draw[-latex] (t.west) -- (in.left corner);
	\draw[densely dashed] (in2.apex) -- (in.apex);
	\draw[densely dashed] (in.apex) -- (K);
	\draw[densely dashed] (K) -- (out.apex);
	\draw[o-] (P) -- (in.apex);
	\draw[o-] (Q) -- (in.apex);
	\end{tikzpicture}
	\caption{Part of the syntax tree  of the example in Fig.~\ref{exampleListSwapSort} affected by the change}
	\label{fig:affect-part}
\end{figure}

\subsubsection{Incremental attribute evaluation}

The change requires  re-evaluating the semantic attributes as
shown in figure~\ref{fig:affect-part}.  The changed part (bounded by
nodes \skiptn{`if'}{44} and \skipnt{stm}{53}) is the subtree rooted in
\textnt{compound_stm}{227}: only the attributes of this subtree
must be recomputed and it is enough to propagate these attributes
through the path to the root of the tree (\textnt{program}{1}).
This way the attributes of other nodes (e.g.,
\textnt{function_def}{29} in Figure~\ref{fig:global-tree}) do not need to be evaluated.

The newly computed attributes and reachability rule instances
are shown in table~\ref{tb:newStuff}.
Notice the use of the
abbreviations defined in figure~\ref{eq:abbrv1}, in addition to those
defined  in figure~\ref{eq:abbrv}.

\begin{figure}[tb]
	\begin{equation*}
		\begin{array}{lll}	
			\mathit{COND2}            & \equiv & \text{\lstinline[basicstyle=\ttfamily]*x != 0 && *} ~\mathit{read}(\text{\lstinline[basicstyle=\ttfamily]*x->next*})\text{\lstinline[basicstyle=\ttfamily]* != 0*} \\
			\mathit{SWAP}^\prime			& \equiv & \text{\lstinline[basicstyle=\ttfamily]!struct listNode* p; !~}\text{\lstinline[basicstyle=\ttfamily]!if(!}\mathit{COND2}\text{\lstinline[basicstyle=\ttfamily]!)!~}\mathit{OTH2}
		\end{array}
	\end{equation*}
	\caption{New abbreviations (in addition to the ones introduced
          in Fig.~\ref{eq:abbrv})}
	\label{eq:abbrv1}
\end{figure}

\begin{table}[tb]
	\caption{Values of the attributes, reachability rule instances,
		and configuration patterns resulting from the change at
		line~\ref{line:validListComp} of the example in
		Fig.~\ref{exampleListSwapSort} \label{tb:newStuff}}
	\centering
	\footnotesize	
		\begin{tabular}{c}
			\subfloat[Attributes: $K$ (list of code tokens), $R$
			(set of reachability rules), $V$ (set of verification
			tasks), $C$ (verification failed flag)]{
				\begin{tabular}{clllllll}
					\toprule
					Node & $K$ & $R$ & $V$ & $C$ \\
					\midrule
					\tablent{eq_exp}{230} & $\{\mathit{read}(\texttt{x->next})\texttt{ != 0}\}$ & 
					$\{\eqref{eq:neqTrue}, \eqref{eq:neqFalse}, \eqref{eq:evalX}, \eqref{eq:memRead}\}$ & $\emptyset$ & $\mathit{true}$\\
					\tablent{log_and_exp}{228} & $\{\texttt{x!=0 \&\& }\mathit{read}(\texttt{x->next}) \texttt{!=0}\}$ & $R_{230} \cup R_{46} \cup \{\eqref{eq:landFalse}, \eqref{eq:landTrue}\}$ & $\emptyset$ & $\mathit{true}$\\			
					\tablent{compound\_stm}{227} & $\begin{array}{l}
					\{
					\texttt{if(}K_{228}\texttt{)\{}\mathit{THEN1}\texttt{\}}\texttt{return x;}\}\end{array}$ & $\begin{array}{l}
					\{\eqref{eq:if1True}, \eqref{eq:if1False}, \eqref{eq:return}, \eqref{eq:evalX}\}\\\cup R_{228} \cup R_{49}
					\end{array}$ & $\emptyset$ & $\mathit{true}$\\			
					\tablent{compound\_decl}{28} & $\{\texttt{struct listNode* p; }K_{227}\}$ & $\{\eqref{eq:defineP}\} \cup R_{227}$ & $\emptyset$ & $\mathit{true}$\\					
					\tablent{function_def}{9} & $\{\mathit{SWAP}\}$ & $\{\eqref{eq:SwapCall1}\} \cup R_{28} \cup R_{29}$ & $V_{29}$ & $\mathit{true}$\\							
					\tablent{global_decl}{2} & $\{\}$ & $\{\eqref{eq:structVal}, \eqref{eq:structNext}, \eqref{eq:sizeof}\} \cup R_9$ & $\emptyset$ & $\mathit{true}$\\
					\bottomrule
				\end{tabular}	
			}
			\\
			\subfloat[Reachability rule instances]{
				\begin{tabular}{cll}
			\toprule
			Node & Reachability rule & \\ \midrule			
			\tablent{log_and_exp}{228} & 
			$			
			\left\langle
			\left\langle \frac{i\texttt{ \&\& }\mathit{read}(\texttt{x->next}) \texttt{ != 0}}{i} \dots \right\rangle _{k}
			\right\rangle
			\land i = 0
			$ & \tagarray\label{eq:landFalse}\\[15pt]	
			\tablent{log_and_exp}{228} & 
			$			
			\left\langle
			\left\langle \frac{i\texttt{ \&\& }\mathit{read}(\texttt{x->next}) \texttt{ != 0}}{\mathit{read}(\texttt{x->next}) \texttt{ != 0}} \dots \right\rangle _{k}
			\right\rangle
			\land i \neq 0
			$ & \tagarray\label{eq:landTrue}\\[15pt]	
			\tablent{function_def}{9} & 
			$
			\left\langle \left\langle \frac{\texttt{swap(}(\mathit{listNode*}, i)\texttt{)}}{\mathit{SWAP^\prime}} \dots \right\rangle _{k}
\left\langle \frac{\cdot}{\texttt{x} \mapsto i} \dots \right\rangle _{\mathit{env}}
			\right\rangle
			$ & \tagarray\label{eq:SwapCall1}\\[15pt]		
			\bottomrule
                        \end{tabular}
        
 			}
			\\
			\subfloat[Configuration patterns]{
				\begin{tabular}{ll}
			\toprule
			Configuration pattern & \\ \midrule						
			$			
			\exists a,v,d,B .
			\left\langle
			\left\langle \texttt{return }(\text{\texttt{if(}}\mathit{COND2}\text{\lstinline!)!~}\mathit{OTH2}) \right\rangle_{k}
			\left\langle e \right\rangle _{\mathit{env}}
			\left\langle m^{\prime\prime} \dots \right\rangle _{mem}
			\right\rangle
			\land p^\prime
			$ & \tagarray\label{eq:step15b1}\\[15pt]			
			$
			\exists a,v,d,f,B.			
			\left\langle
			\left\langle \texttt{return }(\texttt{if(}f~\mathit{AND}\texttt{)}\mathit{OTH2}) \right\rangle_{k}
			\left\langle e \right\rangle _{\mathit{env}}
			\left\langle m^{\prime\prime} \dots \right\rangle _{mem}
			\right\rangle
			\land p^\prime \land f=1
			$ & \tagarray\label{eq:step18b1}\\[15pt]
			$
			\exists a,v,d,B .
			\left\langle
			\left\langle \texttt{return }(\texttt{if(}(\mathit{listNode*}, d) \texttt{ != } 0~\mathit{}\texttt{)}\mathit{OTH2}) \right\rangle_{k}
			\left\langle e \right\rangle _{\mathit{env}}
			\left\langle m^{\prime\prime} \dots \right\rangle _{mem}
			\right\rangle
			\land p^\prime
			$ & \tagarray\label{eq:step23b1}\\[15pt]
			$
			\exists a,v,d,B .			
			\left\langle
			\left\langle \texttt{return }(\texttt{return x;}) \right\rangle_{k}
			\left\langle e \right\rangle _{\mathit{env}}
			\left\langle m^{\prime\prime} \dots \right\rangle _{mem}
			\right\rangle
			\land p^\prime
			$ & \tagarray\label{eq:step24b1}\\[15pt]		
			$
			\exists a,v,d,B .			
			\left\langle
			\left\langle (\texttt{listNode}*, a) \right\rangle_{k}
			\left\langle e \right\rangle _{\mathit{env}}
			\left\langle m^{\prime\prime} \dots \right\rangle _{mem}
			\right\rangle
			\land p^\prime
			$ & \tagarray\label{eq:step27b1}\\[15pt]			
			\bottomrule
		\end{tabular} 			}		
		\end{tabular}		
\end{table}

\subsubsection{Incremental symbolic execution}
 We analyze the symbolic execution of the \texttt{min} verification task
after the change; 
for brevity, we only consider the part of the verification task analyzed
in Section~\ref{sec:exampleVerificationTask}.

\begin{figure}	
	\centering
	\begin{tikzpicture}[scale=0.4]
	\tikzstyle{every node}+=[inner sep=0pt]
	\draw [black] (24.7,34.2) circle (1.5);
	\draw (24.7,34.2) node {\eqref{eq:step15b1}};	
	\draw [black] (31.5,34.2) circle (1.5);
	\draw (31.5,34.2) node {\eqref{eq:step18b1}};
	\draw [black] (42.3,34.2) circle (1.5);
	\draw (42.3,34.2) node {\eqref{eq:step23b1}};
	\draw [black] (47.8,34.2) circle (1.5);
	\draw (47.8,34.2) node {\eqref{eq:step24b1}};
	\draw [black] (56.5,34.2) circle (1.5);
	\draw (56.5,34.2) node {\eqref{eq:step27b1}};
	\draw [black] (26.2,34.2) -- (30,34.2);
	\fill [black] (30,34.2) -- (29.2,34.7) -- (29.2,33.7);
	\draw (26,35) node [right] {\eqref{eq:evalX},\eqref{eq:neqFalse}};
	\draw [black] (33,34.2) -- (40.8,34.2);
	\fill [black] (40.8,34.2) -- (40,34.7) -- (40,33.7);
	\draw (40,35) node [left] {\eqref{eq:landTrue},\eqref{eq:evalX},\eqref{eq:structNext},\eqref{eq:memRead}};
	\draw [black] (43.8,34.2) -- (46.3,34.2);
	\fill [black] (46.3,34.2) -- (45.5,34.7) -- (45.5,33.7);
	\draw (45.5,35) node [left] {\eqref{eq:if1False}};
	\draw [black] (49.3,34.2) -- (55,34.2);
	\fill [black] (55,34.2) -- (54.2,34.7) -- (54.2,33.7);
	\draw (54.4,35) node [left] {\eqref{eq:evalX},\eqref{eq:return},\eqref{eq:return}};
	\end{tikzpicture}
	\caption{Steps of the symbolic execution for the example in
          Fig.~\ref{exampleListSwapSort} modified with the change at line~\ref{line:validListComp}  (nodes are labeled with the
          configuration patterns from
          Tables~\ref{tb:configurationPatterns} and~\ref{tb:newStuff}, edges are labeled with
        the reachability rule instances from
        Tables~\ref{tb:reachabilityRules} and \ref{tb:newStuff})}
	\label{fig:incrementalSymbolicExecutionGraph}
\end{figure}
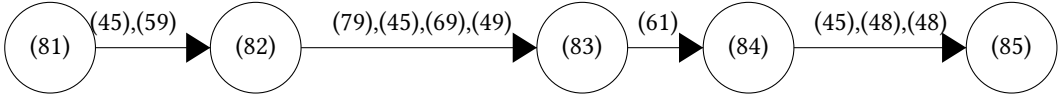

The execution proceeds as described in
Section~\ref{sec:exampleVerificationTask} until reaching configuration
pattern~\eqref{eq:step12b} (see table~\ref{tb:configurationPatterns});
the configuration patterns reached during the evaluation of the
verification task are shown in
table~\ref{tb:newStuff}.  From configuration
pattern~\eqref{eq:step12b}, the execution
continues as depicted in
figure~\ref{fig:incrementalSymbolicExecutionGraph} (nodes represent
the traversed configuration patterns; edges are labeled with the
applied reachability rule instances).  Configuration
pattern~\eqref{eq:step15b1} is obtained from configuration
pattern~\eqref{eq:step12b} by executing the new \texttt{swap}
function through reachability rules~\eqref{eq:SwapCall1} and by
defining variable \texttt{p} through \eqref{eq:defineP}.  Then, the
first condition of the outer \texttt{if} statement is executed through
reachability rule instances~\eqref{eq:evalX} and~\eqref{eq:neqFalse},
yielding configuration pattern~\eqref{eq:step18b1}.  At this point, the
execution of the program is affected by the change, reflected in the
second condition of the \texttt{if}.  Reachability rule
instances~\eqref{eq:landTrue}, \eqref{eq:evalX},
\eqref{eq:structNext}, and~\eqref{eq:memRead} retrieve the values
required by the expression, reaching configuration
pattern~\eqref{eq:step23b1}.  In this case, the \texttt{then} branch
cannot be taken, since the condition evaluates to $\mathit{false}$
(because $d=0$); the execution of the \texttt{false} branch yields
configuration pattern~\eqref{eq:step24b1}, through reachability rule
instance~\eqref{eq:if1False}.  The value to be returned is retrieved
through rules~\eqref{eq:evalX} and \eqref{eq:return} (the latter executed twice) reaching the final configuration pattern
\eqref{eq:step27b1}.  The invocation of algorithm \textsc{Check} 
checks whether pattern~\eqref{eq:step27b1} matches the target
configuration pattern~\eqref{eq:stopMin}.  Logical variable $a$ in
pattern~\eqref{eq:step27b1} is equal to $a^\prime$ in
pattern~\eqref{eq:stopMin}; $A$ and $A^\prime$ are equal, i.e., they
contain the same elements; since $A$ contains only one element, the
expression $v$, $\mathit{leq}(\mathit{head}(A^\prime),A^\prime)$ is
equivalent to $v \leq v$, and evaluates to true.

\subsection{Improving performance of incremental evaluation}
\label{sec:improveIncrementality}

The technique described above allows for efficiently re-evaluating only the nodes affected by the changes.
The speed-up that can be achieved depends on the distribution of the
computational effort along the syntactic tree, since the nodes close
to the root of the tree are those re-evaluated more often.
Furthermore, most of the computational time is spent inside the
verification tasks, whose evaluation can be started only in nodes
containing a loop invariant or a function definition (the latter are
close to the root).
 When function definitions or loop invariants are missing,
 the verification tasks can be
started even closer to the root, if not at the root at all.

Performance can be improved by making rule propagation along the tree more efficient, balancing
the distribution of the computation efforts.  Within each node,
instead of simply copying the semantic rules of the children, we
generate intermediate state transitions that summarize the behavior of
the program fragment contained in the node subtree.  These
intermediate state transitions are obtained by collapsing the
reachability rules propagated from the children nodes.
The generation starts with expanding the program
code related to the node into a symbolic configuration pattern, from
which the traditional verification procedure can be applied.  When a
final configuration pattern is reached, the relation between the final
and the initial patterns is used to generate a new reachability rule
that summarizes the execution.

To illustrate this optimization, consider the previous example referring to the symbolic interpretation attached to \textnt{compound_stm}{227} (see Figure \ref{fig:affect-part}): instead of considering the sequence of reachability rules
\eqref{eq:evalX}, \eqref{eq:neqFalse}, \eqref{eq:landTrue},
\eqref{eq:evalX}, \eqref{eq:structNext} as in Figure
\ref{fig:incrementalSymbolicExecutionGraph}, the generation starts
from the code part of \textnt{compound_stm}{227} belonging to
configuration pattern~\eqref{eq:step15b1}, and first yields configuration pattern~\eqref{eq:genRule}:
\begin{equation}
\left\langle
\left\langle \text{\texttt{if(}}\mathit{COND2}\text{\lstinline!)!~}\mathit{OTH2} \right\rangle_{k}
\right\rangle
\label{eq:genRule}
\end{equation}
When the new reachability rules are generated, the initial
configuration is not fully known. This incomplete knowledge may make
the verification procedure unfeasible; for example, it could be
impossible to evaluate the loop termination conditions within the appropriate execution context.  Therefore,
when loops or function calls occur, the generation process ought to be
suspended to prevent indefinite looping; instead, we do execute loops but we restrict their execution to only one iteration and we allow functions to be called only once, forbidding further iterations
(respectively, calls) inside the same loop (respectively, function).
This way, we are still able to improve the individual execution of
a loop or a function; this improvement will be later on reflected on the execution of the loop/function in the appropriate context.

Another case of incomplete knowledge is the access to a variable which has not been declared (i.e., a variable whose 
 declaration lies in another part of the program).
In this case it is sufficient to add the missing variable to the initial patterns.
Continuing the  example above, the application of reachability rule~\eqref{eq:evalX} to
configuration pattern~\eqref{eq:genRule} yields configuration pattern~\ref{eq:genRule1}: 
\begin{equation}
\exists t .
\left\langle
\left\langle \text{\texttt{if(}}\mathit{COND2}\text{\lstinline!)!~}\mathit{OTH2} \right\rangle_{k}
\left\langle  \texttt{x} \mapsto t, \dots\right\rangle _{\mathit{env}}
\right\rangle
\label{eq:genRule1}
\end{equation}
However, there is no variable \texttt{x} in the $\mathit{env}$ cell.
Thus, we add a symbolic variable \texttt{x}  and continue the generation process from
configuration pattern~\eqref{eq:genRule1} by propagating the logical constraints as shown below:
\begin{equation}
\exists t .
\left\langle
\left\langle \text{\texttt{if(}}\mathit{read}((\mathit{listNode*}, t+1))\texttt{!=0}\text{\lstinline!)!~}\mathit{OTH2} \right\rangle_{k}
\left\langle  \texttt{x} \mapsto t, \dots\right\rangle _{\mathit{env}}
\right\rangle
\land t \neq 0
\label{eq:genRuleEnd0}
\end{equation}
\begin{equation}
\exists t,u .
\left\langle
\left\langle \text{\texttt{if(}}u\text{\lstinline!)!~}\mathit{OTH2} \right\rangle_{k}
\left\langle  \texttt{x} \mapsto t\right\rangle _{\mathit{env}}
\right\rangle
\land t = 0 \land u = 0
\label{eq:genRuleEnd1}
\end{equation}

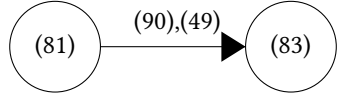
\begin{figure}	
	\centering
	\begin{tikzpicture}[scale=0.4]
	\tikzstyle{every node}+=[inner sep=0pt]
\draw [black] (34.5,34.2) circle (1.5);
	\draw (34.5,34.2) node {\eqref{eq:step15b1}};
	\draw [black] (42.3,34.2) circle (1.5);
	\draw (42.3,34.2) node {\eqref{eq:step23b1}};
\draw [black] (36,34.2) -- (40.8,34.2);
	\fill [black] (40.8,34.2) -- (40,34.7) -- (40,33.7);
	\draw (40,35) node [left] {\eqref{eq:genRule0},\eqref{eq:memRead}};
\end{tikzpicture}
	\caption{Excerpt of the reduced symbolic execution for the example in
		Fig.~\ref{exampleListSwapSort} modified with the change at line~\ref{line:validListComp} (nodes are labeled with the
		configuration patterns from
		Tables~\ref{tb:configurationPatterns} and~\ref{tb:newStuff}, edges are labeled with
		the reachability rule instances from
		Tables~\ref{tb:reachabilityRules} and \ref{tb:newStuff})}
	\label{fig:incrementalShortSymbolicExecutionGraph}
\end{figure}

Let us assume that symbolic execution stops upon reaching configuration patterns~\eqref{eq:genRuleEnd0} and \eqref{eq:genRuleEnd1}. In this case the available information does not allow us to
conclude whether \texttt{x} is always different from zero; we have to
generate two different configuration patterns, covering both cases.
These configuration patterns, paired with the initial pattern~\eqref{eq:genRule1} produce
reachability rules~\eqref{eq:genRule0} and~\eqref{eq:genRule0bis}:
\begin{equation}
\left\langle
\left\langle \frac{\mathit{COND2}}{\mathit{read}((\mathit{listNode*}, t+1))\texttt{!=0}} \dots \right\rangle _{k}
\left\langle  \texttt{x} \mapsto t, \dots\right\rangle _{\mathit{env}}
\right\rangle
\land t \neq 0
\label{eq:genRule0}
\end{equation}
\begin{equation}
\left\langle
\left\langle \frac{\mathit{COND2}}{u} \dots \right\rangle _{k}
\left\langle  \texttt{x} \mapsto t, \dots\right\rangle _{\mathit{env}}
\right\rangle
\land t = 0 \land u = 0
\label{eq:genRule0bis}
\end{equation}
As a result of this optimization (not fully explained here for space reasons), the original fragment from
pattern~\eqref{eq:step15b1} to \eqref{eq:step23b1} of the path
depicted in Figure~\ref{fig:incrementalSymbolicExecutionGraph} is
replaced by the one  in Figure~\ref{fig:incrementalShortSymbolicExecutionGraph}, where the single rule~\eqref{eq:genRule0} replaces the sequence \eqref{eq:evalX}, \eqref{eq:neqFalse}, \eqref{eq:landTrue}, \eqref{eq:evalX}, \eqref{eq:structNext}. Since the verification of every reachability rule requires one or more calls to the SMT solver, we can expect a gain proportional to the reduction of the number of reachability rules.

Another  performance improvement can be obtained by exploiting the natural parallelism enabled by the local parsability
property and the use of synthesized attributes, as already hinted in
Section~\ref{secSidecar}. Parallelism may come in various forms:
several computations are launched in different fragments of the source
code; the corresponding syntactic subtrees are built in parallel and
in conjunction with semantic attributes evaluation; when some
verification tasks are ready, instances of the constraint solver are
launched in parallel.

The two optimization techniques described above aim at increasing the
overall efficiency of the whole verification process, both when
carried on from scratch and when performed incrementally. In the
latter case, however, their benefits are amplified:
\begin{itemize}
\item 
Collapsing
several symbolic interpretation steps (and therefore decreasing the
calls to the solver) may positively impact not only the analysis of
the portion of syntax tree modified by the change, but also the
unavoidable verification that is replayed identically when revisiting
some nodes close to the root.
\item
   Exploiting the intrinsic parallelism enabled by the local parsability
property and the bottom-up attribute evaluation schema not only
improves the performance of the global evaluation but also allows for
managing multiple changes in a natural way.
\end{itemize}

 \section{Implementation and Evaluation}
\label{sec:evaluation}

We implemented our approach in a prototype tool, called \scmc (\emph{SiDECAR
  Matching logic verifier}).

In this section we evaluate our tool along two dimensions. First we show that the
harness needed to support incremental verification does not
significantly alter the performance of non-incremental
verification.  We then investigate how different aspects of the
program being verified (e.g., presence and type of annotations, size of the original program and of its
modified version, size, position, and type of changes) affect the
benefits of our approach for incremental verification.  More
specifically, we evaluate our approach by answering the following
research questions:
\begin{compactenum}[RQ1:]
\item \label{rq1} \emph{What is the overhead introduced by the
    machinery needed to support our syntax-based incremental
    verification approach when compared with non-incremental tools for
    the verification of matching logic properties in non-incremental
    settings?}
\item  \label{rq2} \emph{How effective is the incremental speedup
    provided by our approach and which major factors influence it?}

\end{compactenum}

\subsection{Implementation and Experimental Settings}

\scmc  is developed in Scala 2.11.6 and Java
8 and is rooted in \sidecar, our general framework for
implementing syntax-based incremental verification procedures
\cite{bfgm-scp2015}.

\begin{figure}[tb]
	\centering
	\includegraphics[scale=0.35]{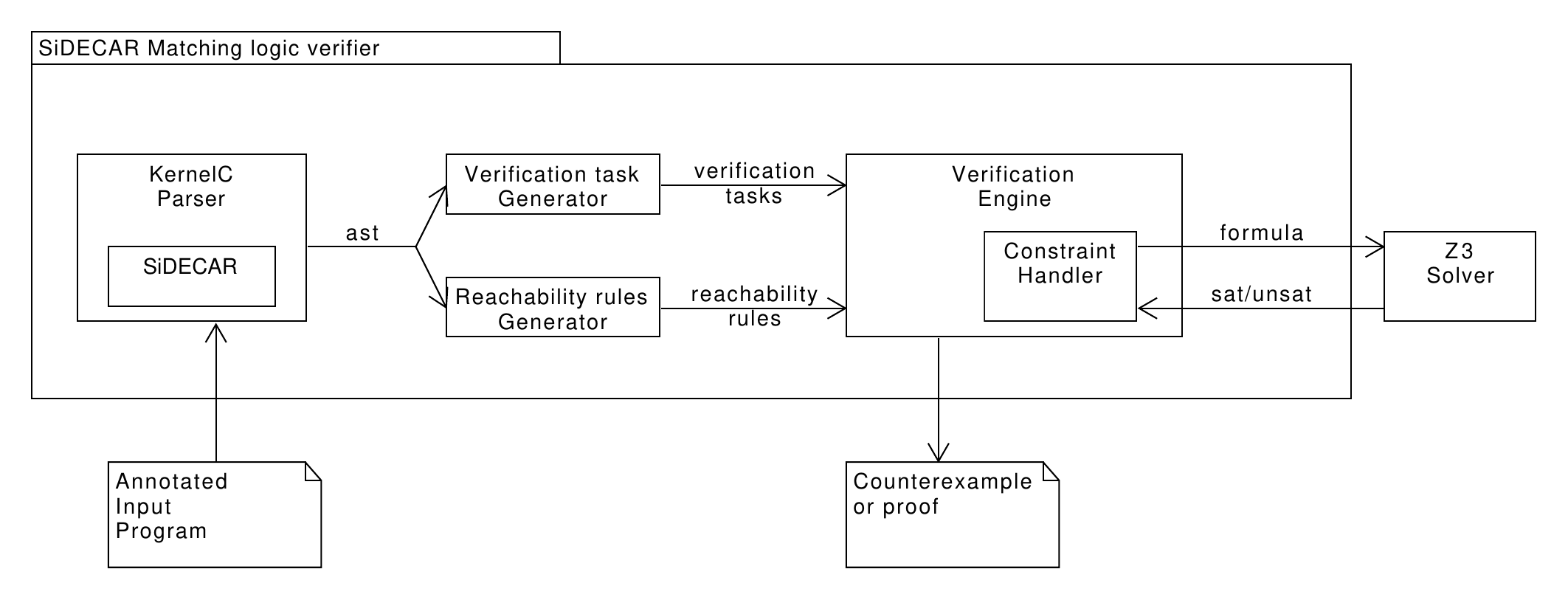}
	\caption{The architecture of \scmc}
	\label{fig:implementation}
\end{figure}

The architecture of \scmc is shown in
figure~\ref{fig:implementation}. \scmc takes as input the KernelC
program to be verified, possibly annotated with matching logic specifications.
The input program is parsed by an incremental KernelC
operator-precedence parser built on top of \sidecar.  The resulting
abstract syntax tree is processed by two generator components, which generate
the reachability rule instances describing the semantics of the
program, as well as the verification tasks; these generators work
according to the \textsc{CompAttribute} algorithm described in
section~\ref{sec:reach-check-thro} (figure~\ref{fig:algo2}).  The
verification tasks are then performed by the \textit{Verification
  Engine}, which also takes as input the reachability rule instances; this
engine implements the \textsc{Process} and \textsc{Check} algorithms
(figure~\ref{fig:algo}).  During the verification process, the
reachability of configurations is checked by querying an
external SMT solver (Z3~\cite{DeMoura:2008:ZES:1792734.1792766}); the
\textit{Constraint Handler} component translates configuration
patterns into SMT formulae.  The SMT solver is also used to check
whether the final configuration pattern satisfies the verification
task contract. 
We chose the Z3 solver because it is a fairly well-known, widely adopted and available tool; notice, however, that it is an "external black box" in our architecture, so that it could be easily replaced by, or integrated with, other possible solvers with complementary features.

The experiments are conducted on a Docker container run on a machine equipped with an Intel
Core i5 \SI{3.2}{\giga\hertz} processor, \SI{32}{\giga\byte} memory,
running  a \SI{64}{\Bit} GNU/Linux OS distribution. We use 
 \matchc revision 565\footnote{Taken from
   \url{https://code.google.com/archive/p/matching-logic/source/default/commits}.}
 and Z3 version 4.4.
The execution times reported in the following sections are the average
over 5 executions, which we deemed as enough to produce fairly stable average values.
Due to the warm-up of the Docker containers involved,
the results of the first run of the experiments in a session
have been discarded. We also point out that the parallel execution
features available in the tool (see
Section~\ref{sec:improveIncrementality}) have been purposely disabled
to obtain data that depend exclusively on incrementality, even at the
expenses of some loss in terms of absolute performance; we
will comment on the further benefits that can be obtained by combining
incrementality with parallelism in the conclusions (Section~\ref{sec:conclusion}).

\subsection{Benchmarks and Evaluation Settings}

We built a suite of six benchmarks to evaluate \scmc, each of them
containing various KernelC programs and modified versions;
Table~\ref{table:Benchmarks used} indicates, for each benchmark, the
number of programs, the number of lines of code (LOC) and the number
of tokens per program.

\begin{asparaenum}[\itshape MB1]
  \item \label{mb1} contains four groups of sample programs (``undefined'', ``simple'',
    ``list'', ``sorting'') taken from the original
    MatchC benchmark. 
  \item \label{mb2} has been built starting from  the seven programs of the ``simple''
    group from benchmark MB\ref{mb1}. For each program,
    we generated mutations according to an operator mutation schema
    and pruned those that could not be verified (i.e., those with
    the non-linear constraints not supported by the current SMT solver).
  \item \label{mb3} contains artificially created programs, obtained by
    concatenating different programs from benchmarks MB\ref{mb1} and
    MB\ref{mb2}. More precisely, each program in MB\ref{mb3} is the
    textual concatenation of the seven programs from benchmark
    MB\ref{mb1} used for generating benchmark~MB\ref{mb2}, where six
    programs are the original version from MB\ref{mb1} and the seventh
     is one of the corresponding mutations contained in
    MB\ref{mb2}.  This benchmark was created to compensate for the
    relatively small size of the programs in benchmark MB\ref{mb2}, so
    that the size of the modified part of each program in MB\ref{mb2}
    is a significant fraction of the whole code.
    
  \item \label{mb4} is obtained by artificially adding to a each program of
    MB\ref{mb3}  a new  function that
    calls all the functions contained in the program. This way a semantic dependency is introduced between the different functions and the incremental approach is more stressed, since it is
    not possible to reason anymore on different modules
    separately. Furthermore, by controlling the position of the
    newly inserted function, this benchmark allows us to correlate the
    performance of our approach with the characteristics of the
    changes.

    \begin{table}[t]
	\caption{Benchmarks used \label{table:Benchmarks used}}
	\centering
		\begin{tabular}{clll}
			\toprule
			Name & Number of programs & LOCs per program & Tokens per program\\ \midrule			
			{MB1} & 29  & 6--56 & 12--269\\	
			{MB2} & 228 & 6--10 & 12--269\\	
			{MB3} & 201 & 54 & 276\\	
			{MB4} & 201 & 60 & 316\\	
			{MB5} & 42  & 180 & 961--975\\	
			{MB6} & 5   & 434--450 & 2675--2808\\	
			\bottomrule
	\end{tabular}
\end{table}

  \item \label{mb5} contains programs derived from the well known Siemens
    collection\footnote{Available at \url{http://sir.unl.edu/}.}.  We
    chose the TCAS benchmark because it contains programs from a
    safety-critical domain (avionics) where the programming style is
    based on the typical features of KernelC. This benchmark has
    been widely used in the literature; e.g., ~\cite{Birch2015,Bloem2013,Konighofer2014,Chaki:2004:MVS:998488.998622}.
    We slightly adapted it to replace the C language features not supported by
    KernelC, i.e., global variables, \texttt{for} loops, and static
    arrays.  Each program in the Siemens collection comes with several
    versions; for some programs, some versions have multiple changes
    in different locations.  Although \scmc can handle multiple
    changes, we chose to modify these versions so that any change
    affects only a single point of the original program; hence the effect of each individual change
   can be analyzed independently.   This benchmark was selected to evaluate our
    approach when dealing with real-world changes generated in an
    industrial environment.  Furthermore the programs in this
    benchmark are significantly larger than those in the previous
    benchmarks; hence this benchmark can  assess the
    scalability of our approach with respect to the size of the
    program.

  \item \label{mb6} contains five versions of a program manipulating red-black trees,
    inspired by classic algorithms as those presented
    in~\cite{cormen2009introduction}, with additional functions for
    testing purpose. Besides the base version, we created two new
    versions.  In the first
    version, we unified the functions performing the left and right
    tree rotations into a single function with a parameter that
    specifies the rotation to be applied. In the second version, we
    modified a function that extracts the minimum element of a subtree,
    by using recursion instead of a  while-loop.
    During the development of these new versions, we unintentionally
    introduced a few errors in the code: two versions containing a bug
    were also added to the benchmark.
    
    This benchmark addresses yet another dimension to our empirical assessment: it
    stresses the effects of incrementality when changes are applied to
    fairly sophisticated algorithms. Indeed, the algorithms to manage
    red-black trees are cleverly designed and rather complex;
    moreover, their functions use heap objects extensively.

\end{asparaenum}

Benchmark MB\ref{mb1} is the only one used for answering RQ\ref{rq1},
while benchmarks MB\ref{mb2}--MB\ref{mb6} are used for answering
RQ\ref{rq2}.  Another major distinguishing feature among the various
benchmarks concerns the use of annotations.  MB\ref{mb1}--MB\ref{mb4},
being derived from the original MatchC benchmark, contain programs
annotated with matching logic specifications in the classical Hoare
style: functions' semantics is specified by pre- and
postconditions and loops are annotated with invariants.  Instead, MB\ref{mb5} and MB\ref{mb6} did not contain annotations, in the original repositories where they have been found. We added, however, a few annotations in the form of debugging assertions, to check correctness of the state at selected critical points.
Thus, for MB\ref{mb1}--MB\ref{mb4}, program verification amounts to  proving or disproving that the program
execution guarantees satisfying the program's postcondition; whereas, in case of MB\ref{mb5} and MB\ref{mb6}, the tool
considers the program as correct iff its execution does not
violate the intermediate assertions (see also the discussion at the beginning of Section \ref{sec:overview}.)

\subsection{Non-incremental verification: comparison between \scmc and MatchC}
\label{sec:comp-non-incr}

To answer RQ\ref{rq1}, we executed \scmc on benchmark MB\ref{mb1} in a
non-incremental setting, i.e., performing the verification from
scratch. We compared our execution time with \matchc~\cite{Stefanescu2014183},
the state-of-the-art tool for the verification of KernelC programs
annotated with matching logic specifications. The results are shown in Table~\ref{table:matchC}\footnote{Some preliminary results of a comparison
  between a previous version of our tool and \matchc were published earlier
  in~\cite{formalise2015}.}.
  
\result{results/matchC.csv}{matchC}{Comparison of the execution time
  of \scmc and \matchc on benchmark MB\ref{mb1} (Column ``Ratio'' is
  the ratio between the execution time of \scmc and the one 
  of \matchc)}

The results show that in most cases \scmc performs similarly to, and in some cases 
better than, \matchc. We further investigated the cases in which \scmc is slower
than \matchc (i.e.,
\texttt{7 comm\_assoc\_1\_0.c}, \texttt{13 filter\_3\_0.c}, and the
five programs of the \texttt{sorting} group): we notice that a large amount of the execution time is
spent in constraint solving (indicated in column \emph{\scmc
  Solver}). Indeed, the type of constraints generated for these programs can
be solved more easily through a rewrite-based approach, as in \matchc
with Maude, than through an SMT solver, as in \scmc.

The answer to RQ\ref{rq1} is that in the majority of the cases the
overhead introduced by the machinery to support our syntax-based
incremental approach is negligible, yielding execution times for 
verification \emph{comparable} to a state-of-the-art non-incremental
approach. However, for programs whose verification entails
solving complex constraints, the execution time can be negatively
affected by the chosen SMT solver.

\subsection{Assessing the benefits of incremental verification}

To answer RQ\ref{rq2}, we executed \scmc on benchmarks
MB\ref{mb2}--MB\ref{mb6}. For each program in these benchmarks, we
consider the  initial version $v_0$ and several variations thereof,
$v_1,\dots,v_i, \dots, v_n$;  for each program version $v_i$,
we compared the time needed to verify $v_i$ from scratch with the time
spent for the incremental verification (based on the original verification of
$v_0$).

\subsubsection*{Benchmark MB\ref{mb2}}

\aincrementalResult{results/singleSimple.csv}{singleSimple}{Execution
  time and speedup for the verification of the programs in benchmark MB\ref{mb2}}

Table~\ref{table:singleSimple} reports the results obtained for
MB\ref{mb2}.  Column \textit{avg.\ initial} indicates the average
execution time for the (non-incremental) verification of the initial
version $v_0$ of the program; column \textit{avg.\ w/o incr.} indicates
the average execution time for the verification from scratch (i.e.,
without using incrementality) of the various modified versions of the
original program; column \textit{avg.\ w/incr.} indicates the average
execution time for the incremental verification (starting from the
results of the initial version) of the various modified versions of
the original program; columns \textit{avg.\ solv.\ initial},
\textit{avg.\ solv.\ w/o incr.}, and \textit{avg.\ solv.\ w/incr.} indicate the average time taken by the SMT solver during the
verification of the initial version, the non-incremental verification,
and the incremental verification, respectively; column
\textit{speedup} indicates the ratio between the average time for
non-incremental verification and the one for incremental verification.

For the first three programs the speedup is  limited and the
incremental verification time is not significantly different from the
verification time of $v_0$.  This is not surprising since these
programs consist of few lines of code (corresponding to circa 30 tokens);
this means that for any change almost the entire program has to be
re-evaluated (about 10--20 tokens were re-parsed for every change),
minimizing the benefits of incremental verification.  For the next
three programs, however, we can already see a significant speedup
despite their relatively small size.  The last program,
\texttt{7_comm_assoc}, has the lowest speedup. One can
notice that most of the verification time is spent by the SMT solver,
to verify both the original version and also the
modified versions; this is due to the complex constraints generated
for this program, especially those containing non-linear operators.

\subsubsection*{Benchmark MB\ref{mb3}}
Table~\ref{table:multiSimple} shows the execution time and the speedup
for the verification of the programs in benchmark MB\ref{mb3}. Each
row, except the first one indicating the initial version,
corresponds to a group of programs containing an operator mutation in
the function indicated in the row name; the columns have the same
meaning as in Table~\ref{table:singleSimple}.

The results show that the incremental verification time is always
lower than the non-incremental one by an order of magnitude, with an average speed-up of about
15. It is interesting to remark the sharp difference between the
verification time for the programs with changes in function
\texttt{7_comm_assoc} and programs with changes in other
functions (rows 2--7).  Since most of the verification time is spent
in verifying function \texttt{7_comm_assoc}, when the change does not
affect this function, the incremental verification is extremely
fast. On the other hand, when the change affects function
\texttt{7_comm_assoc}, the non-incremental verification time is often
reduced, while the incremental verification time is increased.  The
non-incremental verification time is reduced because a random mutation
in function \texttt{7_comm_assoc} can frequently introduce a violation
that is easy to detect, although the verification of programs with
changes to this function is generally heavier than the verification of
programs with changes to other functions.  In fact, the global average
non-incremental verification time and the solver time are mainly
determined by the few cases in which changes to function
\texttt{7_comm_assoc} make the verification non-trivial.

\incrementalResult{results/MB3.csv}{multiSimple}{Execution
  time and speedup for the verification of the programs in benchmark MB\ref{mb3}}

\subsubsection*{Benchmark MB\ref{mb4}}
The different functions in the programs in benchmark MB\ref{mb3} are
independent: if a function is modified, there is no need to
re-evaluate the others. This is the best-case scenario for incremental
verification and indeed we obtained a high speedup value.
To assess the performance of incremental verification also in a less
favorable scenario, we verified the programs in benchmark MB\ref{mb4},
in which a new function calls all subprograms.  In this new scenario,
the overall correctness depends on all subprograms; this dependency
necessarily penalizes the incremental verification time.  The results
are shown in Table~\ref{table:multiSimple1}: the
incremental approach is still quite effective --- achieving  an
average speedup of 4.69 --- despite the intra-procedural dependencies
in the programs.

\incrementalResult{results/MB4.csv}{multiSimple1}{Execution
  time and speedup for the verification of the programs in benchmark MB\ref{mb4}}

\tcasIncrementalResult{results/tcas.csv}{tcas}{Characterization of the
  change, execution time, and speedup for the verification of the
  programs in benchmark MB\ref{mb5}}

\subsubsection*{Benchmark MB\ref{mb5}}
Table~\ref{table:tcas} shows the results for the verification of the
programs in benchmark MB\ref{mb5}. In addition to the columns used in
Table~\ref{table:singleSimple}, column \textit{Error} indicates
whether a program version introduces an error; column \textit{Change
  Type} denotes the type of change applied to each version (see legend
at the bottom of the table for details); column \textit{StCh Position}
represents the position (expressed in percentage) where the program
state differs from the original run (not to be confused with the
position of the change in the source code).

In this case, the overall result in terms of speedup is much lower than
in the previous benchmarks, although still appreciable and in some
cases quite relevant.  The explanation for this performance gap lies
in the most distinctive difference between MB\ref{mb5} and
MB\ref{mb1}--\ref{mb4}: the former does not contain any Hoare-style
annotation and thus the verification is rather in the debugging style.   In
fact, the lack of separate and well-structured annotations for key
program fragments, such as functions and loops, makes the global
verification not compositional and constrains most of it to be
performed in a purely sequential way, which restarts from scratch upon
any change. Under this scenario, only the initial phase, i.e., parsing and attribute
evaluation, is incremental (to a first
approximation\footnote{More precisely, some optimizations already
  produce incremental benefits even in the following phases of
  symbolic interpretation and constraints solving; as we will discuss
  in Section~\ref{sec:conclusion}, additional optimizations are the
  object of further investigation.}).

The effects of incrementality are determined by various factors, such
as the type of the change (e.g., which operator is replaced), the
percentage of time taken by the SMT solver, whether the change
resulted into an error, and---in case of an error---its type (e.g., a
memory access violation or an assertion violation), the position of
the error in the code, and the position where the program state
differs from the original one. We do no
report a detailed analysis on the impact of these factors on the
resulting speedup, because it would be cumbersome and not conclusive
due to the interplay among all these factors; however, we enriched Table~\ref{table:tcas} with the three columns mentioned above, referring to three major factors which certainly have significant impact on the overall performances of the incremental tool.

\incrementalResultSec{results/rbtree-simple.csv}{rbtree-simple}{Incremental evaluation of \textit{MB6}}

\subsubsection*{Benchmark MB\ref{mb6}}
Table~\ref{table:rbtree-simple} shows the results for the verification
of the programs in benchmark MB\ref{mb6}. The results confirm the
observation made for the previous benchmark: the lack of Hoare-style
annotations leads to a decreased, but still relevant, effectiveness of
the incremental verification approach.  The increased algorithmic
complexity, specific to this benchmark, produces a combinatorial
explosion of the execution paths to be verified and, consequently, a
substantial increase of the execution time, both for the case of
incremental verification and for the non-incremental one. Indeed,
differently from all other benchmarks, in this case the execution
times reported in Table~\ref{table:rbtree-simple} are expressed
\emph{in seconds rather than in milliseconds}.  The lack of
compositionality in the whole verification process has also
constrained us to analyze trees with a minimum
amount\footnote{Although minimum, the size of the analyzed tree is
  still big enough so that all possible code branches are exercised.}
of nodes.

The remarkable difference between the speedup obtained for version v1
and the one obtained for version v2 can be explained in terms of the
changes applied to the two program versions and their effects. In v1,
the change is the merge of the functions for performing left and right
tree rotations.  In red-black tree algorithms, rotations are invoked
very often; when program annotations are not provided, the rotation
function has to be (symbolically) re-executed at every call. This
impacts on the total execution time, yielding a modest speedup of
1.41.  In v2, the change is the replacement of a loop by a recursive
call in function \texttt{find-minimum}. This change has a limited
impact on the overall (symbolic) execution of the program, yielding a
remarkable speedup of 28.55.

The (unintended) errors introduced in versions v3 and v4\footnote{In
  v3, the lack of a \texttt{return} statement resulted in changing the
  color of both children of a node, instead of only one; v4 does not
  handle correctly the case in which the same key is inserted several
  times.} were detected by the violation of debugging assertions that
we introduced in the code; such violations triggered the immediate
interruption of the execution but also resulted in repeating a
significant part thereof even in the case of incremental verification.

\subsubsection*{Discussion}
The above results show that our approach for incremental verification
can provide a speedup in program verification ranging from 1.06 to
109.64.  The main explanation for the wide gap between these extreme values
is the fundamental difference between the two types of verification
supported by the Matching logic approach: the
compositional, Hoare-style verification and the
more traditional, but essentially sequential, verification carried on
through symbolic execution.
Among the additional factors determining this gap we recall:
\begin{itemize}
\item \textit{Size of the program}. For small
programs, for any change almost the entire program has to be
re-evaluated, minimizing the speedup. 
\item \textit{Constraint solver}. The verification of some programs
  involves the solution of complex constraints (e.g., with non-linear
  operators), which affects the overall checking time.
\item \textit{Inter-procedural dependencies}. When a change occurs in
  a function with strong inter-procedural dependencies, these
  dependencies penalize the speedup.
\item \textit{Position, type, and consequences of the change.}  The
  position of the change (e.g., whether it is located at the beginning
  or at the end of the code), its type, and its consequences
  (e.g., whether it produces an early or a late deviation from the
  original execution flow; whether it introduces an error or a
  different but still correct behavior) certainly affect the
  verification performance; however, such an impact may be hardly
  abstracted into general rules.
\end{itemize}

Overall, the results of incremental verification derive from a
combination of these factors; as a consequence, often their effects
cannot be easily analyzed separately.  Finally, let us remark that
highly-sophisticated algorithms dramatically increase the global
complexity of verification, both for the case of non-incremental
verification and for the incremental one; in the latter, however, even
small speedups may result into high gain in absolute execution time.

 \section{Related Work}
\label{sec:related-work}

The approach presented in this paper is mainly related to work performed in
the context of matching logic and incremental verification.

The verification of programs annotated
with matching logic specifications has been investigated
in~\cite{rosu-stefanescu-2012-oopsla}, which provides a
language-independent program verification framework and an
instantiation of this framework for KernelC (as implemented in
MatchC~\cite{Stefanescu2014183}). Differently from our approach, this
work does not support incremental verification.

The problem of incremental verification has been tackled from
different angles~\cite{paper1}.  One of the most common approaches is
based on the \emph{assume-guarantee}
paradigm~\cite{Jones:1983:TST:69575.69577}, which considers the system
as a collection of cooperating modules, each of them annotated with
properties that are guaranteed to the other modules. Verification is
done in a compositional way, first by reasoning about each module
separately and then by deducing properties about the integration of
the different modules. If a change impacts only one module,
verification can be performed only on it, with no need for verifying
the other modules. This approach has been used both in the context of
model checking~\cite{cobleigh2003learning}, program analysis~\cite{calcagno11:compos:shape:analy:means:bi:abduc}, and also for the
quantitative analysis of models~\cite{KNPQ10}; it has also been
extended to support the substitutability problem
\cite{sharygina2005dynamic}. The main weakness of compositional
verification methods is their reliance on the modular structure of the
system, which cannot deal with cross-module changes and intrinsically
entails a coarse granularity at the model level. By contrast, our
approach does not require any a priori decomposition of the system
into a modular structure, since it leverages the natural structure
encoded in parse tree, which is available at any granularity level.
Furthermore, we have seen that our approach can achieve the benefits
of compositionality even independently from a modular structure by
exploiting suitable annotations.

\emph{Parametric analysis} is an approach for incremental verification
that allows for a certain degree of uncertainty (e.g., due to future
changes) in the system to verify, by replacing some actual system
parameters with symbolic placeholders. The output of parametric
analysis is an expression having as unknowns those parameters. If a
change in the system can be conveniently reflected by a new assignment
to the parameters, one can just re-evaluate the symbolic expression with
the new parameters, instead of re-executing the verification procedure
from scratch. This technique has been mainly used for the analysis of
quantitative properties~\cite{ daws2005symbolic, Hahn2010Param,
  Filieri2011a,Filieri2012FormSera}.  Parametric analysis requires
anticipated knowledge of both the global structure of the system and
the location of possible changes; furthermore, it requires an ability
to cast changeable information as system parameters. Our approach,
instead, does not require to know in advance the location of possible
changes, and supports any kind of change (not only the one related to
system parameters).

Another approach for incremental verification is based on \emph{model
  reuse}, where the idea is to reuse ``models'' constructed during the
verification of a base version for the verification of a new
version. For example, extreme model checking~\cite{HenzingerJMS03:0}
proposes to reuse a proof of correctness obtained from a previous run
and to start the new state-space exploration from the point in the
abstract state space in which the old proof cannot be further followed.  Changes are
detected by comparing the
control flow automata of the new program with the abstract
reachability tree of the previous version.
Reference~\cite{Conway05incrementalalgorithms} proposes two
incremental algorithms for automaton-based safety analysis of C/C++
programs. The idea is to first compute the synchronous product of the
control-flow graphs (CFGs) of the program functions and the automaton
of the property to verify; this product is represented as a derivation
graph. Upon a change (represented as  a list
of additions, deletions, and modifications of the program CFGs), the previous derivation graph is inspected and
the synchronous product is recomputed only for the parts affected by
the changes.  In the context of explicit-state model checking,
reference~\cite{Lauterburg:2008:ISE:1368088.1368128} proposes
incremental state-space exploration (ISSE), which saves the
state-space graph of a previous program version and examines it during
the next exploration for a new program version, to check which
transitions need to be re-executed and explored, based on the code changes made since the previous exploration.  The basic idea of
ISSE is to avoid executing unnecessary transitions and the related
computations.  Another approach for explicit-state model checking is
Regression Model Checking~\cite{conf/icsm/YangDR09} (RMC), which also
first saves the state space from the analysis of a previous version;
upon a change, it performs (dynamic) impact analysis  to identify that parts affected by
a change, identifies ``dangerous elements'' in the graph,
i.e., elements whose behavior can be affected by the change, and
prunes the safe ones.  eVolCheck~\cite{eVolCheck} is a bounded model
checker for C programs that (re)uses over-approximating summaries of
all program functions, obtained through Craig interpolation.
eVolCheck compares two versions for identifying the set of
syntactic changes, function by function, using the longest common
sub-sequence algorithm. Upon a
change, it checks whether the summaries of the functions
affected by the change are still valid over-approximations; if it is
the case the change is considered safe and the verification stops.
Otherwise, either the function summaries have to be refined or one of
the summaries is violated and the check has to be propagated up
through the call tree.
The main differences of these approaches with
ours lie in:
\begin{inparaenum}[1)]
  \item change detection: our approach exploits the
mechanism of incremental parsing for OPG;
\item granularity of the information that can be reused:  our
approach maintains information relevant to the verification (in the
form of grammar attributes) at a very fine level of granularity, for
each node of the syntax tree.
\end{inparaenum}

Other approaches for incremental verification are based on
\emph{equivalence checking} between a new program version and an old
one. For example, regression
verification~\cite{godlin2012:regression-veri} uses bounded model
checking to prove the partial equivalence of two programs (i.e., of
two sets of functions), by the analysis of the call graph. An
extension of regression verification for multi-threaded programs,
including a notion of partial equivalence for non-deterministic
programs, is presented in~\cite{chaki2012:regression-veri}.  Another
way of checking the equivalence of programs is checking only the
fragments impacted by changes, for example using change-impact
analysis; this method has been exploited in the context of incremental
symbolic execution~\cite{PersonYRK11,6405261,Guo:2016:CIS:2970276.2970332}.  The main difference of these
approaches with ours is that the latter is agnostic from the point of
view of equivalence between two versions, since our algorithm is
driven by changes at the attribute level (as determined by changes in
the parse tree).

Also related is the work addressing general approaches for defining
incremental program analyses integrated into IDEs, such as the IncA
language~\cite{Szabo:2016:IDD:2970276.2970298}. IncA is a
domain-specific language for defining efficient incremental program
analyses, which are then translated into graph patterns; the analyses
are performed using incremental graph pattern matching. This approach
is conceptually similar to our general approach
\sidecar~\cite{bfgm-scp2015,sidecar-report}, at the basis of the work
presented in this paper.

Finally, the benefits of our incremental approach can certainly be affected by suitable caching
strategies of verification results, such as those used in the context
of proof obligations for the Dafny verifier~\cite{Leino2015}; we have
not addressed this issue in this paper, but plan to investigate it
as part of future work.

 \section{Conclusions}
\label{sec:conclusion}
In this paper we applied the \sidecar technology---a general
syntactic-semantic framework for incremental verification---to make  matching
logic-based \cite{rosu2010:matching-logic:} verification of KernelC
\cite{Stefanescu2014183} programs incremental. 
More precisely, we encoded a verification procedure for KernelC
programs (annotated with matching logic properties) as (synthesized)
semantic attributes  associated with an operator precedence grammar
that defines the syntax of KernelC. In this representation, 
every attribute of an internal node of the syntax tree is a collection
of:
\begin{inparaenum}[1)]
  \item semantic rules associated with the code fragment that is the
frontier of the tree;
\item  verification tasks,  to be combined later with those of
  other subtrees.
\end{inparaenum}
Verification tasks are performed as soon as possible during the
construction of the tree until its root is reached and the whole
verification is completed (unless an error is detected before that).

The chosen application field is particularly challenging because, even
if parsing (and the associated evaluation of synthesized attributes)
can be performed locally in the case of operator precedence languages,
software execution is necessarily sequential. This peculiarity is well
caught by the matching logic formalism, which is a nice synthesis of
two major formal verification techniques: 1) Hoare's style, based on
invariants and function input-output specifications, which is
compositional but requires human ingenuity; 2) symbolic execution,
which is more sequential in nature and subject to state explosion.

We have implemented this approach in the \scmc tool and
evaluated it through a rich and structured benchmark. The evaluation
results show that our syntactic-semantic approach is at least
comparable, in terms of efficiency, with the state-of-the-art,
non-incremental tool \matchc~\cite{Stefanescu2014183}.  Furthermore,
when compared to redoing the whole verification from scratch, our
approach outperforms program re-verification after changes if the
program is fully annotated, as in Hoare' style proofs; when instead
the verification is based on sequential symbolic interpretation the
effectiveness of incrementality depends on various factors (e.g., the
position of the change and its type), often in an interdependent way,
but in any case the benefits range from minimum but still meaningful
to quite relevant.

The results obtained by this new application of \sidecar can be
empowered along two main directions.
\paragraph{Generalization.} The first experience with KernelC, which is a
significant subset of C including cumbersome language features such as
dynamic memory management, argues in favor of the possibility of
replicating it with other languages. Furthermore, the fairly large and
heterogeneous benchmark adopted gives evidence that the benefits of
incrementality can be exploited in many program categories and
application domains, although, necessarily, one must expect strong
variations in the performance of the tool depending on program
features, type of changes, and adopted verification techniques (i.e.,
whether programs are annotated).

\paragraph{Taming complexity and scalability by means of parallelism.}
There is no doubt that (semi)automatic verification suffers from
scalability problems due to the intrinsic complexity of the necessary
algorithms --- essentially the combinatorial explosion of execution
paths (see, for example, the difference between the execution times of
benchmark MB\ref{mb6} and the other benchmarks). These problems can be
alleviated by traditional approaches such as modularity and the
compositionality that can be achieved through program annotations.
Incrementality is another powerful tool which can be further endowed
by allowing for processing multiple, scattered changes simultaneously
and by pairing it with parallelism. The latter in fact is another way
of exploiting the local parsability property which has already been
applied successfully to compiler
generation~\cite{barenghi2015parallel}.
\\
The current version of \sidecar actually already integrates the
parallel parser proposed in~\cite{barenghi2015parallel} with parallel
attribute evaluation, It also supports parallel calls to the solver,
which are anticipated during attribute evaluation of the internal
nodes of the syntax tree. The combination of these techniques enables
the simultaneous support of multiple changes in parallel and in an
incremental way (as described in
Section~\ref{sec:improveIncrementality}).  Such features already
significantly improve the overall performance of the tool; however,
they have been switched off during the experimentation reported in
Section~\ref{sec:evaluation} because our goal was to evaluate the
benefits of incrementality in isolation. Nevertheless, by switching on
these parallel-enabling features, the overall ratio between \scmc and
\matchc would decrease from $117.19\%$ to $70.41\%$.  A more ambitious
way to exploit parallelism even during the---not completely
necessarily---sequential symbolic interpretation of program code is
the object of further investigation.

\bibliographystyle{ACM-Reference-Format-Journals}

\appendix
\section*{APPENDICES}
\setcounter{section}{1}
In these appendices we first provide various details about the formalization of KernelC semantics in terms of matching logic and of its reshaping to make formulas processable by our SMT solver and to increase the efficiency of incremental verification. In Appendix~\ref{sec:construct-details}, we provide the rules to manage the stack of function calls and input/output operations; then,   Appendix~\ref{sec:heapFormalization} deals with the complex operations to manage the heap, which have been presented in a fairly sketched way in the main body of the paper to make it not too cumbersome.  Appendix~\ref{sec:memIncremental} describes another technique, complementing those presented in Section~\ref{sec:improveIncrementality}, to increase the level of incrementality when verifying code with no annotations.
Finally, in Appendix~\ref{sec:attributeEvaluation}, we illustrate the attribute evaluation step for the example shown in
figure~\ref{exampleListSwapSort} in section~\ref{sec:running-example}.

\subsection{Construct-specific details}
\label{sec:construct-details}

\subsubsection{Functions}
\label{sec:functionsAppendix}

The cell $\mathit{stack}$
(see Section~\ref{sec:functions})
contains the list of active subroutines which is represented as a list of integers.
For instance, function \texttt{foo} is expressed by constant $\mathit{foo}$ and 
the configuration pattern~\eqref{eq:functionProperty} specifies that function \texttt{bar} can be called only from function \texttt{foo}.

\begin{equation}
\label{eq:functionProperty}
\left\langle
\left\langle \texttt{bar()}\ldots \right\rangle _{k}
\left\langle S\ldots \right\rangle _{\mathit{stack}} \land \mathit{head}(S)=\mathit{foo}
\right\rangle
\end{equation}

Logical function $\mathit{head}$ is used to retrieve the first element of the list,
thus \texttt{foo} must be the active function when the call to function \texttt{bar} is performed.

Stack properties are part of function contracts which are checked when the function  is called:
if the stack does not match the function specification, the function call cannot be performed, thus execution stops into a sink error configuration.

We stress the fact that a theory of lists is needed to handle stack-related properties.
In fact the call stack is seen as a list of active functions; stack properties can be expressed on such list.
It is possible to provide different theories for lists which can handle any program stack.
Our SMT solver was already equipped with the basic LISP list theory, composed of the empty list, the constructor $\mathit{cons}$, and the selectors $\mathit{car}$ and $\mathit{cdr}$ ~\cite{Oppen:1980:RRD:322203.322204}.
We have introduced additional operations/axioms to check more specific stack properties. Nevertheless, neither our extensions nor other approaches (e.g., \cite{MR0470107} ) allow to deal with list theory in its full generality.

\subsubsection{Input/output}
\label{sec:io}

KernelC provides two functions for reading and writing integer values from the input and output streams \texttt{scanf} and \texttt{printf}. The two streams are represented as lists of integers stored in the cells $\mathit{in}$ and $\mathit{out}$, respectively.

Function \texttt{scanf} takes as argument the address of a variable
\texttt{v}. This variable will  be assigned the value of the first element of  list $\mathit{in}$; such element will be removed from the list. This is formalized in rule~\eqref{eq:scanf}:

\begin{equation}
\label{eq:scanf}
\left\langle
\left\langle \frac{\texttt{scanf(}\texttt{"\%i"},\texttt{\&v}\texttt{)}}{\cdot} \dots \right\rangle _{k}
\left\langle \frac{j}{\cdot}\ldots \right\rangle _{in}
	\left\langle \texttt{v} \mapsto \frac{i}{j} \dots \right\rangle _{\mathit{env}}
\right\rangle
\end{equation}

Similarly, function \texttt{printf} takes as argument an integer value and appends it to  list  $\mathit{out}$, as described in rule~\eqref{eq:printf}:

\begin{equation}
\label{eq:printf}
\left\langle
\left\langle \frac{\texttt{printf(}\texttt{"\%i"},~i\texttt{)}}{\cdot} \dots \right\rangle _{k}
\left\langle \frac{\cdot}{\mathit{i}}\ldots \right\rangle _{out}
\right\rangle
\end{equation}

Both $\mathit{in}$ and $\mathit{out}$ contain lists of integers; thus,
configuration patterns involving input/output
can be managed by our SMT solver as explained above.

\subsection{Complete Heap formalization}
\label{sec:heapFormalization}

We now specify heap encoding and its properties.
First we formally define the basic operations which can be performed on the heap
in terms of reachability rules.
Then, we present the logical constraints that a valid heap must satisfy,
in a way manageable by an SMT solver.
Finally we show how more complex memory objects (such as lists or user-defined structures) can 
be handled by our heap model.

\subsubsection{Reachability rules for Heap operations}
\label{sec:heapOperations}
The four memory primitives available in KernelC to manipulate the heap are
$\mathit{malloc}$, $\mathit{free}$, $\mathit{write}$, and $\mathit{read}$.

Operation $\mathit{malloc}(i)$ allocates $i$ contiguous memory blocks
starting from a new location $l$. Rule~\eqref{eq:genMalloc} shows the
semantics of this primitive: $\textit{mem}$ is extended by mapping $i$
consecutive locations, starting from $l$, to $i$ empty blocks
($\bot$); the mapping $l\mapsto i$ is added to $\textit{ma}$ to record
the number of blocks allocated starting from $l$.

\begin{equation}
\label{eq:genMalloc}
\left\langle
\left\langle \frac{\texttt{malloc(} i \texttt{)}}{\texttt{(void*)} l} \dots \right\rangle _{k}
\left\langle \frac{\cdot}{l \mapsto i} \dots \right\rangle _{\textit{ma}}
\left\langle \frac{\cdot}{l \mapsto\mathit{\bot}, \ldots, l+i-1\mapsto\mathit{\bot}} \dots \right\rangle _{mem}
\right\rangle
\end{equation}

Operation $\mathit{free}(l)$ removes the sequence of consecutive
locations allocated starting from the initial address $l$; the number
of locations to free is determined according to the content of cell $\mathit{ma}$.
Rule~\eqref{eq:genFree} shows the semantics of the $\mathit{free}$ primitive.

\begin{equation}
\label{eq:genFree}
\left\langle
\left\langle \frac{\texttt{free(} l \texttt{)}}{\cdot} \dots \right\rangle _{k}
\left\langle \frac{l \mapsto i}{\cdot} \dots \right\rangle _{\mathit{ma}} \right.\\\left.
\left\langle \frac{l \mapsto\mathit{\bot}, \ldots, l+i-1\mapsto\mathit{\bot}}{\cdot} \dots \right\rangle _{mem}
\right\rangle
\end{equation}

Operation $\mathit{write}(l, j)$ stores the value $j$ at location
$l$. If $l$ is not allocated before the write operation (i.e., it does
not appear in cell $\textit{mem}$), the sink error configuration is reached.
The semantics of $\mathit{write}$ is described in
rule~\eqref{eq:genMemWrite}, where the value $h$ (which can possibly be
the empty block $\bot$) stored at location $l$ is replaced by $j$.

\begin{equation}
\label{eq:genMemWrite}
\left\langle
\left\langle \frac{\mathit{write}(l,j)}{\cdot} \dots \right\rangle _{k}
\left\langle l \mapsto \frac{h}{j} \dots \right\rangle _{mem}
\right\rangle
\end{equation}

Operation $\mathit{read}(l)$ retrieves the content of the block at 
location $l$. If the requested address is not allocated, the execution
will move to a sink error configuration. The semantics of $\mathit{read}$ is shown
in rule~\eqref{eq:genMemRead}.

\begin{equation}
\label{eq:genMemRead}
\left\langle
\left\langle \frac{\mathit{read}(l)}{j} \dots \right\rangle _{k}
\left\langle l \mapsto j \dots \right\rangle _{mem}
\right\rangle
\end{equation}

When the $\mathit{read}$ and $\mathit{write}$ operations are executed through the C dereferencing operator~\texttt{*},
the corresponding reachability rules become rules~\eqref{eq:readOp} and~\eqref{eq:writeOp}
respectively.
\begin{equation}
\label{eq:readOp}
\left\langle
\left\langle \frac{\texttt{*}i}{\mathit{read}(i)} \dots \right\rangle _{k}
\right\rangle
\end{equation}

\begin{equation}
\label{eq:writeOp}
\left\langle
\left\langle \frac{\texttt{*}i\texttt{ = }j\texttt{;}}{\mathit{write}(i, j)} \dots \right\rangle _{k}
\right\rangle
\end{equation}

\subsubsection{Expressing memory-related constraints through the SMT solver}
\label{sec:memoryConstraints}

The contents of a configuration pattern $\mathit{mem}$ 
must satisfy certain regularity properties to be a valid heap state.
These properties are enforced as additional constraints,
which are added to the logic formula passed to the SMT solver.
In fact the solver evaluates a formula $F_{S}$ composed by two parts:
$F_{S}=F_c\land C_h$ where $F_c$ is the original configuration formula and $C_h$ is the conjunction of the constraints
derived from the heap.
The constraints required for heaps are given below:

\paragraph{NULL value}
The C \texttt{NULL} constant points to a non-existing memory cell.
For consistency with C semantics we use for it the value $0$. Of course, it should never happen that a location with value $0$ is allocated in the heap. To ensure this property we need the following logical constraint:

\begin{equation}
\label{eq:C0}
C_0=\forall ((i \mapsto j) \in m:\textit{mem}). i\neq 0
\end{equation}

Since every $\textit{mem}$ structure has a finite number of elements, constraint~\eqref{eq:C0} is 
instantiated in our solver as a finite logical $\land$; e.g.,
in a $\mathit{mem}$ cell containing $n$ locations with addresses $h_1,\ldots,h_n$ , 
constraint~\eqref{eq:C0} would be:
\begin{equation}
C_0 = \bigwedge\limits_{i=1}^n h_i\neq 0
\end{equation}

\paragraph{Heap separation}
We also have to ensure that the heap cannot have different entries at the same location, as specified by the logical constraint

\begin{equation}
\label{eq:C1}
C_{hs}=\forall ((i \mapsto a) \in m:\mathit{mem}), ((j \mapsto b) \in m:\mathit{mem}). 
(i = j \implies (i \mapsto a) \equiv (j \mapsto b))
\end{equation}

The $\mathit{distinct}$ function provided by the Z3 SMT solver,
which assesses that its arguments are distinct,
has been exploited to join in one constraint  both logical constraints~\eqref{eq:C0} and~\eqref{eq:C1}:

\begin{equation}
C_{0is} = \mathit{distinct}(0, h_1,\ldots,h_n)
\end{equation}

\paragraph{Positive memory locations}

Usually the memory locations allowed by the C semantics 
are only positive integers.
Thus, it may be useful to include the following constraint (for convenience, formulated in an abstract version making use of a quantifier):

\begin{equation}
C_{+}=\forall ((i \mapsto a) \in m:\mathit{mem}). i > 0
\end{equation}

 \subsubsection{Arrays and structs}
\label{sec:structs}

Arrays in C are just ''syntactic sugar'' for pointer arithmetics and have not been included in the original KernelC syntax; our version of the language, instead, does include syntactic constructs to use arrays but they are limited to the case where their elements are integers and therefore have unitary size: consistently, the access to an array element is formalized through rule~\eqref{eq:arrayReadOp}:

\begin{equation}
\label{eq:arrayReadOp}
\left\langle
\left\langle \frac{\texttt{array[index]}}{*(\texttt{array} + \texttt{index})} \dots \right\rangle _{k}
\right\rangle
\end{equation}

Structs too are implemented as contiguous memory locations.
In such a case we must provide access to a location with
a certain offset using an identifier. For instance, with reference to the typical \texttt{->} operator,
reachability rule~\eqref{eq:structExample} computes the actual address of a struct field, whereas rule~\eqref{eq:writeStruct} formalizes the assignment to a struct's field.

\begin{equation}
\label{eq:structExample}
\left\langle
\left\langle \frac{\mathit{p}\texttt{->field}}{l} \dots \right\rangle _{k}
\right\rangle \land l=\mathit{p}+\mathit{fieldOffset}
\end{equation}

\begin{equation}
\label{eq:writeStruct}
\left\langle
\left\langle \frac{\texttt{ptr->field=}i\texttt{;}}{\mathit{write}(\texttt{ptr->field}, i)} \dots \right\rangle _{k}
\right\rangle
\end{equation}

\subsubsection{Struct and Typing}

Since KernelC has only one integer type it seems perfectly reasonable
to avoid storing the type of each logical variable.
To handle structs, however, this need arises,
since, e.g., there could be multiple struct definitions
with the same field at different positions.
On the other hand, there is no way for reachability rule~\eqref{eq:structExample}
to associate a particular struct definition to pointer $p$,
unless extending logical variables with a type.

To handle these situations,
we introduced an optional type for integer logical variables,
forming a pair $(\mathit{type},\mathit{value})$ which can be used 
along traditional $\mathit{value}$ integer logical variables.
A typed value can discard its type to match an untyped variable,
but not vice-versa.
The only reachability rules which are affected by types
are variable declaration, evaluation, assignment~(see Table~\ref{eq:basicConstructs})
and $\mathit{read}$ operation (see reachability rule~\eqref{eq:readOp}).

Types are associated with local variables in cell $\mathit{env}$ for any
variable, as shown in reachability rule~\eqref{eq:defineStruct},
which associates type $\textit{foo}*$ to program variable \texttt{ptr}.

\begin{equation}
\label{eq:defineStruct}
\left\langle
\left\langle \frac{\texttt{struct foo* ptr;}}{\cdot} \dots \right\rangle _{k}
\left\langle \frac{\cdot}{\texttt{ptr} \mapsto (\textit{foo}*, \textit{undef})} \dots \right\rangle _{\mathit{env}}
\right\rangle
\end{equation}

Of course, during program variable evaluation the associated type
(if present) must be retrieved,
as shown in reachability rule~\eqref{eq:evalType}.

\begin{equation}
\label{eq:evalType}
\left\langle
\left\langle \frac{\texttt{x}}{(t,a)} \dots \right\rangle _{k}
\left\langle \texttt{x} \mapsto \mathit{(t,a)}\dots \right\rangle _{\mathit{env}}
\right\rangle
\end{equation}

The type of local program variables in $\mathit{env}$ 
is preserved, as shown in reachability rule~\eqref{eq:assignmentType}.
\begin{equation}
\label{eq:assignmentType}
\left\langle
\left\langle \frac{\texttt{x = }(s,a)}{(s,a)} \dots \right\rangle _{k}
\left\langle \texttt{x} \mapsto \frac{(t,b)}{(t,a)}\dots \right\rangle _{\mathit{env}}
\right\rangle
\end{equation}
In this case even though type $s$ of expression $a$ is maintained in cell $k$,
type $t$ is not overwritten in cell $\mathit{env}$.

Logical variable types are not stored in the $\mathit{mem}$ cell.
However in $\mathit{read}$ operation (reachability rule~\eqref{eq:genMemRead}),
the type of the address is maintained as the type of the value read,
as shown in rule~\eqref{eq:genMemTypeRead}.
\begin{equation}
\label{eq:genMemTypeRead}
\left\langle
\left\langle \frac{\mathit{read}((t,l))}{(t,j)} \dots \right\rangle _{k}
\left\langle l \mapsto j \dots \right\rangle _{mem}
\right\rangle
\end{equation}

Logical variable types are used to trigger a specific reachability rule to access struct fields: rule~\eqref{eq:structExample1} extends  rule~\eqref{eq:structExample} to 
provide access to field \texttt{field} of struct \texttt{structPtr} in the case of a typed variable.
\begin{equation}
\label{eq:structExample1}
\left\langle
\left\langle \frac{(\mathit{structPtr},\mathit{p})\texttt{->field}}{(\mathit{fieldType},l)} \dots \right\rangle _{k}
\right\rangle \land l=\mathit{p}+\mathit{fieldOffset}
\end{equation}
The type of the field value $\mathit{fieldType}$
is set for the field location,
which, after applying the $\mathit{read}$ operation of
reachability rule~\eqref{eq:genMemTypeRead} will be
transferred to the read value.

We also introduced a casting operator,
allowing to change the type of an object.
Its semantics is shown in reachability rule~\eqref{eq:castExample}.
\begin{equation}
\label{eq:castExample}
\left\langle
\left\langle \frac{\texttt{(otherType)}o}{(\mathit{otherType},o)} \dots \right\rangle _{k}
 \right\rangle
\end{equation}
In this case $o$ type is changed to $\mathit{otherType}$.

Finally, we define the behavior of
operator \texttt{sizeof}, through reachability rule~\eqref{eq:genericSizeOf}.
\begin{equation}
\label{eq:genericSizeOf}
\left\langle
\left\langle \frac{\texttt{sizeof(struct structName)}}{S} \dots \right\rangle _{k}
\right\rangle
\end{equation}

The rule computes the number of memory blocks
required by a \texttt{structName} object.
The value $S$ is a constant computed from the struct declaration
as the sum of the sizes of its fields.

\subsubsection{Storing data structures}
\label{sec:complexObjects}

Arrays and records are stored in memory as sequences of contiguous memory cells; thus we define a reachability rule that associates to a variable identifier a finite sequence of contiguous locations, starting from a base address as shown in cell~\eqref{eq:complexObj}.
\begin{equation}
\label{eq:complexObj}
\left\langle x \mapsto [a,b,c] \right\rangle_\textit{mem}
\end{equation}
The contents of cell~\eqref{eq:complexObj} is equivalent to cell~\eqref{eq:complexObjEquivalent}.
\begin{equation}
\label{eq:complexObjEquivalent}
\left\langle x \mapsto a, x+1 \mapsto b, x+2 \mapsto c \right\rangle_\textit{mem}
\end{equation}
Cell~\eqref{eq:complexObj} is more abstract and useful to write specifications;
cell~\eqref{eq:complexObjEquivalent} is lower-level and useful to execute a program.

As usual, a $\mathit{list}$ element consists of a struct with two fields:
the first field is the element's value, the second one is the pointer to the next element.
To illustrate how lists are managed in our implementation, consider cell~\eqref{eq:complexList}
\begin{equation}
\label{eq:complexList}
\left\langle x \mapsto \mathit{list}(A) \right\rangle_\textit{mem}
\end{equation}
where $A$ is a logical variable representing a list of elements, which
abstracts away from implementation details such as the precise address of each element.
This information, however, is necessary to execute a program using such data structure;
thus, to access a single element of the list,  an equivalent representation of the list must be built as a sequence of locations.

A naive solution would be to perform this conversion
at the beginning of a verification task by ``statically allocating'' a fixed number of  cells for any list.
In the example of cell~\eqref{eq:complexList}, if $A$ is composed by elements $[4,2]$,
the resulting heap for $\textit{mem}$~\eqref{eq:complexList} would be: 
\begin{equation}
\label{eq:complexListUnrolled}
\left\langle x \mapsto 4, x+1 \mapsto i, i \mapsto 2, i+1 \mapsto 0  \right\rangle_\textit{mem}
\end{equation}
In this case read/write operations can be easily implemented but this approach has an obvious drawback: since lists are unbounded dynamic objects, we would be compelled to allocate a fixed maximum number of locations for each of them.
To overcome this limitation the unrolling of such objects is delayed to the moment in which an operation on the object is required.
In the above example we have the structure:
\begin{equation}
\left\langle x \mapsto \mathit{list}([4,2])  \right\rangle_\textit{mem}
\end{equation}
When an operation on the locations $x$ or $x+1$ is required,
the heap is expanded as follows:
\begin{equation}
\left\langle x \mapsto 4, x+1 \mapsto i, i \mapsto \mathit{list}([2])  \right\rangle_\textit{mem}
\end{equation}
One step of unrolling is performed: the first location is expanded;
the rest of the list is pointed by $i$.
In this case $i$ is a fresh logical variable, distinct from any other location in the heap.

A subsequent access to $i$ or $i+1$ locations
would cause the complete extension of the list as follows:
\begin{equation}
\left\langle x \mapsto 4, x+1 \mapsto i, i \mapsto 2, i+1 \mapsto 0 \right\rangle_\textit{mem}
\end{equation}
In this way infinite objects can be handled by performing successive unrolling steps 
only when needed,
requiring only the minimum number of locations in the heap.

A second problem arises when a final configuration pattern is reached: in order to verify the list contents against a possible postcondition it may be necessary to fold the heap back, if the final configuration pattern contains
list object in the more abstract format.

As a particular case let us consider the original example of $\textit{mem}$ cell~\ref{eq:complexList},
when $A$ is (or becomes) the empty list.
In this situation the entry $x \mapsto \mathit{list}(A)$
should be removed from the heap since the list does not 
occupy any cell and a constraint stating $x=0$ should be added (to be consistent with C semantics).

\subsection{Increasing compositionality in memory management}
\label{sec:memIncremental}
In Section \ref{sec:improveIncrementality} we have emphasized that the intrinsically sequential program execution may compel to concentrate most of the computational effort in the syntactic nodes close to the root, with serious risks of loss of efficiency, mainly during incremental re-evaluation. In that section we mentioned a few techniques to ``compact'' reachability rules propagated from children nodes onto a sort of ``macro reachability rules'' forwarded to the father node.
In this appendix we detail a further technique to avoid delaying many
execution steps towards the root of the syntax tree. Precisely, notice
that the previous memory access operations need a complete knowledge
of the memory contents at the time of their execution; such a
knowledge is available if we are executing a function provided with
suitable preconditions or a loop provided with an invariant, but it is lacking if such annotations are missing.

In such a case, we model the initial memory contents of the code fragment with a dummy array term~\cite{Mccarthy62towardsa} $e$
added to the $\textit{mem}$ cell,
which can be used to handle memory reads and writes.
The $\textit{mem}$ cell will also contain the 
map between memory addresses and contents of the 
memory locations allocated during the execution,
which would have a known address independent
on the rest of the heap. Thus, memory operations are specified along the following reachability rules.

\paragraph{Read/write}
We handle the $\mathit{read}$ operation in two ways depending on whether the address belongs to the newly allocated part or not.
In the former case,
as specified by semantic rule \eqref{eq:unkMemRead}, a normal read is performed.
\begin{equation}
\label{eq:unkMemRead}
\left\langle
\left\langle \frac{\mathit{read}(i)}{j} \dots \right\rangle _{k}
\left\langle e, i \mapsto j \dots \right\rangle _{mem}
\right\rangle
\end{equation}
Otherwise the result of such read is the term $\mathit{read}({j},{e})$ which depends on the dummy environment variable $e$: 
\begin{equation}
\label{eq:unkMemRead1}
\left\langle
\left\langle \frac{\mathit{read}(i)}{j} \dots \right\rangle _{k}
\left\langle e \dots \right\rangle _{\mathit{mem}}
\land i \neq 0 \land j=\mathit{read}({i},{e})
\right\rangle
\end{equation}
Note that the read array  term ($\mathit{read}({i},{e})$) is an assumption we
make on the state of the heap at the beginning of the execution.
This assumption must be discarded at the end of execution by adding a
suitable precondition to the generated rules.

Similarly, $\mathit{write}$ operations are specified by rule \eqref{eq:unkMemWrite} when they are applied to newly allocated variables:
\begin{equation}
\label{eq:unkMemWrite}
\left\langle
\left\langle \frac{\mathit{write}(i,j)}{\cdot} \dots \right\rangle _{k}
\left\langle e, i \mapsto \frac{h}{j} \dots \right\rangle_{\mathit{mem}}
\right\rangle
\end{equation}
Otherwise the operation is stored inside the term $\mathit{write}(i,j,e)$, which
keeps track of the sequence of operations on memory.
\begin{equation}
\label{eq:unkMemWrite1}
\left\langle
\left\langle \frac{\mathit{write}(i,j)}{\cdot} \dots \right\rangle _{k}
\left\langle \frac{e}{\mathit{write}(i,j,e)} \dots \right\rangle _{mem}
\right\rangle
\end{equation}

When performing read and write operations on the heap, the functional composition of
terms related to the read and write operations are always simplified
by applying the array theory proposed in reference~\cite{Mccarthy62towardsa}.
\begin{equation}
\label{eq:sameWrite}
\forall i,j,l,e.\mathit{write}(i,l,\mathit{write}(i,j,e)) = \mathit{write}(i,l,e)
\end{equation}
\begin{equation}
\label{eq:orthogonalWrites}
\forall i,j,l,e.i\neq j \rightarrow 
\mathit{write}(i,l,\mathit{write}(j,l,e)) = \mathit{write}(j,l,\mathit{write}(i,l,e))
\end{equation}
\begin{equation}
\label{eq:read}
\forall i,j,e.\mathit{read}(i, \mathit{write}(i,j,e)) = j
\end{equation}
Equation~\eqref{eq:sameWrite} states that if two nested writes are performed over the same address,
only the last one must be considered, while the previous one can be removed.
Equation~\eqref{eq:orthogonalWrites} states that $\mathit{write}$ occurring at different addresses can be switched
and equation~\eqref{eq:read} states that the value of a $\mathit{read}$ at the same address of a $\mathit{write}$ holds the last written value.

\paragraph{Malloc/free}
The $\mathit{malloc}(i)$ is handled by adding a new mapping between memory addresses and location contents
(using reachability rule~\eqref{eq:genMalloc}),
since the newly generated addresses will never collide with the older ones.
The $\mathit{free}$ operation, instead, can only be handled if the address to be freed belongs to the
new part of memory.

\subsubsection{Fixing the heap}

Once the final configuration patterns have been obtained,
the precise conditions of the heap state at the beginning must be met.
First, all the memory locations affected by a read or a write must be disambiguated,
obtaining the set of distinct memory locations by splitting undefined addresses,
potentially generating new configuration pattern.
Then it is possible to build the full heap of the initial configuration pattern.
The unknown heap obtained in the final configuration pattern is taken and
all the $\mathit{write}$ operations are removed,
by transforming a heap in the form $\mathit{write}(i,j,e)$
to $e$ and producing a pair $i \mapsto j$ until the resulting heap
does not contain any $\mathit{write}$ operations.
All this pairs are added to the initial configuration pattern $\mathit{heap}$ cell.
Finally all the $\mathit{read}$ operation terms in the final configuration
can be resolved as before providing a symbolic value.

Notice that the technique illustrated in this appendix introduces some
level of compositionality even in case of code fragments partially
annotated or with no annotations.
In particular, portions of identical computations, where the unknown initial memory contents are represented by symbolic parameters, can be saved and further reused when instantiating such parameters to concrete values.

 \subsection{Attribute evaluation}
\label{sec:attributeEvaluation}
We illustrate how attributes are evaluated for the most significant
subtrees of the example shown in figure~\ref{exampleListSwapSort} in
section~\ref{sec:running-example}.  We recall that throughout the
example we use the abbreviations for code/environment snippets shown
in figure~\ref{eq:abbrv} on page~\pageref{eq:abbrv}; the reachability
rules used for attribute evaluation are shown in
Table~\ref{tb:reachabilityRules} on
page~\pageref{tb:reachabilityRules}, with a reference to the nodes
where they are generated.

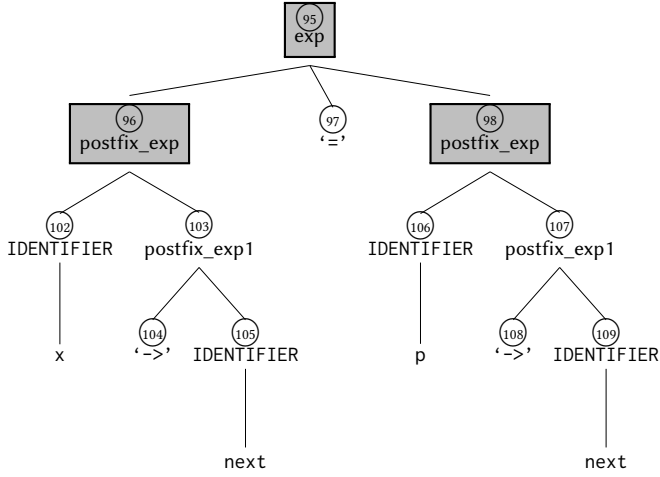
\begin{figure}
	\centering
	
	\begin{tikzpicture}[scale=0.8,level distance=50pt]	
	\Tree[.\refnt{exp}{95}
	[.\refnt{postfix_exp}{96}
	[.\tn{IDENTIFIER}{102} \texttt{x} ] [.\nt{postfix_exp1}{103}
	\tn{`->'}{104} [.\tn{IDENTIFIER}{105} \texttt{next} ]
	]]
	\tn{`='}{97} 
	[.\refnt{postfix_exp}{98}
	[.\tn{IDENTIFIER}{106} \texttt{p} ] [.\nt{postfix_exp1}{107}
	\tn{`->'}{108} [.\tn{IDENTIFIER}{109} \texttt{next} ]
	]]
	]	
	\end{tikzpicture}	
	\caption{Subtree representing the statement at line~\ref{line:xNextAssign} of the \texttt{swap} function}
	\label{fig:evalAssign}
\end{figure}

\paragraph{\parnt{exp}{95} (figure~\ref{fig:evalAssign}): variable evaluation and assignment}
The semantics is expressed by reachability rule instance~\eqref{eq:assignXnext},
which performs the assignment of value \texttt{p->next} to \texttt{x->next}.
This rule instance relies on both the computation of the actual 
destination address and on the evaluation of \texttt{p->next}.

Since pointers are involved, memory operations must be considered; we
use memory primitives\footnote{ In this example we show a simplified
  version of the memory write rules; see
  appendices~\ref{sec:construct-details}
  and~\ref{sec:heapFormalization} for the complete
  formalization.}~$\mathit{read}$ and $\mathit{write}$ to access and
modify memory.  Operation $\mathit{read}(i)$ retrieves the value
stored at address $i$ in cell $\mathit{mem}$; its semantics is defined
by  reachability rule instance~\eqref{eq:memRead} in
table~\ref{tb:reachabilityRules}.  Note that the type $t$ of address
$i$ is used to specify the type of the value $j$, to guarantee that
the memory location contains a value of the expected type.  Operation
$\mathit{write}(i, j)$ modifies the $\mathit{mem}$ cell by writing
value $j$ at address $i$; its semantics is defined by reachability
rule instance~\eqref{eq:memWrite}, which writes to memory by replacing
the previous value stored at address $i$ with value $j$.

Reachability rule instance~\eqref{eq:evalX} in
table~\ref{tb:reachabilityRules} describes the evaluation of variable
\texttt{x}.  It looks up for variable \texttt{x} in memory and places
it as a variable in cell $k$.  The value of \texttt{x} is a tuple
consisting of the variable type and its actual value.  Both tuple
elements are used in the evaluation, since they are needed to compute
the actual addresses of fields \texttt{val} and \texttt{next}.
Reachability rule instance~\eqref{eq:evalP} works in the same way for
variable \texttt{p}.
Reachability rule instance~\eqref{eq:evalX}~(and \eqref{eq:evalP}) are
propagated from nodes \skipnt{IDENTIFIER}{102} and \skipnt{IDENTIFIER}{106},
respectively.

The reachability rule instance responsible for computing the offset of
the \texttt{struct} field \texttt{next} in variable \texttt{x} is
generated inside the \texttt{struct listNode} definition.

\begin{figure}
	\centering
	\begin{tikzpicture}[scale=0.8,level distance=40pt]
	\Tree[.\refnt{stm}{101}
	\tn{`return'}{120} [.\nt{id}{121}
	[.\tn{IDENTIFIER}{122} \texttt{p} ]
	] 
	]			
	\end{tikzpicture}
	\caption{Subtree representing the statement at line~\ref{line:retP} of the \texttt{swap} function}
	\label{fig:returnSemantics}
\end{figure}
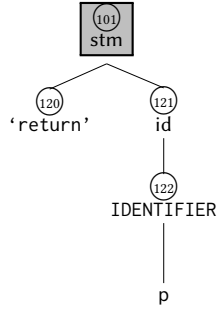

\paragraph{\parnt{stm}{101} (figure~\ref{fig:returnSemantics} ): return}
Its semantics  is defined by reachability rule instance~\eqref{eq:return} in table~\ref{tb:reachabilityRules}.
The rule leaves in cell $k$ the address pointed by \texttt{p}
(represented by the tuple $(t,i)$).
The evaluation of \texttt{p} is handled by reachability
rule instance~\eqref{eq:evalP}, which is received from \textnt{id}{121}.

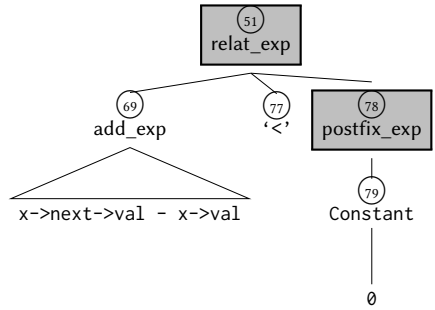
\begin{figure}
	\centering	
	\begin{tikzpicture}[scale=0.8,level distance=40pt]
	\Tree[.\refnt{relat_exp}{51}
	[.\nt{add_exp}{69} \edge[roof]; \texttt{x->next->val - x->val} ]
	\tn{`<'}{77} [.\refnt{postfix_exp}{78}
	[.\tn{Constant}{79} \texttt{0} ]
	]
	]
	\end{tikzpicture}
	\caption{Subtree representing the statement at line~\ref{line:ltComp} of the \texttt{swap} function}
	\label{fig:ifCondition}
\end{figure}

\paragraph{\parnt{relat_exp}{51} (figure~\ref{fig:ifCondition}): if statement}
The evaluation of the condition of this statement requires to retrieve
the values of \texttt{x->next->val} and \texttt{x->val}, which are  obtained through reachability rule instances~\eqref{eq:evalX} and~\eqref{eq:memRead} in table~\ref{tb:reachabilityRules}.
The fields \texttt{val} and \texttt{next} are
evaluated with the reachability rules from the \texttt{struct}
definition.
Reachability rule instance \eqref{eq:doSub} performs the actual subtraction between the two variables.

\begin{figure}
	\centering	
	\begin{tikzpicture}[scale=0.8,level distance=40pt]
	\Tree[.\nt{parameter}{36}
	[.\nt{type}{37}
	\tn{`struct'}{38} [.\tn{IDENTIFIER}{39} \texttt{listNode} ]
	] \tn{`*'}{40} [.\nt{id}{41}
	[.\tn{IDENTIFIER}{42} \texttt{p} ]
	]
	]
	\end{tikzpicture}
	\caption{Subtree representing the statement at line~\ref{line:pDecl} of the \texttt{swap} function}
	\label{fig:pDeclaration}
\end{figure}
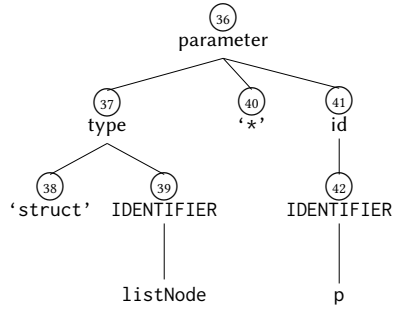

\paragraph{\parnt{parameter}{36} (figure~\ref{fig:pDeclaration}):
  declaration of  variable \texttt{p}}
The variable declaration  is performed
through reachability rule instance~\eqref{eq:defineP}, which
adds to cell $\mathit{env}$ variable \texttt{p} as a 
pointer to a \texttt{struct listNode} holding an undefined value.

\begin{figure}
	\centering	
	\begin{tikzpicture}[scale=0.8,level distance=40pt]
	\Tree[.\refnt{struct_field_list}{5}
	[.\refnt{parameter}{6}  [.\nt{type}{10}
	\tn{`int'}{11}
	] \tn{IDENTIFIER}{12} ] [.\nt{struct_field_list}{7}
	[.\refnt{parameter}{13}
	[.\nt{type}{14}
	\tn{`struct'}{15} \tn{IDENTIFIER}{16}
	] \tn{`*'}{17} [.\nt{id}{18}
	\tn{IDENTIFIER}{19}
	]
	] 		
	]	
	]			
	\end{tikzpicture}
	\caption{Excerpt from the  abstract syntax tree in
          Fig.~\ref{fig:global-tree}, corresponding to the
          declaration of \texttt{struct listNode}
          (lines~\ref{line:beginStruct}--\ref{line:endStruct} in Fig.~\ref{exampleListSwapSort})}
	\label{fig:structDefinition}
\end{figure}
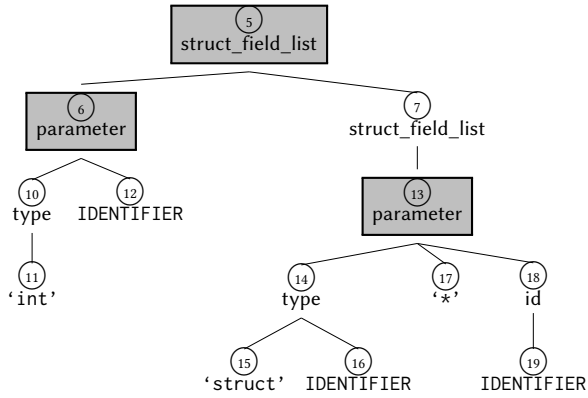

\paragraph{\parnt{struct_field_list}{5}
  (figure~\ref{fig:structDefinition}):  definition of a struct}

To handle \texttt{struct} fields data, nodes of type \snt{parameter}{}
have an attribute $N$ containing a pair
$(\mathit{name}, \mathit{type})$. Nodes of type
\snt{struct_field_list}{} have two additional attributes: $S$, an
integer containing the size of the \texttt{struct}; $F$, a list of
tuples (one for each field of the \texttt{struct}) of the form
$(\mathit{name}, \mathit{type}, \mathit{offset})$.  The former is used
to generate the rule instances that evaluate the \texttt{sizeof}
operator for the \texttt{struct}. The latter is used to build the rule
instances that retrieve the field offsets.

Reachability rule instances~\eqref{eq:structVal} and~\eqref{eq:structNext}
compute, respectively, the offsets of fields \texttt{val} and
\texttt{next}  of the \texttt{listNode struct}.
Reachability rule instance~\eqref{eq:sizeof}  is used to evaluate the \texttt{sizeof}
operator for the \texttt{listNode struct}.

The complete attribute evaluation is available in
table~\ref{tb:KattributeValues} on page~\pageref{tb:KattributeValues}.

\end{document}